\documentclass[a4paper,11pt]{article}

\usepackage{jheppub_mod} 
\usepackage[T1]{fontenc} 

\usepackage{amsmath,booktabs}
\usepackage{graphicx}
\usepackage{xspace}
\usepackage{color}
\pagecolor{white}
\usepackage[dvipsnames]{xcolor}
\usepackage{url}
\usepackage[utf8]{inputenc}
\usepackage{epstopdf}
\usepackage{hyperref}
\usepackage{multirow}
\usepackage{gensymb}

\newcommand{\mpi}{M_\pi}
\newcommand{\mpii}{M_{\pi^0}}
\newcommand{\beq}{\begin{equation}}
\newcommand{\eeq}{\end{equation}}
\newcommand{\diff}{\text{d}}
\newcommand{\eps}{\epsilon}

\newcommand{\Order}{\mathcal{O}}

\newcommand{\mw}{M_\omega}
\newcommand{\Gw}{\Gamma_\omega}
\newcommand{\mr}{M_\rho}
\newcommand{\Gr}{\Gamma_\rho}
\newcommand{\mphi}{M_\phi}
\newcommand{\Gphi}{\Gamma_\phi}

\newcommand{\epsrw}{\eps_{\omega}}
\newcommand{\gwg}{g_{\omega\gamma}}
\newcommand{\grg}{g_{\rho\gamma}}

\newcommand{\GeV}{\,\text{GeV}}
\newcommand{\MeV}{\,\text{MeV}}

\renewcommand{\Im}{\text{Im}\,}
\renewcommand{\Re}{\text{Re}\,}

\newcommand{\F}{\mathcal{F}}
\newcommand{\A}{\mathcal{A}}
\newcommand{\dilog}{\mathrm{Li}_2}

\allowdisplaybreaks[1]

\title{Isospin-breaking effects in the three-pion contribution to hadronic vacuum polarization}

\author[a]{Martin Hoferichter,}
\author[a]{Bai-Long Hoid,}
\author[b]{Bastian Kubis,}
\author[b]{and Dominic Schuh}

\affiliation[a]{Albert Einstein Center for Fundamental Physics, Institute for Theoretical Physics, University of Bern, Sidlerstrasse 5, 3012 Bern, Switzerland}

\affiliation[b]{Helmholtz-Institut f\"ur Strahlen- und Kernphysik (Theorie) and \\
Bethe Center for Theoretical Physics, Universit\"at Bonn, 53115 Bonn, Germany}

\preprint{INT-PUB-23-021}

\emailAdd{hoferichter@itp.unibe.ch}
\emailAdd{longbai@itp.unibe.ch}
\emailAdd{kubis@hiskp.uni-bonn.de}
\emailAdd{schuh@hiskp.uni-bonn.de}

\abstract{Isospin-breaking (IB) effects are required for an evaluation of hadronic vacuum polarization at subpercent precision. While the dominant contributions arise from the $e^+e^-\to\pi^+\pi^-$ channel, also IB in the subleading channels can become relevant for a detailed understanding, e.g., of the comparison to lattice QCD. Here, we provide such an analysis for $e^+e^-\to 3\pi$ by extending our dispersive description of the process, including estimates of final-state radiation (FSR) and $\rho$--$\omega$ mixing. In particular, we develop a formalism to capture the leading infrared-enhanced effects in terms of a correction factor $\eta_{3\pi}$ that generalizes the analog treatment of virtual and final-state photons in the $2\pi$ case. 
The global fit to the $e^+e^-\to 3\pi$ data base, subject to constraints from analyticity, unitarity, and the chiral anomaly,  gives $a_\mu^{3\pi}|_{\leq 1.8\GeV}=45.91(53)\times 10^{-10}$ for the total $3\pi$ contribution to the anomalous magnetic moment of the muon, of which $a_\mu^\text{FSR}[3\pi]=0.51(1)\times 10^{-10}$ and $a_\mu^{\rho\text{--}\omega}[3\pi]=-2.68(70)\times 10^{-10}$ can be ascribed to IB. We argue that the resulting cancellation with $\rho$--$\omega$ mixing in $e^+e^-\to 2\pi$ can be understood from a narrow-resonance picture, and provide
updated values for the vacuum-polarization-subtracted vector-meson parameters $\mw=782.70(3)\MeV$, $\mphi=1019.21(2)\MeV$, $\Gw=8.71(3)\MeV$, and $\Gphi=4.27(1)\MeV$.}

\begin{document}
\maketitle
	
\section{Introduction}
\label{sec:intro}

A detailed understanding of hadronic vacuum polarization (HVP) is critical for the interpretation of the anomalous magnetic moment of the muon $a_\mu=(g-2)_\mu/2$~\cite{Muong-2:2021ojo,Muong-2:2021ovs,Muong-2:2021xzz,Muong-2:2021vma,Muong-2:2006rrc},
\begin{equation}
 a_\mu^\text{exp}=116\,592\,061(41)\times 10^{-11},
\end{equation}
given that the uncertainty in the Standard-Model (SM) prediction~\cite{Aoyama:2020ynm,Aoyama:2012wk,Aoyama:2019ryr,Czarnecki:2002nt,Gnendiger:2013pva,Davier:2017zfy,Keshavarzi:2018mgv,Colangelo:2018mtw,Hoferichter:2019gzf,Davier:2019can,Keshavarzi:2019abf,Hoid:2020xjs,Kurz:2014wya,Melnikov:2003xd,Colangelo:2014dfa,Colangelo:2014pva,Colangelo:2015ama,Masjuan:2017tvw,Colangelo:2017qdm,Colangelo:2017fiz,Hoferichter:2018dmo,Hoferichter:2018kwz,Gerardin:2019vio,Bijnens:2019ghy,Colangelo:2019lpu,Colangelo:2019uex,Blum:2019ugy,Colangelo:2014qya},
\begin{equation}
\label{amuSM}
a_\mu^\text{SM}[e^+e^-]=116\,591\,810(43)\times 10^{-11},
\end{equation}
is dominated by the uncertainties propagated from $e^+e^-\to\text{hadrons}$ cross sections. Moreover, while for the second-most-important hadronic contribution, hadronic light-by-light scattering, subsequent studies in lattice QCD~\cite{Chao:2021tvp,Chao:2022xzg,Blum:2023vlm,Alexandrou:2022qyf,Gerardin:2023naa} and using data-driven methods~\cite{Hoferichter:2020lap,Ludtke:2020moa,Bijnens:2020xnl,Bijnens:2021jqo,Zanke:2021wiq,Danilkin:2021icn,Colangelo:2021nkr,Holz:2022hwz,Leutgeb:2022lqw,Bijnens:2022itw,Ludtke:2023hvz,Hoferichter:2023tgp} point towards a consistent picture in line with the evaluation from Ref.~\cite{Aoyama:2020ynm} and on track to meet the precision requirements of the Fermilab experiment~\cite{Muong-2:2015xgu,Colangelo:2022jxc}, various tensions persist for the case of HVP.\footnote{Higher-order hadronic corrections~\cite{Calmet:1976kd,Kurz:2014wya,Colangelo:2014qya,Hoferichter:2021wyj} are already under sufficient control.} 

First, the global HVP integral from the lattice-QCD evaluation of Ref.~\cite{Borsanyi:2020mff}
differs from $e^+e^-$ data~\cite{Aoyama:2020ynm} by $2.1\sigma$. Confirmation by other lattice-QCD collaborations for the entire integral is still pending, but
the stronger tension in
a partial quantity, the intermediate window~\cite{RBC:2018dos}, has been established by several independent calculations~\cite{Ce:2022kxy,ExtendedTwistedMass:2022jpw,FermilabLatticeHPQCD:2023jof,Blum:2023qou,Colangelo:2022vok}.
Second, new $e^+e^-\to\text{hadrons}$ data have become available since Ref.~\cite{Aoyama:2020ynm}, including the crucial $e^+e^-\to 2\pi$ channel. Here, the measurement by SND20~\cite{Achasov:2020iys} comports with previous experiments, but the result  
by CMD-3~\cite{CMD-3:2023alj} differs from  CMD-2~\cite{Akhmetshin:2006bx}, SND05~\cite{Achasov:2006vp},  BaBar~\cite{Lees:2012cj}, KLOE~\cite{Anastasi:2017eio}, and BESIII~\cite{Ablikim:2015orh},  at a combined level of $5\sigma$.

Neither of these tensions are currently understood, and the patterns in which the deviations occur do not point to a simple solution.\footnote{Explanations in terms of physics beyond the SM
have been considered~\cite{DiLuzio:2021uty,Darme:2021huc,Crivellin:2022gfu,Coyle:2023nmi}, but rather elaborate constructions would be required to evade other constraints on the parameter space.} That is, the relative size of the deviations in the intermediate window and the total HVP integral, together with the consequences for the hadronic running of the fine-structure constant~\cite{Passera:2008jk,Crivellin:2020zul,Keshavarzi:2020bfy,Malaescu:2020zuc,Colangelo:2020lcg,Ce:2022eix}, indicates that the changes in the cross section cannot be contained to the $2\pi$ channel alone, but that some component at intermediate energies beyond $1\GeV$ in center-of-mass energy is required. Besides further scrutiny of $e^+e^-\to\pi^+\pi^-$~\cite{Colangelo:2022lzg,Chanturia:2022rcz,Colangelo:2022prz}, 
this motivates the consideration of subleading channels such as $e^+e^-\to 3\pi$~\cite{Hoferichter:2019gzf} and $e^+e^-\to \bar K K$~\cite{Stamen:2022uqh}, which, in combination with more calculations of window observables or related quantities~\cite{ExtendedTwistedMassCollaborationETMC:2022sta}, could help locate the origin of the tensions. 

In addition, the detailed comparison to lattice QCD requires the calculation of isospin-breaking (IB) effects and other subleading corrections to the isosymmetric, quark-connected correlators. The sum of the dominant IB effects from $2\pi$, $\bar K K$, and the radiative channels $\pi^0\gamma$, $\eta\gamma$~\cite{Hoferichter:2022iqe} agrees reasonably well with Ref.~\cite{Borsanyi:2020mff}, but in the context of strong IB a larger result was observed, more in line with an inclusive estimate from chiral perturbation theory (ChPT)~\cite{James:2021sor}. Albeit consistent within uncertainties, it thus seems prudent to extend the analysis to IB in $e^+e^-\to 3\pi$, especially, since the BaBar analysis~\cite{BABAR:2021cde} reports a signal for $\rho$--$\omega$ mixing, an effect dominated by strong IB.
Apart from the direct comparison to lattice QCD, such IB corrections are also of interest for an indirect, data-driven determination of quark-disconnected contributions~\cite{Boito:2022rkw,Boito:2022dry,Benton:2023dci}. 

In this work, we address the two main IB effects in $e^+e^-\to 3\pi$. First, we study the role of radiative corrections, with the aim to provide a correction factor analogous to $\eta_{2\pi}(s)$ in $e^+e^-\to 2\pi$~\cite{Hoefer:2001mx,Czyz:2004rj,Gluza:2002ui,Bystritskiy:2005ib} that quantifies the combined correction due to virtual photons and final-state radiation (FSR) as a function of the center-of-mass energy of the process. Given that already the leading order in ChPT is determined by the Wess--Zumino--Witten (WZW) anomaly~\cite{Wess:1971yu,Witten:1983tw}, whose derivative structure mandates the inclusion of contact terms to render loop corrections UV finite~\cite{Ametller:2001yk,Ahmedov:2002tg,Bakmaev:2005sg}, 
a complete treatment that captures all low-energy terms at $\Order(\alpha)$ while, at the same time, accounting for the resonance physics of the process becomes a formidable challenge. Instead, we make use of the observation from Ref.~\cite{Moussallam:2013una} in the context of $e^+e^-\to\pi\pi(\gamma)$, i.e., that by far the most relevant numerical effect arises from the infrared (IR) enhanced contributions that survive after the cancellation of IR singularities between virtual and bremsstrahlung diagrams. We set up a framework that allows us to evaluate these corrections using as input basis function for a Khuri--Treiman (KT) treatment of $\gamma^*\to 3\pi$~\cite{Khuri:1960zz}, and extrapolate the resulting correction factor $\eta_{3\pi}(s)$ to the $3\pi$ threshold by means of a non-relativistic (NR) expansion. After a review of our dispersive representation for $e^+e^-\to 3\pi$ in Sec.~\ref{sec:disp}, this formalism is presented in Sec.~\ref{sec:EM}. 

The second major IB effect is generated by $\rho$--$\omega$ mixing. On the one hand, this effect is expected to be enhanced compared to $e^+e^-\to 2\pi$ since the $\omega$ couples more weakly to the electromagnetic current than the $\rho$---in a vector-meson-dominance (VMD) picture by a relative factor $3$, which thus translates to almost an order of magnitude in the $\rho$--$\omega$ mixing contribution. On the other hand, the large width of the $\rho$ makes the effect much less localized than in the $2\pi$ system, in such a way that the resulting integral becomes more sensitive to the assumed line shape, and care is required to differentiate an IB effect from background contributions to the cross section. To parameterize the line shape in a way consistent with the definition of the mixing parameter $\epsrw$ as a residue in $e^+e^-\to 2\pi$, we follow the coupled-channel formalism from Ref.~\cite{Holz:2022hwz}, with the main features summarized in Sec.~\ref{sec:rhoomega}. Updated fits to the $e^+e^-\to 3\pi$ data base are presented in Sec.~\ref{sec:fits}, the consequences for $a_\mu$ in Sec.~\ref{sec:amu}. We summarize our results in Sec.~\ref{sec:conclusions}.

\section{Dispersive parameterization of $\boldsymbol{e^+e^-\to 3\pi}$}
\label{sec:disp}

As starting point for the study of IB effects, we use the dispersive representation of the $e^+e^-\to 3\pi$ cross section from Ref.~\cite{Hoferichter:2019gzf}, first derived in the context of the pion transition form factor~\cite{Hoferichter:2014vra,Hoferichter:2018dmo,Hoferichter:2018kwz}. The key idea amounts to combining the normalization from the WZW anomaly in terms of the pion decay constant $F_\pi$~\cite{Adler:1971nq,Terentev:1971cso,Aviv:1971hq} with a calculation of the $\pi\pi$ rescattering corrections in the KT formalism, generalizing work on $\omega,\phi\to3\pi$ decays~\cite{Aitchison:1977ej,Niecknig:2012sj,Schneider:2012ez,Hoferichter:2012pm,Danilkin:2014cra,Dax:2018rvs} to arbitrary photon virtualities $\gamma^*\to 3\pi$.

The general expression of the matrix element for $\gamma^*(q)\rightarrow \pi^+(p_+)\pi^-(p_-)\pi^0(p_0)$  is given by
\begin{equation}
    \langle0|j_\mu(0)|\pi^+(p_+)\pi^-(p_-)\pi^0(p_0)\rangle = -\epsilon_{\mu\nu\alpha\beta}p_+^\nu p_-^\alpha p_0^\beta \F(s,t,u;q^2),
\end{equation}
with $q=p_++p_-+p_0, s=(p_++p_-)^2, t=(p_-+p_0)^2,u=(p_++p_0)^2$, and $s+t+u=3M_\pi^2+q^2$. We further decompose the invariant function $\F$ as
\begin{equation}
\label{reconstruction}
    \F(s,t,u;q^2)
= \F(s,q^2)+\F(t,q^2)+\F(u,q^2),
\end{equation}
and perform a partial-wave expansion, where due to Bose symmetry only odd partial waves contribute~\cite{Jacob:1959at} 
\beq
\F(s,t,u;q^2)=\sum \limits_{\ell \text{\,odd}} f_\ell(s,q^2) P'_\ell(z_s).
\eeq
The kinematic quantities are
\begin{align}
        z_s&=\cos\theta_s =\frac{t-u}{\kappa(s,q^2)}, \qquad
        \kappa(s,q^2)=\sigma_\pi(s)\lambda^{1/2}(q^2,M_\pi^2,s),\notag\\
     \lambda(x,y,z)&=x^2+y^2+z^2-2(xy+yz+xz),\qquad  \sigma_\pi(s)=\sqrt{1-\frac{4\mpi^2}{s}},  
\end{align}
and $P_\ell'(z)$ denotes the derivatives of the Legendre polynomials. The decomposition~\eqref{reconstruction} strictly applies as long as the discontinuities of $F$- and higher partial waves are negligible, as well justified below the onset of the $\rho_3(1690)$ resonance~\cite{Niecknig:2012sj,Hoferichter:2017ftn,Hoferichter:2019gzf}. The resulting cross section is expressed as the integral 
\beq
\label{eq:epemcross1}
\sigma_{e^+ e^- \to 3\pi}(q^2) = \alpha^2\int_{s_\text{min}}^{s_\text{max}} \diff s \int_{t_\text{min}}^{t_\text{max}} \diff t \,
\frac{s[\kappa(s,q^2)]^2(1-z_s^2)}{768 \, \pi \, q^6}  \, |\F(s,t,u;q^2)|^2, 
\eeq
with integration boundaries
\begin{align}
s_\text{min} &= 4 M_\pi^2, \qquad\qquad \,s_\text{max} = \big(\sqrt{q^2}-M_\pi \big)^2, \notag \\ 
t_\text{min/max}&= (E_-^*+E_0^*)^2-\bigg( \sqrt{E_-^{*2}-M_\pi^2} \pm  \sqrt{E_0^{*2}-M_\pi^2} \bigg)^2,
\end{align}
and
\beq
E_-^*=\frac{\sqrt{s}}{2},\qquad E_0^*=\frac{q^2-s-M_\pi^2}{2\sqrt{s}}.
\eeq
The momentum dependence of the partial wave $f_1(s,q^2)$ is then predicted from the KT formalism up to an overall normalization $a(q^2)$, which we parameterize following the same ansatz as in Ref.~\cite{Hoferichter:2019gzf}  
\beq
\label{eq:a-par}
a(q^2)=\alpha_A+\frac{q^2}{\pi}\int_{s_\text{thr}}^{\infty} \diff s'\frac{\Im\A (s')}{s'(s'-q^2)}+C_p(q^2).
\eeq
The three terms correspond to the WZW normalization, resonance contributions (most notably $\omega$ and $\phi$, but also $\omega'(1420)$, $\omega''(1650)$ to be able to describe the data up to $1.8\GeV$), and a conformal polynomial to parameterize non-resonant contributions. For the WZW normalization, the best estimate is still given by~\cite{Bijnens:1989ff,Hoferichter:2012pm} 
\beq
\alpha_A= \frac{F_{3\pi}}{3}\times 1.066(10),\qquad F_{3\pi}=\frac{1}{4\pi^2 F^3_\pi},
\eeq
a low-energy theorem that could be tested with future lattice-QCD calculations~\cite{Briceno:2016kkp,Alexandrou:2018jbt,Niehus:2021iin}. The resonant contributions are described by taking the imaginary part from
\beq
\A(q^2)=\sum_{V}\frac{c_V}{M_V^2-q^2-i\sqrt{q^2}\, \varGamma_V(q^2)}.
\eeq
The energy-dependent widths $\varGamma_V(q^2)$ for $V=\omega,\phi$ include the main decay channels, in particular, $\omega\to \pi^0\gamma$ sets the integration threshold to $s_\text{thr}=\mpii^2$. For the $3\pi$ channel, the partial width accounts for the $3\pi$ rescattering as well, and the remaining tiny effects from the neglected channels $\omega\to 2\pi$ and $\phi\to\eta\gamma$ are corrected by a rescaling of the partial widths. $\omega'$ and $\omega''$ are assumed to exclusively decay to $3\pi$ for simplicity. As before, we fix the $\omega''$ parameters to the PDG values~\cite{ParticleDataGroup:2022pth}, but for $\omega'$ we observe that our fits do become sensitive to the assumption for the mass parameter, and thus introduce $M_{\omega'}$ as an additional degree of freedom in our representation.  The conformal polynomial, 
\beq
\label{Cp}
C_p(q^2)=\sum_{i=1}^p c_i\big(z(q^2)^i-z(0)^i\big), \qquad z(q^2)=\frac{\sqrt{s_\text{inel}-s_1}-\sqrt{s_\text{inel}-q^2}}{\sqrt{s_\text{inel}-s_1}+\sqrt{s_\text{inel}-q^2}},
\eeq
is unchanged compared to Ref.~\cite{Hoferichter:2019gzf}: $s_\text{inel}=1\GeV^2$, $s_1=-1\GeV^2$, and the absence of an $S$-wave cusp as well as the sum rule for $\alpha_A$ are imposed as additional constraints on $C_p(q^2)$.

\section{Electromagnetic corrections to $\boldsymbol{e^+e^-\to 3\pi}$}
\label{sec:EM}

In the $2\pi$ channel, the effect of radiative corrections on the total cross section is often estimated using ``FsQED,'' i.e., scalar QED dressed with the pion form factor $F_\pi^V$. In a dispersive picture, this approach amounts to isolating pion-pole contributions and replacing the constant $\pi\pi\gamma$ coupling as predicted in scalar QED by the full matrix element. This procedure captures the IR-enhanced contributions, which provide the dominant effect compared to non-pion-pole $\pi\pi\gamma$ states~\cite{Moussallam:2013una},  
leading to a universal correction factor
\begin{equation}
    \sigma_{e^+e^-\to2\pi(\gamma)}(q^2) = \sigma_{e^+e^-\to2\pi}^{(0)}(q^2)\Big(1+\frac{\alpha}{\pi}\eta_{2\pi}(q^2)\Big),\qquad 
 \sigma_{e^+e^-\to2\pi}^{(0)}(q^2)=\frac{\pi\alpha^2}{3q^2}\sigma_\pi^3(q^2)\big|F_\pi^V(q^2)\big|^2,   
\end{equation}
with~\cite{Hoefer:2001mx,Czyz:2004rj,Gluza:2002ui,Bystritskiy:2005ib} 
\begin{align}
\label{eta2pi}
 \eta_{2\pi}(s) &= \frac{3(1+\sigma_\pi^2(s))}{2\sigma_\pi^2(s)} - 4 \log\sigma_\pi(s) + 6 \log \frac{1+\sigma_\pi(s)}{2} + \frac{1+\sigma_\pi^2(s)}{\sigma_\pi(s)} F(\sigma_\pi(s)) \notag\\
				& - \frac{(1-\sigma_\pi(s))\big(3+3\sigma_\pi(s)-7\sigma_\pi^2(s)+5\sigma_\pi^3(s)\big)}{4\sigma_\pi^3(s)} \log\frac{1+\sigma_\pi(s)}{1-\sigma_\pi(s)}, \notag\\
			F(x) &= -4\dilog(x)+4\dilog(-x)+2\log x \log\frac{1+x}{1-x} + 3 \dilog\Big(\frac{1+x}{2} \Big) - 3\dilog\Big(\frac{1-x}{2} \Big) + \frac{\pi^2}{2}, \notag\\
			\dilog(x) &= - \int_0^x dt \frac{\log(1-t)}{t}.
\end{align}
For the $2\pi$ channel, the role of radiative corrections beyond the FsQED approximation is an active subject of discussion~\cite{Campanario:2019mjh,Colangelo:2022lzg,Ignatov:2022iou,JMPhDThesis,Abbiendi:2022liz}, especially in view of the CMD-3 measurement~\cite{CMD-3:2023alj}, but for $e^+e^-\to 3\pi$ so far no robust estimates of radiative corrections are available at all, which strongly motivates the focus on the IR-enhanced effects as the numerically dominant contribution.   

To isolate these effects, we proceed as follows: even when neglecting the discontinuities of $\ell\geq 3$ partial waves, the full amplitude receives contributions beyond $P$-waves from the projection of the crossed-channel amplitudes, i.e., 
\beq
f_1(s,q^2)=\F(s,q^2)+\hat \F(s,q^2),
\eeq
where 
\beq
\hat \F(s,q^2)=\frac{3}{2}\int_{-1}^1\diff z_s\big(1-z_s^2\big)\F\big(t(s,q^2,z_s),q^2\big).
\eeq
For the pure $P$-wave subsystem, the combination of the IR-enhanced virtual-photon and bremsstrahlung diagrams reproduces the functional form of $\eta_{2\pi}$ as given in Eq.~\eqref{eta2pi}, with the momentum not determined by the $e^+e^-$ invariant mass $q^2$, but by the invariant mass of the $\pi^+\pi^-$ subsystem. Accordingly, we can capture this effect by writing
\begin{align}
\label{sigma_SP_1}
    \sigma_{e^+e^-\to 3\pi(\gamma)} (q^2) &\propto 
    \int_{s_\text{min}}^{s_\text{max}} \diff s \int_{t_\text{min}}^{t_\text{max}} \diff t \,s\big[\kappa(s,q^2)\big]^2(1-z_s^2)\notag\\
    &\times \bigg|\big(\underbrace{\F(s,q^2)+\hat\F(s,q^2)}_{f_1(s,q^2)}\big)\sqrt{1+\frac{\alpha}{\pi}\eta_{2\pi}(s)}\notag\\
&\qquad    +\big(\underbrace{\F(t,q^2)+\F(u,q^2)-\hat \F(s,q^2)}_{f_3(s,q^2)+\cdots}\big)\sqrt{1+\frac{\alpha}{\pi}\eta_{2\pi}(s_\text{PT})}\bigg|^2,
\end{align}
where all factors that drop out in
\beq
\label{eta3pi}
1+\frac{\alpha}{\pi}\eta_{3\pi}(q^2)=\frac{\sigma_{e^+e^-\to 3\pi(\gamma)} (q^2)}{\sigma^{(0)}_{e^+e^-\to 3\pi} (q^2)}
\eeq
have been ignored and $s_\text{PT}=(\sqrt{q^2}-\mpi)^2$ denotes the position of the pseudothreshold. This kinematic point is critical, since $\hat \F(s,q^2)$ diverges at $s_\text{PT}$, in such a way that the crossed-channel contribution starting at $\ell=3$ needs to be multiplied by a function that ensures that the cancellation at $s_\text{PT}$ is maintained in the presence of radiative corrections. The choice of this kinematic function is not unique, in Eq.~\eqref{sigma_SP_1} we show the minimal variant in which a constant correction is assumed. However, the ambiguity in this correction only affects higher partial waves, and by definition cannot contribute to the IR-enhanced effects in the $P$-wave subsystem. For this reason, we may choose to evaluate this correction factor at $s$ instead of $s_\text{PT}$, which simplifies the result to 
\begin{align}
\label{sigma_SP_final}
    \sigma_{e^+e^-\to 3\pi(\gamma)} (q^2) &\propto 
    \int_{s_\text{min}}^{s_\text{max}} \diff s \int_{t_\text{min}}^{t_\text{max}} \diff t \,s\big[\kappa(s,q^2)\big]^2(1-z_s^2)\notag\\
    &\times
    \Big|\F(s,q^2)+\F(t,q^2)+\F(u,q^2)\Big|^2\Big(1+\frac{\alpha}{\pi}\eta_{2\pi}(s)\Big).
\end{align}
We checked that both variants indeed lead to minor differences, and will continue to work with Eq.~\eqref{sigma_SP_final} in the following. For the numerical evaluation of Eq.~\eqref{eta3pi} we use the KT basis functions from Ref.~\cite{Stamen:2022eda}. 

\begin{figure}[t]
    \centering
    \includegraphics[width=0.8\textwidth]{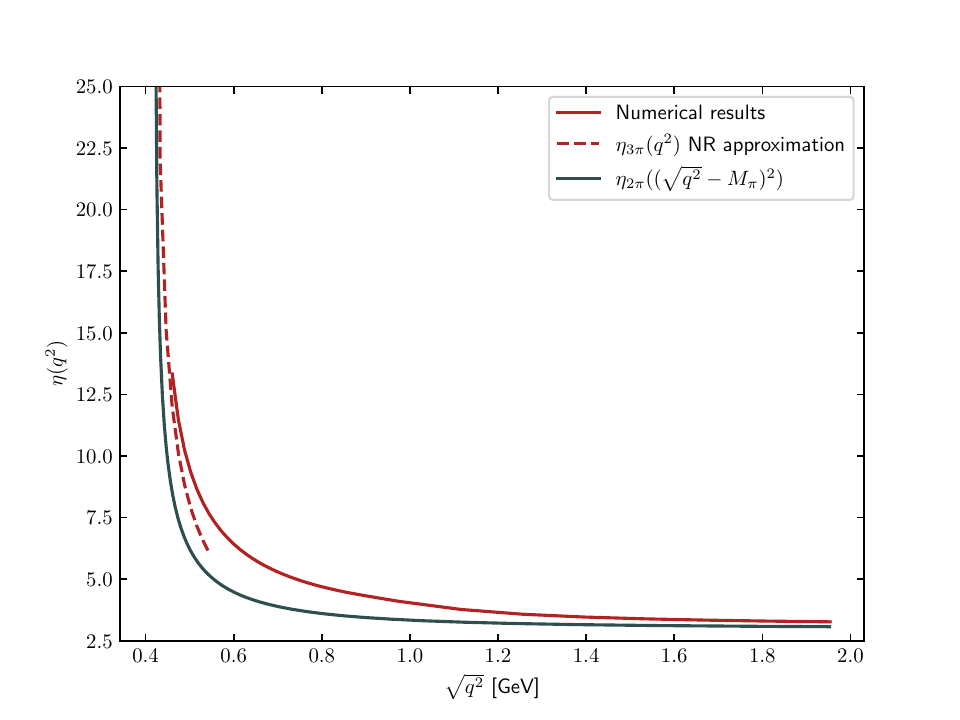}
    \caption{Comparison of the numerical result for $\eta_{3\pi}$ (red solid), its NR approximation (red dashed), and $\eta_{2\pi}$ shifted to the $3\pi$ threshold (blue solid).}
    \label{fig:eta3pi}
\end{figure}

\begin{figure}[t]
    \centering
    \includegraphics[width=0.8\textwidth]{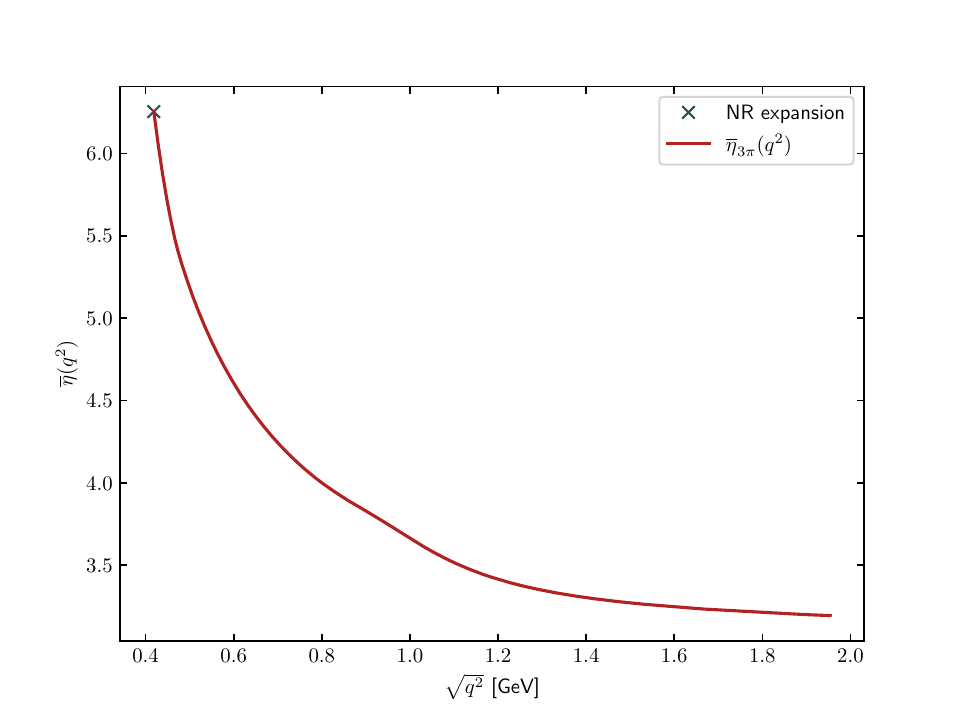}
    \caption{$3\pi$ FSR factor $\bar \eta_{3\pi}$, with the threshold divergence removed according to Eq.~\eqref{bareta} (the result is available as text file in the supplementary material).}
    \label{fig:bareta3pi}
\end{figure}

In analogy to $\eta_{2\pi}$, the correction factor $\eta_{3\pi}(q^2)$ involves a Coulomb divergence at threshold $\propto (q^2-9\mpi^2)^{-1/2}$,\footnote{We emphasize that the Coulomb divergence $\propto (s-4\mpi^2)^{-1/2}$ is present in Eq.~\eqref{sigma_SP_final} for every $q^2$.  However, after integration over $s$ and $t$, this translates into a divergence in $q^2$ only at the three-pion threshold.} so that for the application in fits to $e^+e^-\to 3\pi$ cross-section data it is convenient to provide numerical results for 
\beq
\label{bareta}
\bar\eta_{3\pi}(q^2)=\eta_{3\pi}(q^2)\sqrt{1-\frac{9\mpi^2}{q^2}}. 
\eeq
Moreover, the numerical solution of the KT equations becomes unstable close to threshold, making the determination of the coefficient of the Coulomb divergence by other means valuable to be able to interpolate to the range starting at $\sqrt{q^2}\approx 3.3\mpi$ where a  numerical solution is feasible. This can be achieved by a NR expansion. Starting from 
\beq
\eta_{2\pi}(s)=\frac{\pi^2}{2\sigma_\pi(s)}-2+\Order(\sigma_\pi),
\eeq
we perform the substitution $s=4\mpi^2(1-x)+(\sqrt{q^2}-\mpi)^2 x$ and expand $\sqrt{q^2}=3\mpi(1+\eps)$ around threshold. The $t$ integration and $\F(s,q^2)+\F(t,q^2)+\F(u,q^2)$ can be ignored as they cancel in the ratio, while the remaining kinematic dependence leads to
\beq
 \eta_{3\pi}(\eps)=
 \frac{256\pi}{105\sqrt{3\eps}}-2+\Order(\sqrt{\eps}),
\eeq
and therefore 
\beq
\bar \eta_{3\pi}(9\mpi^2)=\frac{256\pi}{105}\sqrt{\frac{2}{3}}. 
\eeq
The numerical result for $\eta_{3\pi}$ is shown in Fig.~\ref{fig:eta3pi}, in comparison to the NR approximation and $\eta_{2\pi}$ shifted to the $3\pi$ threshold. From this comparison it follows that $\eta_{3\pi}$ is really distinctly different from $\eta_{2\pi}$, reflected by the increase that is observed in addition to the change of threshold. In Fig.~\ref{fig:bareta3pi}, we also show the result for $\bar\eta_{3\pi}$, as we will use in the implementation together with the threshold factor in Eq.~\eqref{bareta}.

\section{$\boldsymbol{\rho}$--$\boldsymbol{\omega}$ mixing in $\boldsymbol{e^+e^-\to 3\pi}$}
\label{sec:rhoomega}

$\rho$--$\omega$ mixing in $e^+e^-\to 2\pi$ can be implemented via a correction factor~\cite{Colangelo:2022prz} 
\begin{align}
\label{Gomega}
	G_\omega(s) = 1 &+ \frac{s}{\pi} \int_{9\mpi^2}^\infty \diff s^\prime \frac{\Re\epsrw}{s^\prime(s^\prime-s)} \Im\left[ \frac{s'}{(\mw - \frac{i}{2} \Gw)^2 - s'} \right]  \left( \frac{1 - \frac{9\mpi^2}{s^\prime}}{1 - \frac{9\mpi^2}{\mw^2}} \right)^4 \nonumber\\
		&+ \frac{s}{\pi} \int_{\mpii^2}^\infty \diff s^\prime \frac{\Im\epsrw}{s^\prime(s^\prime-s)} \Re\left[ \frac{s'}{(\mw - \frac{i}{2} \Gw)^2 - s'} \right]  \left( \frac{1 - \frac{\mpii^2}{s^\prime}}{1 - \frac{\mpii^2}{\mw^2}} \right)^3 ,
\end{align}
which amounts to a dispersively improved variant of a Breit--Wigner ansatz
\beq
\label{VFF_rw}
g_\omega(s)=1+\frac{\epsrw s}{\mw^2-s-i\mw\Gw}
\eeq
that, besides removing the unphysical imaginary part below the $3\pi$ threshold, also allows for the $\pi^0\gamma$ cut and thus an IB phase in $\epsrw$. In particular, Eq.~\eqref{Gomega} shows that $\rho$--$\omega$ mixing in $e^+e^-\to 2\pi$ is intimately related to the residue at the $\omega$ pole, since the small width of the $\omega$, together with the threshold and asymptotic constraints on the line shape, leaves little freedom in the construction of $G_\omega(s)$.

In contrast, $\rho$--$\omega$ mixing in $e^+e^-\to 3\pi$ is far less localized, and due to the large width of the $\rho$ a significant sensitivity to the assumed line shape off the resonance is expected. In particular, the absence of sharp interference features in the cross section could potentially lead one to misidentify a non-resonant background as an IB contribution. To mitigate such effects, we use the line shape as predicted by the coupled-channel formalism from Ref.~\cite{Holz:2022hwz} for $e^+e^-$, $\pi^+\pi^-$, and $3\pi$, constructed for a consistent implementation of $\rho$--$\omega$ mixing in $e^+e^-\to 2\pi$, $\eta'\to\pi^+\pi^-\gamma$, and the $\eta'$ transition form factor. The main idea follows
Ref.~\cite{Hanhart:2012wi} (see also Refs.~\cite{Ropertz:2018stk,VonDetten:2021rax}): the full multichannel scattering amplitude arises from iterating a scattering potential via self energies, combined with elastic $\pi\pi$ rescattering as described by the Omn\`es function~\cite{Omnes:1958hv} in a way that is consistent with analyticity and unitarity. Including the photon and $\omega$ poles in the resonance potential, the formalism then predicts the shape of the amplitudes in the various channels that follows from a dispersive representation of the self energies together with the multichannel dynamics.  

As a first step, the formalism reproduces the vacuum polarization (VP) function
\beq
\label{VP}
\Pi(s)=\Pi_e(s)+\Pi_\pi(s)\bigg(1+\frac{2s\epsrw}{\mw^2-s-i\mw\Gw}\bigg)+\Pi_\omega(s)+\Order(\epsrw^2),
\eeq
where the $\omega$ contribution is represented in a narrow-width approximation
\beq
\label{Piw}
\Pi_\omega(s)=\frac{1}{\gwg^2}\frac{e^2s}{s-\mw^2+i\mw\Gw},
\eeq
while the $\rho$ remains resolved as a $2\pi$ resonance
\beq
\label{Pi_pi_disp}
\Pi_\pi(s)=-\frac{e^2 s}{48\pi^2}\int_{4\mpi^2}^\infty \diff s'\frac{\sigma_\pi^3(s')|F_\pi^V(s')|^2}{s'(s'-s-i\eps)}.
\eeq
$\Pi_e(s)$ in Eq.~\eqref{VP} gives the leptonic VP, and $\rho$--$\omega$ mixing is represented by a product of $\Pi_\pi(s)$ with the narrow-width $\omega$ propagator. For $\Gr\to 0$, $\Pi_\pi(s)$ collapses to 
\beq
\label{Pir}
\Pi_\pi(s)\to \Pi_\rho(s)=\frac{1}{\grg^2}\frac{e^2s}{s-\mr^2+i\mr\Gr}. 
\eeq
The couplings introduced in Eqs.~\eqref{Piw} and~\eqref{Pir} are related to the dilepton decay $V\to e^+e^-$ via $\Gamma_{V\to e^+e^-}=4\pi\alpha^2 M_V/(3|g_{V\gamma}|^2)$; numerically,  we will use $|\gwg|=16.2(8)$~\cite{Holz:2022hwz} and $|\grg|=4.9(1)$~\cite{Hoferichter:2017ftn}, where the latter determination invokes an analytic continuation to the $\rho$ pole instead of a narrow-resonance estimate on the real axis. The deviation of $|\gwg|/|\grg|=3.3(2)$ from $3$ quantifies the deviation from the VMD expectation in these couplings. 

Most importantly, the same formalism also reproduces Eq.~\eqref{VFF_rw} for $\rho$--$\omega$ mixing in $e^+e^-\to 2\pi$, and predicts the analog correction for $e^+e^-\to 3\pi$~\cite{Holz:2022hwz}
\beq
\label{3pi_rw}
g_\pi(s)=1-\frac{\gwg^2\epsrw}{e^2}\Pi_\pi(s). 
\eeq
In the narrow-width limit~\eqref{Pir} one thus finds the exact same form apart from $\epsrw\to \epsrw\gwg^2/\grg^2$, and thus the VMD enhancement factor expected from the smaller photon coupling of the $\omega$.\footnote{In the typical conventions~\cite{Sakurai:1969,Klingl:1996by}, the coupling strength is proportional to $1/g_{V\gamma}$.} However, instead of having to rely on a narrow-width approximation for the $\rho$, we can use the full result~\eqref{3pi_rw}, which ensures that the mixing parameter $\epsrw$ is defined in a way consistent with $e^+e^-\to 2\pi$, and that the line shape correctly implements the dispersion relation for the two-pion self energy. In practice, we will use Eq.~\eqref{Pi_pi_disp} with $\rho$--$\omega$ mixing in $F_\pi^V(s)$ switched off, given that such higher-order IB effects cannot be described in a consistent manner. 

From similar arguments, we can glean some intuition about the size of IB to be expected in the different channels when inserted into the HVP integral. To this end, we write the HVP master formula as~\cite{Bouchiat:1961lbg,Brodsky:1967sr}
\beq
\label{master}
	a_\mu^\text{HVP} = \Big( \frac{\alpha m_\mu}{3\pi} \Big)^2
        \int_{s_\text{thr}}^\infty \diff s \frac{\hat K(s)}{s^2}
        R_\text{had}(s)
  = -\Big( \frac{\alpha m_\mu}{3\pi} \Big)^2
        \int_{s_\text{thr}}^\infty \diff s \frac{\hat K(s)}{s^2}\frac{12\pi}{e^2}\Im \Pi(s),
\eeq
with
\begin{align}
		\hat K(s) &= \frac{3s}{m_\mu^2} \bigg[
		 \frac{x^2}{2} (2-x^2) + \frac{(1+x^2)(1+x)^2}{x^2} \Big( \log(1+x) - x + \frac{x^2}{2} \Big) + \frac{1+x}{1-x} x^2 \log x \bigg] , \notag\\
		x &= \frac{1-\sigma_\mu(s)}{1+\sigma_\mu(s)},\quad \sigma_\mu(s)=\sqrt{1-\frac{4m_\mu^2}{s}}. 
\end{align}
In the narrow-width limit, one thus finds
\begin{align}
 a_\mu^\rho&=\Big( \frac{\alpha m_\mu}{3\pi} \Big)^2\frac{\hat K(\mr^2)}{\mr^2}\frac{12\pi^2}{|\grg|^2}\simeq 482\times 10^{-10},\notag\\
 a_\mu^\omega&=\Big( \frac{\alpha m_\mu}{3\pi} \Big)^2\frac{\hat K(\mw^2)}{\mw^2}\frac{12\pi^2}{|\gwg|^2}\simeq 43.6\times 10^{-10},
\end{align}
both within a few percent of the expected contribution when integrating around the $\rho$ and $\omega$ resonances in the $2\pi$ and $3\pi$ cross sections, respectively. Based on Eq.~\eqref{VP}, the $\rho$--$\omega$ mixing contribution becomes 
\beq
\label{amurhow}
a_\mu^{\rho\text{--}\omega}=\Big(\frac{\alpha m_\mu}{3\pi} \Big)^2\int_{s_\text{thr}}^\infty\diff s\,\hat K(s)\frac{24\pi\epsrw}{|\grg|^2}\Im\bigg[\frac{1}{s-\mr^2+i\mr\Gr}\frac{1}{s-\mw^2+i\mw\Gw}\bigg].
\eeq
This expression is of course rather sensitive to integration range and line shape, clearly, for such a subtle interference a narrow-width approximation for the $\rho$ is not adequate. Still, it is striking that the numerical evaluation of Eq.~\eqref{amurhow} produces $|a_\mu^{\rho\text{--}\omega}|\lesssim 0.5\times 10^{-10}$, while even a simple narrow-width formula such as Eq.~\eqref{Pir} for $F_\pi^V(s)$ multiplied with Eq.~\eqref{VFF_rw} gives results for the $\rho$--$\omega$ contribution in the $2\pi$ channel much closer to the detailed analysis of Ref.~\cite{Colangelo:2022prz}. Ultimately, this behavior seems to arise because the entire $\rho$--$\omega$ mixing effect from Eq.~\eqref{VP} should not be attributed to the $2\pi$ channel alone, instead, a partial-fraction decomposition 
\begin{align}
\label{partial_fraction}
 \frac{1}{s-\mr^2+i\mr\Gr}\frac{1}{s-\mw^2+i\mw\Gw}&=\frac{1}{\mr^2-\mw^2-i\mr\Gr+i\mw\Gw}\notag\\
 &\times\bigg(\frac{1}{s-\mr^2+i\mr\Gr}-\frac{1}{s-\mw^2+i\mw\Gw}\bigg)
\end{align}
suggests that a $\rho$--$\omega$ mixing contribution should arise in both the $2\pi$ and $3\pi$ channel, and while the detailed phenomenology will again crucially depend on the line shape, evaluating both terms in Eq.~\eqref{partial_fraction} separately in the integral~\eqref{amurhow} does produce sizable cancellations. From this perspective, at least a partial cancellation of the $\rho$--$\omega$ mixing contributions in the actual $e^+e^-\to2\pi$ and $e^+e^-\to 3\pi$ cross sections would not appear surprising.

\section{Fits to $\boldsymbol{e^+e^-\to 3\pi}$ data}
\label{sec:fits}

\subsection{Fits to data base prior to BaBar 2021}
\label{sec:old_data_base}

As a first step, we update the combined fit presented in Ref.~\cite{Hoferichter:2019gzf}, to reflect several recent developments and provide a first demonstration of the consequences of the IB corrections included in the new fit function. Regarding the data base, the SND data set~\cite{Aulchenko:2015mwt} has been superseded by the update from Ref.~\cite{SND:2020ajg}, and likewise Ref.~\cite{BaBar:2004ytv} has been superseded by Ref.~\cite{BABAR:2021cde}, which we will consider in Sec.~\ref{sec:BaBar}.  
All other data sets from SND~\cite{Achasov:2000am,Achasov:2002ud,Achasov:2003ir} and CMD-2~\cite{Akhmetshin:1995vz,Akhmetshin:1998se,Akhmetshin:2003zn,Akhmetshin:2006sc} are treated as described in Ref.~\cite{Hoferichter:2019gzf}, while the old data from DM1~\cite{Cordier:1979qg}, DM2~\cite{DM2:1992zkc}, and ND~\cite{Dolinsky:1991vq} are no longer included. The motivation for this cut is given by inconsistencies that exist especially in the energy range between the $\omega$ and $\phi$ resonances compared to the modern data sets. Previously, these tensions in the data base  essentially resulted in a slightly worse $\chi^2/\text{dof}$, but, as expected,  the size of the $\rho$--$\omega$ mixing contribution depends more strongly on the line shape between the resonances, in such a way that such inconsistencies can no longer be tolerated without distorting the $\rho$--$\omega$ mixing signal. We also update the resonance parameters of the excited $\omega$ states~\cite{ParticleDataGroup:2022pth}
\begin{align}
\label{omegap}
M_{\omega'}&=1410(60)\MeV,& \Gamma_{\omega'}&=290(190)\MeV,\notag\\
M_{\omega''}&=1670(30)\MeV,& \Gamma_{\omega''}&=315(35)\MeV.
\end{align}
However, given that the $\omega'$ parameters are rather uncertain, we also considered variants in which $M_{\omega'}$, $\Gamma_{\omega'}$ are allowed to vary, revealing in some cases a relevant sensitivity to $M_{\omega'}$, which will therefore be added as a free parameter.  A D'Agostini bias~\cite{DAgostini:1993arp} from correlated systematic errors is avoided by an iterative procedure~\cite{Ball:2009qv}, see Ref.~\cite{Hoferichter:2019gzf} for more details. 

The new fit function decomposes as follows: the normalization $a(q^2)$ is parameterized as in Eq.~\eqref{eq:a-par}, with the $\omega$ contribution multiplied by $g_\pi(q^2)$ defined as in Eq.~\eqref{3pi_rw}. Furthermore, all data sets are assumed to contain FSR corrections, which we remove using $\eta_{3\pi}(q^2)$ prior to the fit. Only in the final step, the calculation of the HVP loop integral~\eqref{master}, are the FSR corrections added back. This procedure follows the same approach as for $e^+e^-\to\pi^+\pi^-$, since the dispersive representation, strictly speaking, only applies for the amplitudes from which virtual-photon corrections have been removed.  

\begin{table}[t]
	\centering
	\footnotesize
	\renewcommand{\arraystretch}{1.3}
	\begin{tabular}{lcccccc}
	\toprule
	& \multicolumn{3}{c}{$n_\text{conf}=0$} & \multicolumn{3}{c}{$n_\text{conf}=1$}\\
	$p_\text{conf}$ & $2$ & $3$ & $4$ & $2$ & $3$ & $4$\\
	$\chi^2/\text{dof}$ & $274.4/228$ & $271.2/227$ & $270.4/226$ & $287.5/228$ & $284.5/227$& $271.6/226$\\
	& $=1.21$ & $=1.19$ & $=1.20$ & $=1.26$ & $=1.25$& $=1.20$\\
	$p$-value & $0.02$ & $0.02$ & $0.02$ & $0.005$ & $0.006$& $0.02$\\
	$\mw \ [\text{MeV}]$ & $782.70(3)$ & $782.69(3)$ & $782.70(3)$ & $782.70(3)$ & $782.70(3)$& $782.69(3)$\\
	$\Gw \ [\text{MeV}]$ & $8.73(3)$ & $8.74(3)$ & $8.74(4)$  & $8.73(3)$ & $8.73(3)$& $8.71(4)$\\
	$\mphi \ [\text{MeV}]$& $1019.20(1)$ & $1019.19(1)$ & $1019.19(1)$ & $1019.20(1)$ & $1019.20(1)$& $1019.21(1)$\\
	$\Gphi \ [\text{MeV}]$ & $4.25(3)$ & $4.24(3)$ & $4.24(3)$  & $4.25(3)$ & $4.25(3)$& $4.26(3)$\\
	$M_{\omega'} \ [\text{GeV}]$ & $1.433(17)$ & $1.416(24)$& $1.408(22)$& $1.383(11)$& $1.392(10)$& $1.425(29)$\\
	$c_\omega \ [\text{GeV}^{-1}]$ & $2.91(2)$ & $2.92(2)$ & $2.93(3)$ & $2.93(2)$ & $2.92(2)$& $2.88(3)$\\
	$c_\phi \ [\text{GeV}^{-1}]$ & $-0.388(3)$ & $-0.388(3)$ & $-0.387(3)$ & $-0.388(3)$ & $-0.388(3)$& $-0.390(3)$\\
	$c_{\omega'} \ [\text{GeV}^{-1}]$ & $-0.22(4)$ & $-0.12(6)$ & $-0.15(7)$  & $-0.23(5)$ & $-0.29(7)$& $0.13(7)$\\
	$c_{\omega''} \ [\text{GeV}^{-1}]$ & $-1.64(7)$ & $-1.54(8)$ & $-1.51(10)$ & $-0.89(6)$ & $-0.93(7)$& $3.37(7)$\\
	$c_1 \ [\text{GeV}^{-3}]$ & $-0.34(8)$ & $-0.25(9)$ & $-0.16(14)$ & $-1.31(9)$ & $-1.30(9)$& $-2.14(6)$\\
	$c_2 \ [\text{GeV}^{-3}]$ & $-1.26(5)$ & $-1.34(7)$ & $-1.40(10)$ & $-0.29(10)$ & $-0.32(10)$& $-1.81(5)$\\
	$c_3 \ [\text{GeV}^{-3}]$ & --- & $-0.53(7)$ & $-0.50(8)$ & --- & $-0.23(8)$& $-0.62(6)$\\
	$c_4 \ [\text{GeV}^{-3}]$ & --- & --- & $1.38(9)$ & --- & --- & $2.80(12)$ \\
	$10^4\times \xi_\text{CMD-2}$ & $1.3(5)$ & $1.2(5)$ & $1.2(5)$ & $1.3(5)$ & $1.3(5)$ & $1.3(5)$ \\
	$10^3\times \Re\epsrw$ & $1.48(28)$ & $1.42(28)$ & $1.46(28)$ & $1.62(28)$ & $1.61(29)$ & $1.39(30)$\\
	$10^{10}\times a_\mu^{3\pi}|_{\leq 1.8\GeV}$ & $45.68(48)$ & $45.84(49)$ & $46.01(54)$ & $45.83(51)$ & $45.77(51)$ & $45.18(51)$\\ 
	$10^{10}\times a_\mu^\text{FSR}[3\pi]$ & $0.51(1)$ & $0.51(1)$ & $0.51(1)$ & $0.51(1)$ & $0.51(1)$ & $0.50(1)$\\
	$10^{10}\times a_\mu^{\rho\text{--}\omega}[3\pi]$ & $-2.62(49)$ & $-2.54(49)$ & $-2.61(49)$ & $-2.95(50)$ & $-2.91(51)$ & $-2.32(49)$\\
	\bottomrule
	\renewcommand{\arraystretch}{1.0}
	\end{tabular}
	\caption{Fits to the combination of SND~\cite{Achasov:2000am,Achasov:2002ud,Achasov:2003ir,SND:2020ajg} and CMD-2$'$~\cite{Akhmetshin:1995vz,Akhmetshin:1998se,Akhmetshin:2003zn,Akhmetshin:2006sc} (the prime indicates the data selection as detailed in Ref.~\cite{Hoferichter:2019gzf}, including the energy calibration $\xi_\text{CMD-2}$). $p_\text{conf}$ denotes the number of degrees of freedom in the conformal polynomial $C_p(q^2)$, $n_\text{conf}$ refers to the asymptotic behavior $\Im C_p(q^2)\sim q^{-(2n_\text{conf}+1)}$. All couplings are given in units of $1/e=1/\sqrt{4\pi\alpha}$, in accordance with Ref.~\cite{Hoferichter:2019gzf}. The uncertainties refer to fit errors only, they are not yet rescaled by $S=\sqrt{\chi^2/\text{dof}}$.}
	\label{tab:fits_old_combination}
\end{table}

The results for this updated fit are shown in Table~\ref{tab:fits_old_combination}. First of all, we observe a clear improvement of the $\chi^2/\text{dof}$ when including $\rho$--$\omega$ mixing, from about $1.4$ to $1.2$, which is reflected by the fact that all fits prefer a non-zero value of $\Re\epsrw$ at high significance (about $5\sigma$ in terms of the fit uncertainty). Remarkably, the resulting value of $\Re\epsrw$ comes out largely consistent with the extraction from $e^+e^-\to\pi^+\pi^-$, $\Re\epsrw=1.97(3)\times 10^{-3}$~\cite{Colangelo:2022prz}. Moreover, since the fits are still not perfect, suggesting residual systematic tensions in the data base, we will inflate the final errors by a scale factor $S=\sqrt{\chi^2/\text{dof}}$. Table~\ref{tab:fits_old_combination} also includes the entire HVP integral $a_\mu^{3\pi}$, the FSR contribution $a_\mu^\text{FSR}$, and the $\rho$--$\omega$ mixing  contribution $a_\mu^{\rho\text{--}\omega}$, all integrated up to $1.8\GeV$. To isolate first-order IB effects, we follow Ref.~\cite{Colangelo:2022prz} and define $a_\mu^\text{FSR}$ for $\epsrw=0$ and $a_\mu^{\rho\text{--}\omega}$ with FSR corrections switched off. A more detailed account of the consequences for IB effects in $a_\mu$ will be given in Sec.~\ref{sec:amu}, but one can already anticipate that $a_\mu^{\rho\text{--}\omega}$ comes out large and negative, leading to a significant cancellation with $a_\mu^{\rho\text{--}\omega}[2\pi]=3.68(17)\times 10^{-10}$~\cite{Colangelo:2022prz}.

\begin{table}[t]
	\centering
	\footnotesize
	\renewcommand{\arraystretch}{1.3}
	\begin{tabular}{lcccccc}
	\toprule
	& \multicolumn{3}{c}{$\delta_\eps=3.5\degree$} & \multicolumn{3}{c}{$\delta_\eps$ free}\\
	$p_\text{conf}$ & $2$ & $3$ & $4$ & $2$ & $3$ & $4$\\
	$\chi^2/\text{dof}$ & $276.4/228$ & $271.4/227$ & $269.9/226$ & $275.0/227$ & $271.2/226$& $269.8/225$\\
	& $=1.21$ & $=1.20$ & $=1.19$ & $=1.21$ & $=1.20$& $=1.20$\\
	$p$-value & $0.02$ & $0.02$ & $0.02$ & $0.02$ & $0.02$& $0.02$\\
	$\mw \ [\text{MeV}]$ & $782.69(3)$ & $782.69(3)$ & $782.70(3)$ & $782.70(3)$ & $782.69(3)$& $782.70(3)$\\
	$\Gw \ [\text{MeV}]$ & $8.71(3)$ & $8.73(3)$ & $8.74(4)$  & $8.74(4)$ & $8.74(4)$& $8.73(3)$\\
	$\mphi \ [\text{MeV}]$& $1019.20(1)$ & $1019.19(1)$ & $1019.20(1)$ & $1019.20(1)$ & $1019.19(1)$& $1019.20(2)$\\
	$\Gphi \ [\text{MeV}]$ & $4.25(3)$ & $4.24(3)$ & $4.24(3)$  & $4.25(3)$ & $4.24(3)$& $4.24(3)$\\
	$M_{\omega'} \ [\text{GeV}]$ & $1.432(16)$ & $1.415(24)$& $1.405(19)$& $1.433(17)$& $1.416(24)$& $1.403(19)$\\
	$c_\omega \ [\text{GeV}^{-1}]$ & $2.92(2)$ & $2.93(2)$ & $2.95(3)$ & $2.90(3)$ & $2.92(3)$& $2.95(4)$\\
	$c_\phi \ [\text{GeV}^{-1}]$ & $-0.389(3)$ & $-0.388(3)$ & $-0.387(3)$ & $-0.388(3)$ & $-0.388(3)$& $-0.387(3)$\\
	$c_{\omega'} \ [\text{GeV}^{-1}]$ & $-0.22(4)$ & $-0.12(6)$ & $-0.16(7)$  & $-0.22(4)$ & $-0.12(6)$& $-0.16(7)$\\
	$c_{\omega''} \ [\text{GeV}^{-1}]$ & $-1.65(6)$ & $-1.55(8)$ & $-1.51(9)$ & $-1.63(7)$ & $-1.54(8)$& $-1.50(9)$\\
	$c_1 \ [\text{GeV}^{-3}]$ & $-0.33(8)$ & $-0.24(9)$ & $-0.12(15)$ & $-0.34(8)$ & $-0.25(10)$& $-0.09(18)$\\
	$c_2 \ [\text{GeV}^{-3}]$ & $-1.26(5)$ & $-1.34(7)$ & $-1.43(10)$ & $-1.26(5)$ & $-1.34(7)$& $-1.45(12)$\\
	$c_3 \ [\text{GeV}^{-3}]$ & --- & $-0.54(7)$ & $-0.50(8)$ & --- & $-0.53(7)$& $-0.49(8)$\\
	$c_4 \ [\text{GeV}^{-3}]$ & --- & --- & $1.40(9)$ & --- & --- & $1.42(10)$ \\
	$10^4\times \xi_\text{CMD-2}$ & $1.3(5)$ & $1.2(5)$ & $1.3(5)$ & $1.3(5)$ & $1.2(5)$ & $1.3(5)$ \\
	$10^3\times \Re\epsrw$ & $1.49(29)$ & $1.45(29)$ & $1.51(29)$ & $1.43(29)$ & $1.43(29)$ & $1.54(30)$\\
	$\delta_\eps \ [\degree]$ & $3.5$ & $3.5$ & $3.5$ & $-4.2(7.1)$ & $0.5(7.2)$ & $5.4(7.5)$\\
	$10^{10}\times a_\mu^{3\pi}|_{\leq 1.8\GeV}$ & $45.56(48)$ & $45.73(49)$ & $45.95(53)$ & $45.82(49)$ & $45.82(50)$ & $45.94(57)$\\ 
	$10^{10}\times a_\mu^\text{FSR}[3\pi]$ & $0.51(1)$ & $0.51(1)$ & $0.51(1)$ & $0.50(1)$ & $0.51(1)$ & $0.52(1)$\\
	$10^{10}\times a_\mu^{\rho\text{--}\omega}[3\pi]$ & $-2.99(58)$ & $-2.92(58)$ & $-3.08(58)$ & $-2.13(93)$ & $-2.59(95)$ & $-3.33(1.17)$\\
	\bottomrule
	\renewcommand{\arraystretch}{1.0}
	\end{tabular}
	\caption{Same as Table~\ref{tab:fits_old_combination} ($n_\text{conf}=0$), with $\delta_\eps=3.5\degree$ (left) and a free fit parameter (right).}
	\label{tab:fits_old_combination_phase}
\end{table}

\begin{table}[t]
	\centering
	\footnotesize
	\renewcommand{\arraystretch}{1.3}
	\begin{tabular}{lcccccc}
	\toprule
	& \multicolumn{3}{c}{$n_\text{conf}=0$} & \multicolumn{3}{c}{$n_\text{conf}=1$}\\
	$p_\text{conf}$ & $2$ & $3$ & $4$ & $2$ & $3$ & $4$\\
	$\chi^2/\text{dof}$ & $183.7/130$ & $183.7/129$ & $181.7/128$ & $219.7/130$ & $216.7/129$& $214.7/128$\\
	& $=1.41$ & $=1.42$ & $=1.42$ & $=1.69$ & $=1.68$& $=1.68$\\
	$p$-value & $0.001$ & $0.001$ & $0.001$ & $1\times 10^{-6}$ & $2\times 10^{-6}$& $2\times 10^{-6}$\\
	$\mw \ [\text{MeV}]$ & $782.54(1)$ & $782.54(1)$ & $782.54(1)$ & $782.54(1)$ & $782.54(1)$& $782.54(1)$\\
	$\Gw \ [\text{MeV}]$ & $8.72(2)$ & $8.72(2)$ & $8.72(2)$  & $8.70(2)$ & $8.69(2)$& $8.69(2)$\\
	$\mphi \ [\text{MeV}]$& $1019.28(1)$ & $1019.28(1)$ & $1019.28(1)$ & $1019.28(1)$ & $1019.28(1)$& $1019.28(1)$\\
	$\Gphi \ [\text{MeV}]$ & $4.29(1)$ & $4.29(1)$ & $4.29(1)$  & $4.29(1)$ & $4.29(1)$& $4.29(1)$\\
	$M_{\omega'} \ [\text{GeV}]$ & $1.457(8)$ & $1.456(9)$& $1.471(9)$& $1.394(5)$& $1.400(6)$& $1.411(9)$\\
	$c_\omega \ [\text{GeV}^{-1}]$ & $2.96(2)$ & $2.96(2)$ & $2.95(2)$ & $2.94(2)$ & $2.94(2)$& $2.93(3)$\\
	$c_\phi \ [\text{GeV}^{-1}]$ & $-0.380(2)$ & $-0.380(3)$ & $-0.379(2)$ & $-0.378(2)$ & $-0.378(2)$& $-0.378(2)$\\
	$c_{\omega'} \ [\text{GeV}^{-1}]$ & $-0.37(5)$ & $-0.37(8)$ & $-0.48(10)$  & $-0.46(6)$ & $-0.51(7)$& $-0.53(7)$\\
	$c_{\omega''} \ [\text{GeV}^{-1}]$ & $-2.04(7)$ & $-2.03(14)$ & $-2.44(26)$ & $-1.11(5)$ & $-1.16(5)$& $-1.25(8)$\\
	$c_1 \ [\text{GeV}^{-3}]$ & $0.24(11)$ & $0.23(13)$ & $0.29(11)$ & $-1.11(6)$ & $-1.06(6)$& $-1.07(6)$\\
	$c_2 \ [\text{GeV}^{-3}]$ & $-1.15(4)$ & $-1.16(7)$ & $-0.76(25)$ & $-0.16(7)$ & $-0.14(6)$& $-0.09(8)$\\
	$c_3 \ [\text{GeV}^{-3}]$ & --- & $-1.01(15)$ & $-1.45(27)$ & --- & $-0.33(7)$& $-0.31(7)$\\
	$c_4 \ [\text{GeV}^{-3}]$ & --- & --- & $1.17(12)$ & --- & --- & $-0.09(8)$ \\
	$10^3\times \Re\epsrw$ & $1.73(22)$ & $1.73(22)$ & $1.70(22)$ & $1.85(23)$ & $1.83(23)$ & $1.79(24)$\\
	$10^{10}\times a_\mu^{3\pi}|_{\leq 1.8\GeV}$ & $45.98(44)$ & $45.99(44)$ & $45.84(46)$ & $45.78(44)$ & $45.81(42)$ & $45.74(44)$\\ 
	$10^{10}\times a_\mu^\text{FSR}[3\pi]$ & $0.51(1)$ & $0.51(1)$ & $0.51(1)$ & $0.51(1)$ & $0.51(1)$ & $0.50(1)$\\
	$10^{10}\times a_\mu^{\rho\text{--}\omega}[3\pi]$ & $-3.09(38)$ & $-3.09(38)$ & $-3.02(38)$ & $-3.33(41)$ & $-3.31(41)$ & $-3.24(42)$\\
	\bottomrule
	\renewcommand{\arraystretch}{1.0}
	\end{tabular}
	\caption{Same as Table~\ref{tab:fits_old_combination}, but for the BaBar 2021 data set~\cite{BABAR:2021cde}.}
	\label{tab:fits_BaBar}
\end{table}

\begin{table}[t]
	\centering
	\footnotesize
	\renewcommand{\arraystretch}{1.3}
	\begin{tabular}{lcccccc}
	\toprule
	& \multicolumn{3}{c}{$\delta_\eps=3.5\degree$} & \multicolumn{3}{c}{$\delta_\eps$ free}\\
	$p_\text{conf}$ & $2$ & $3$ & $4$ & $2$ & $3$ & $4$\\
	$\chi^2/\text{dof}$ & $179.7/130$ & $179.6/129$ & $178.1/128$ & $176.3/129$ & $174.3/128$& $174.1/127$\\
	& $=1.38$ & $=1.39$ & $=1.39$ & $=1.37$ & $=1.36$& $=1.37$\\
	$p$-value & $0.003$ & $0.002$ & $0.002$ & $0.004$ & $0.004$& $0.004$\\
	$\mw \ [\text{MeV}]$ & $782.54(1)$ & $782.54(1)$ & $782.54(1)$ & $782.54(1)$ & $782.54(1)$& $782.54(1)$\\
	$\Gw \ [\text{MeV}]$ & $8.71(2)$ & $8.71(2)$ & $8.71(2)$  & $8.71(2)$ & $8.70(2)$& $8.70(2)$\\
	$\mphi \ [\text{MeV}]$& $1019.28(1)$ & $1019.28(1)$ & $1019.28(1)$ & $1019.28(1)$ & $1019.28(1)$& $1019.28(1)$\\
	$\Gphi \ [\text{MeV}]$ & $4.29(1)$ & $4.29(1)$ & $4.29(1)$  & $4.29(1)$ & $4.29(1)$& $4.29(1)$\\
	$M_{\omega'} \ [\text{GeV}]$ & $1.457(2)$ & $1.454(12)$& $1.466(10)$& $1.454(9)$& $1.443(9)$& $1.448(17)$\\
	$c_\omega \ [\text{GeV}^{-1}]$ & $2.98(2)$ & $2.98(2)$ & $2.97(2)$ & $3.01(3)$ & $3.03(3)$& $3.02(3)$\\
	$c_\phi \ [\text{GeV}^{-1}]$ & $-0.380(2)$ & $-0.380(2)$ & $-0.379(2)$ & $-0.379(2)$ & $-0.379(2)$& $-0.379(2)$\\
	$c_{\omega'} \ [\text{GeV}^{-1}]$ & $-0.40(4)$ & $-0.37(8)$ & $-0.45(10)$  & $-0.44(6)$ & $-0.37(7)$& $-0.38(8)$\\
	$c_{\omega''} \ [\text{GeV}^{-1}]$ & $-2.14(7)$ & $-2.08(14)$ & $-2.40(24)$ & $-2.30(10)$ & $-2.18(13)$& $-2.27(21)$\\
	$c_1 \ [\text{GeV}^{-3}]$ & $0.36(10)$ & $0.32(13)$ & $0.35(11)$ & $0.57(14)$ & $0.55(14)$& $0.53(14)$\\
	$c_2 \ [\text{GeV}^{-3}]$ & $-1.13(3)$ & $-1.15(7)$ & $-0.84(23)$ & $-1.08(5)$ & $-1.15(7)$& $-1.06(21)$\\
	$c_3 \ [\text{GeV}^{-3}]$ & --- & $-1.06(16)$ & $-1.39(25)$ & --- & $-1.13(13)$& $-1.21(23)$\\
	$c_4 \ [\text{GeV}^{-3}]$ & --- & --- & $1.25(11)$ & --- & --- & $1.45(11)$ \\
	$10^3\times \Re\epsrw$ & $1.84(22)$ & $1.84(22)$ & $1.81(23)$ & $1.93(23)$ & $1.96(22)$ & $1.95(23)$\\
	$\delta_\eps \ [\degree]$ & $3.5$ & $3.5$ & $3.5$ & $9.8(3.1)$ & $12.2(3.2)$ & $11.9(3.5)$\\
	$10^{10}\times a_\mu^{3\pi}|_{\leq 1.8\GeV}$ & $45.92(43)$ & $45.95(45)$ & $45.82(45)$ & $45.77(44)$ & $45.82(45)$ & $45.79(49)$\\ 
	$10^{10}\times a_\mu^\text{FSR}[3\pi]$ & $0.52(1)$ & $0.52(1)$ & $0.52(1)$ & $0.53(1)$ & $0.54(1)$ & $0.54(1)$\\
	$10^{10}\times a_\mu^{\rho\text{--}\omega}[3\pi]$ & $-3.75(45)$ & $-3.75(45)$ & $-3.67(45)$ & $-4.83(72)$ & $-5.31(72)$ & $-5.20(82)$\\
	\bottomrule
	\renewcommand{\arraystretch}{1.0}
	\end{tabular}
	\caption{Same as Table~\ref{tab:fits_old_combination_phase}, but for the BaBar 2021 data set~\cite{BABAR:2021cde}.}
	\label{tab:fits_BaBar_phase}
\end{table}

To understand how robust this cancellation is, it is critical to study the systematic uncertainties in $a_\mu^{\rho\text{--}\omega}[3\pi]$. As argued in Sec.~\ref{sec:rhoomega}, the main effect is expected from the assumptions on the line shape, which we determined in such a way that analyticity and unitarity constraints from the coupled-channel system are incorporated, in order to ensure consistency with the $2\pi$ contribution. Beyond such considerations, the analysis from Ref.~\cite{Colangelo:2022prz} demonstrates that in the $2\pi$ case the biggest impact on the line shape arises from the small IB phase $\delta_\eps$ in $\epsrw$, generated by $\pi^0\gamma$ and other radiative channels. To quantify its impact, we consider three scenarios: (i) $\delta_\eps=0$ (as assumed in Table~\ref{tab:fits_old_combination}), (ii) $\delta_\eps=3.5\degree$ (as expected from narrow-resonance arguments~\cite{Colangelo:2022prz}), and (iii) a free phase $\delta_\eps$ as additional fit parameter. To avoid unphysical imaginary parts below threshold, we implement this phase via
\beq
\label{phase}
\epsrw\to\Re\epsrw + i\Im \epsrw \frac{\Big(1-\frac{\mpii^2}{q^2}\Big)^3}{\Big(1-\frac{\mpii^2}{\mw^2}\Big)^3}\theta\big(q^2-\mpii^2\big), 
\eeq
motivated by the main decay channel $\rho\to\pi^0\gamma\to\omega$ that can generate such a phase. The results for (ii) and (iii) collected in Table~\ref{tab:fits_old_combination_phase} show that the data are not sensitive to $\delta_\eps$, but the variation in $a_\mu^{\rho\text{--}\omega}$ provides some indication for the uncertainty associated with the assumed line shape.  

\subsection{Fits to BaBar 2021}
\label{sec:BaBar}

Next, we perform the same fits as in Sec.~\ref{sec:old_data_base} to the BaBar data~\cite{BABAR:2021cde}. This data set is split into two parts, above and below $\sqrt{q^2}=1.1\GeV$. For the data set below $1.1\GeV$, we use the statistical and systematic covariance matrices as provided in Ref.~\cite{BABAR:2021cde}, for the data set above $1.1\GeV$ we assume the systematic errors to be $100\%$ correlated. In either case  we use the bare cross sections as provided, again interpreted as including soft FSR effects.

\begin{table}[t]
	\centering
	\footnotesize
	\renewcommand{\arraystretch}{1.3}
	\begin{tabular}{lcccccc}
	\toprule
	& \multicolumn{3}{c}{$n_\text{conf}=0$} & \multicolumn{3}{c}{$n_\text{conf}=1$}\\
	$p_\text{conf}$ & $2$ & $3$ & $4$ & $2$ & $3$ & $4$\\
	$\chi^2/\text{dof}$ & $504.9/368$ & $494.7/367$ & $481.9/366$ & $545.9/368$ & $537.8/367$& $537.3/366$\\
	& $=1.37$ & $=1.35$ & $=1.32$ & $=1.48$ & $=1.47$& $=1.47$\\
	$p$-value & $3\times 10^{-6}$ & $1\times 10^{-5}$ & $4\times 10^{-5}$ & $4\times 10^{-9}$ & $1\times 10^{-8}$& $1\times 10^{-8}$\\
	$\mw \ [\text{MeV}]$ & $782.70(3)$ & $782.70(3)$ & $782.70(3)$ & $782.71(3)$ & $782.70(3)$& $782.70(3)$\\
	$\Gw \ [\text{MeV}]$ & $8.70(2)$ & $8.71(2)$ & $8.72(2)$  & $8.71(2)$ & $8.70(2)$& $8.70(2)$\\
	$\mphi \ [\text{MeV}]$& $1019.21(1)$ & $1019.21(1)$ & $1019.21(1)$ & $1019.22(1)$ & $1019.22(1)$& $1019.22(1)$\\
	$\Gphi \ [\text{MeV}]$ & $4.27(1)$ & $4.27(1)$ & $4.27(1)$  & $4.27(1)$ & $4.27(1)$& $4.27(1)$\\
	$M_{\omega'} \ [\text{GeV}]$ & $1.445(10)$ & $1.436(23)$& $1.418(11)$& $1.395(6)$& $1.403(6)$& $1.408(9)$\\
	$c_\omega \ [\text{GeV}^{-1}]$ & $2.93(1)$ & $2.93(1)$ & $2.96(2)$ & $2.95(1)$ & $2.94(1)$& $2.94(2)$\\
	$c_\phi \ [\text{GeV}^{-1}]$ & $-0.380(1)$ & $-0.380(1)$ & $-0.381(2)$ & $-0.381(1)$ & $-0.381(1)$& $-0.381(1)$\\
	$c_{\omega'} \ [\text{GeV}^{-1}]$ & $-0.24(3)$ & $-0.15(4)$ & $-0.23(5)$  & $-0.29(3)$ & $-0.38(4)$& $-0.37(5)$\\
	$c_{\omega''} \ [\text{GeV}^{-1}]$ & $-1.77(4)$ & $-1.67(5)$ & $-1.59(6)$ & $-1.02(4)$ & $-1.09(4)$& $-1.13(7)$\\
	$c_1 \ [\text{GeV}^{-3}]$ & $-0.18(5)$ & $-0.13(6)$ & $0.03(8)$ & $-1.16(5)$ & $-1.11(5)$& $-1.11(5)$\\
	$c_2 \ [\text{GeV}^{-3}]$ & $-1.21(3)$ & $-1.29(4)$ & $-1.43(6)$ & $-0.16(5)$ & $-0.16(4)$& $-0.15(5)$\\
	$c_3 \ [\text{GeV}^{-3}]$ & --- & $-0.65(5)$ & $-0.57(6)$ & --- & $-0.41(5)$& $-0.40(5)$\\
	$c_4 \ [\text{GeV}^{-3}]$ & --- & --- & $1.35(5)$ & --- & --- & $-0.03(5)$ \\
	$10^4\times \xi_\text{CMD-2}$ & $1.4(5)$ & $1.3(5)$ & $1.3(5)$ & $1.4(5)$ & $1.4(5)$ & $1.4(5)$ \\
	$10^3\times \xi_\text{BaBar}$ & $1.3(2)$ & $1.3(2)$ & $1.3(2)$ & $1.3(2)$ & $1.3(2)$ & $1.3(2)$ \\
	$10^3\times \xi_\text{BaBar}' \ [\text{GeV}^{-1}]$ & $-2.3(3)$ & $-2.3(4)$ & $-2.3(3)$ & $-2.3(4)$ & $-2.3(4)$ & $-2.3(4)$ \\
	$10^3\times \Re\epsrw$ & $1.51(18)$ & $1.49(18)$ & $1.60(17)$ & $1.71(18)$ & $1.68(17)$ & $1.65(19)$\\
	$10^{10}\times a_\mu^{3\pi}|_{\leq 1.8\GeV}$ & $45.74(31)$ & $45.91(32)$ & $46.26(33)$ & $45.96(31)$ & $45.92(31)$ & $45.86(32)$\\ 
	$10^{10}\times a_\mu^\text{FSR}[3\pi]$ & $0.51(0)$ & $0.51(0)$ & $0.52(0)$ & $0.51(0)$ & $0.51(0)$ & $0.51(0)$\\
	$10^{10}\times a_\mu^{\rho\text{--}\omega}[3\pi]$ & $-2.70(31)$ & $-2.68(31)$ & $-2.91(30)$ & $-3.14(31)$ & $-3.08(30)$ & $-3.03(33)$\\
	\bottomrule
	\renewcommand{\arraystretch}{1.0}
	\end{tabular}
	\caption{Same as Table~\ref{tab:fits_old_combination}, but for the global fit to SND~\cite{Achasov:2000am,Achasov:2002ud,Achasov:2003ir,SND:2020ajg}, CMD-2$'$~\cite{Akhmetshin:1995vz,Akhmetshin:1998se,Akhmetshin:2003zn,Akhmetshin:2006sc}, and BaBar~\cite{BABAR:2021cde}.}
	\label{tab:fits_global}
\end{table}

In addition to the iterative procedure required to obtain unbiased fit results, another complication for data taken using initial-state radiation (ISR) concerns the energy calibration. In contrast to the energy-scan experiments SND or CMD-2, the cross-section data do not correspond to a set beam energy, but are provided in bins, with events distributed in accordance with the underlying cross section. Accordingly, the actual observable for a bin $\big[q^2_{i, \text{min}},q^2_{i, \text{max}}\big]$ is given by
\beq
  f(x_i) = \frac{1}{q^2_{i, \text{max}}-q^2_{i, \text{min}}} \int_{q^2_{i, \text{min}}}^{q^2_{i, \text{max}}} \diff q^2\,  \sigma_{e^+e^-\to 3\pi(\gamma)}(q^2),
\eeq
or, equivalently, the actual $q_i^2$, replacing the center of the bin, can be obtained by solving  $f(x_i)=\sigma_{e^+e^-\to3\pi(\gamma)}(q_i^2)$. The results of the fits are summarized in Tables~\ref{tab:fits_BaBar} and~\ref{tab:fits_BaBar_phase}. In general, the conclusions regarding $\rho$--$\omega$ mixing are similar as for the previous fits in Sec.~\ref{sec:old_data_base}. While there is some indication that a positive phase is favored, the gain in the $\chi^2$ is marginal, and we conclude that also in this case the data are hardly sensitive to $\delta_\eps$. The real part $\Re\epsrw$ comes out slightly larger, but, within uncertainties,  in agreement with the direct-scan experiments. More problematic is the discrepancy in the pole parameters of $\omega$ and $\phi$, with $\mw$ significantly below the values extracted from the direct-scan experiments, and $\mphi$ significantly above.

\begin{table}[t]
	\centering
	\footnotesize
	\renewcommand{\arraystretch}{1.3}
	\begin{tabular}{lcccccc}
	\toprule
	& \multicolumn{3}{c}{$\delta_\eps=3.5\degree$} & \multicolumn{3}{c}{$\delta_\eps$ free}\\
	$p_\text{conf}$ & $2$ & $3$ & $4$ & $2$ & $3$ & $4$\\
	$\chi^2/\text{dof}$ & $512.7/368$ & $497.2/367$ & $479.4/366$ & $498.4/367$ & $494.1/366$& $478.3/365$\\
	& $=1.39$ & $=1.35$ & $=1.31$ & $=1.36$ & $=1.35$& $=1.31$\\
	$p$-value & $8\times 10^{-7}$ & $7\times 10^{-6}$ & $6\times 10^{-5}$ & $6\times 10^{-6}$ & $9\times 10^{-6}$& $6\times 10^{-5}$\\
	$\mw \ [\text{MeV}]$ & $782.69(2)$ & $782.70(3)$ & $782.70(2)$ & $782.70(3)$ & $782.70(2)$& $782.70(2)$\\
	$\Gw \ [\text{MeV}]$ & $8.69(2)$ & $8.70(2)$ & $8.72(2)$  & $8.72(2)$ & $8.71(2)$& $8.71(2)$\\
	$\mphi \ [\text{MeV}]$& $1019.22(1)$ & $1019.21(1)$ & $1019.21(1)$ & $1019.21(1)$ & $1019.21(1)$& $1019.22(1)$\\
	$\Gphi \ [\text{MeV}]$ & $4.27(1)$ & $4.27(1)$ & $4.27(1)$  & $4.27(1)$ & $4.27(1)$& $4.27(1)$\\
	$M_{\omega'} \ [\text{GeV}]$ & $1.444(9)$ & $1.432(5)$& $1.412(10)$& $1.445(10)$& $1.437(3)$& $1.407(6)$\\
	$c_\omega \ [\text{GeV}^{-1}]$ & $2.93(1)$ & $2.94(2)$ & $2.97(2)$ & $2.91(2)$ & $2.93(2)$& $2.99(1)$\\
	$c_\phi \ [\text{GeV}^{-1}]$ & $-0.380(1)$ & $-0.380(1)$ & $-0.382(1)$ & $-0.381(1)$ & $-0.380(1)$& $-0.381(1)$\\
	$c_{\omega'} \ [\text{GeV}^{-1}]$ & $-0.25(3)$ & $-0.13(4)$ & $-0.23(5)$  & $-0.23(3)$ & $-0.16(4)$& $-0.24(2)$\\
	$c_{\omega''} \ [\text{GeV}^{-1}]$ & $-1.79(4)$ & $-1.68(5)$ & $-1.59(6)$ & $-1.72(5)$ & $-1.67(5)$& $-1.58(1)$\\
	$c_1 \ [\text{GeV}^{-3}]$ & $-0.15(5)$ & $-0.10(5)$ & $0.11(8)$ & $-0.22(6)$ & $-0.15(6)$& $0.19(4)$\\
	$c_2 \ [\text{GeV}^{-3}]$ & $-1.20(3)$ & $-1.30(4)$ & $-1.48(5)$ & $-1.22(3)$ & $-1.28(4)$& $-1.51(3)$\\
	$c_3 \ [\text{GeV}^{-3}]$ & --- & $-0.65(4)$ & $-0.56(5)$ & --- & $-0.65(5)$& $-0.55(1)$\\
	$c_4 \ [\text{GeV}^{-3}]$ & --- & --- & $1.40(5)$ & --- & --- & $1.44(2)$ \\
	$10^4\times \xi_\text{CMD-2}$ & $1.4(5)$ & $1.3(5)$ & $1.3(5)$ & $1.3(5)$ & $1.2(5)$ & $1.3(5)$ \\
	$10^3\times \xi_\text{BaBar}$ & $1.3(2)$ & $1.3(2)$ & $1.3(2)$ & $1.3(2)$ & $1.3(2)$ & $1.3(2)$ \\
	$10^3\times \xi_\text{BaBar}' \ [\text{GeV}^{-1}]$ & $-2.3(3)$ & $-2.3(4)$ & $-2.3(3)$ & $-2.3(3)$ & $-2.3(3)$ & $-2.3(3)$ \\
	$10^3\times \Re\epsrw$ & $1.45(18)$ & $1.49(18)$ & $1.68(17)$ & $1.44(18)$ & $1.47(17)$ & $1.72(10)$\\
	$\delta_\eps \ [\degree]$ & $3.5$ & $3.5$ & $3.5$ & $-7.2(3.3)$ & $-2.2(3.6)$ & $6.8(1.8)$\\
	$10^{10}\times a_\mu^{3\pi}|_{\leq 1.8\GeV}$ & $45.61(30)$ & $45.81(31)$ & $46.22(32)$ & $46.00(31)$ & $45.97(32)$ & $46.17(26)$\\ 
	$10^{10}\times a_\mu^\text{FSR}[3\pi]$ & $0.51(0)$ & $0.51(0)$ & $0.52(0)$ & $0.50(0)$ & $0.51(0)$ & $0.53(0)$\\
	$10^{10}\times a_\mu^{\rho\text{--}\omega}[3\pi]$ & $-2.93(36)$ & $-3.04(37)$ & $-3.47(34)$ & $-1.87(46)$ & $-2.42(51)$ & $-3.97(34)$\\
	\bottomrule
	\renewcommand{\arraystretch}{1.0}
	\end{tabular}
	\caption{Same as Table~\ref{tab:fits_old_combination_phase}, but for the global fit to SND~\cite{Achasov:2000am,Achasov:2002ud,Achasov:2003ir,SND:2020ajg}, CMD-2$'$~\cite{Akhmetshin:1995vz,Akhmetshin:1998se,Akhmetshin:2003zn,Akhmetshin:2006sc}, and BaBar~\cite{BABAR:2021cde}.}
	\label{tab:fits_global_phase}
\end{table}

\subsection{Global fit}
\label{sec:fit_global}

From the fits presented in Secs.~\ref{sec:old_data_base} and~\ref{sec:BaBar} it is clear that some tensions in the data base are present that will prevent a global fit of acceptable fit quality, in fact, already the fits to the BaBar data~\cite{BABAR:2021cde} alone display rather low $p$-values.  
In the end, we will attempt to remedy this shortcoming by introducing scale factors $S=\sqrt{\chi^2/\text{dof}}$ to try and include unaccounted-for systematic effects. More critical than the overall fit quality
is the mismatch in $\mw$ and $\mphi$ into opposite directions, which cannot be resolved via a linear shift in the energy calibration, as was included in Ref.~\cite{Hoferichter:2019gzf} for part of the CMD-2 data~\cite{Akhmetshin:2006sc} and vital for a global analysis of $e^+e^-\to\pi^+\pi^-$~\cite{Colangelo:2018mtw,Colangelo:2022prz}. However, a consistent energy calibration of ISR data covering both the $\omega$ and $\phi$ resonances is challenging,  as reflected by the additional uncertainties $\Delta\mw=0.06\MeV$ and $\Delta\mphi=0.08\MeV$ quoted in
Ref.~\cite{BABAR:2021cde}.\footnote{We thank M.~Davier and V.~Druzhinin for their assessment of the expected accuracy of the energy calibration in the ISR data.} We emphasize that the agreement with PDG parameters found in
Ref.~\cite{BABAR:2021cde} is accidental, relying on including the $\omega$ mass determination from $\bar p p\to \omega\pi^0\pi^0$~\cite{CrystalBarrel:1993gtk} in the average despite being in conflict with $e^+e^-\to 3\pi$, but acknowledge that the associated uncertainties make it appear likely that the energy calibration in the direct-scan data should be considered more robust. To account for the tensions in $\mw$ and $\mphi$ in a minimal fashion, we thus allow for a quadratic energy rescaling
\beq
\label{rescaling}
\sqrt{s}\to\sqrt{s}+\xi\big(\sqrt{s}-3\mpi\big)+\xi'\big(\sqrt{s}-3\mpi\big)^2
\eeq
in the fit to Ref.~\cite{BABAR:2021cde}.\footnote{We apply this rescaling only to the data set below $1.1\GeV$, since no tensions arise in the fit of the data above.} The  results of this global fit are summarized in Tables~\ref{tab:fits_global} and~\ref{tab:fits_global_phase}. In particular, the comparison of the fits with $\delta_\eps=0$, $\delta_\eps=3.5\degree$, and a free $\delta_\eps$ again shows that the sensitivity to this parameter is small, with marginal changes in the fit quality. In contrast, fits with improved asymptotic behavior, $n_\text{conf}=1$, do display a significantly worse $\chi^2/\text{dof}$.

\begin{figure}[t]
	\centering
	\includegraphics[width=\linewidth]{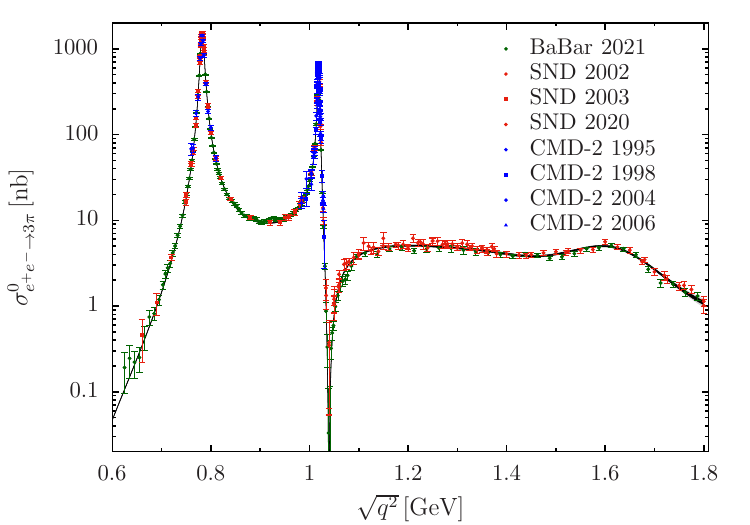}
	\caption{Fit to the bare $e^+e^-\to3\pi$ data sets as detailed in Sec.~\ref{sec:old_data_base}. The gray band shows the total uncertainty, while the black band represents the fit uncertainty only.   The difference between the two is hardly visible on the logarithmic plot, as they are of similar size in most regions. 
	}
	\label{fig:cross_section}
\end{figure}

\begin{figure}[t]
	\centering
	\includegraphics[width=0.49\linewidth,clip]{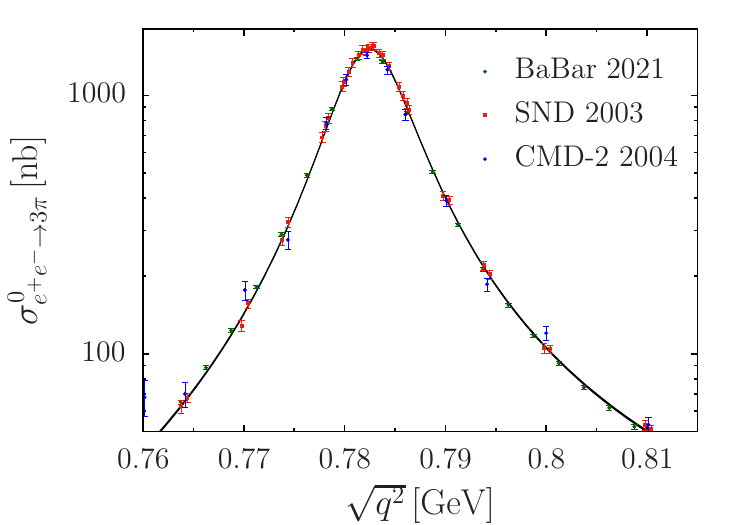}
	\includegraphics[width=0.49\linewidth,clip]{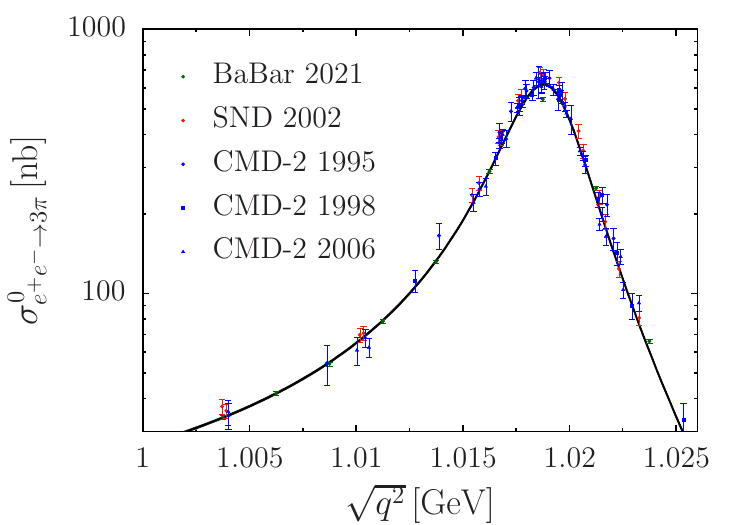}
	\caption{Same as Fig.~\ref{fig:cross_section}, but for the close-up views of  the $\omega$ and $\phi$ resonance regions.}
	\label{fig:cross_section_res}
\end{figure}

To assign uncertainties to our results we thus proceed as follows. First, the statistical errors are inflated by the scale factor $S=\sqrt{\chi^2/\text{dof}}$. Following Ref.~\cite{Hoferichter:2019gzf}, we take the fits with $p_\text{conf}=3$ to define the central values, as several fits with $p_\text{conf}=4$ already display signs of overfitting. The systematic error from the truncation of the conformal polynomial is then estimated as the maximum difference compared to the fit variants with $p_\text{conf}=2,4$. In addition, in Ref.~\cite{Hoferichter:2019gzf} we included the variation to fits with $n_\text{conf}=1$, but given the observations above this recipe no longer appears appropriate with our improved dispersive formalism. The remaining uncertainties are better represented by scanning over the sensitivity to $\delta_\eps$, given that these fits are not distinguished by the $\chi^2$ criterion. In view of the narrow-width arguments in favor of a small phase $\delta_\eps=3.5(1.0)\degree$~\cite{Colangelo:2022prz}, combined with the lack of sensitivity to this phase in the $e^+e^-\to3\pi$ data themselves, we quote the results at $\delta_\eps=0$ as central values, while assigning the change to $\delta_\eps=3.5\degree$ as an additional source of systematic uncertainty.  Our central fit is illustrated in Fig.~\ref{fig:cross_section},  with zoom-in views of the $\omega$ and $\phi$ regions in Fig.~\ref{fig:cross_section_res}.

With this procedure, we find
\begin{align}
\label{resonance_parameters}
 \mw&=782.697(32)(4)(4)[32]\MeV, & \Gw&=8.711(21)(12)(10)[26]\MeV,\notag\\
 \mphi&=1019.211(17)(4)(1)[17]\MeV, & \Gphi&=4.270(13)(3)(1)[13]\MeV,\notag\\
 M_{\omega'}&=1436(26)(17)(6)[32]\MeV, & &\notag\\
 \Re\epsrw&=1.49(21)(11)(8)[25]\times 10^{-3},
\end{align}
where the errors refer to statistics, truncation of the conformal polynomial, dependence on $\delta_\eps$, and quadratic sum, respectively. The mass of the $\omega'$ comes out in agreement with Eq.~\eqref{omegap}, the mixing parameter about $1.9\sigma$ below the expectation from $e^+e^-\to 2\pi$. The comparison of the $\omega$ and $\phi$ resonance parameters to our previous determinations~\cite{Hoferichter:2019gzf,Hoid:2020xjs,Stamen:2022uqh} as well as the PDG values~\cite{ParticleDataGroup:2022pth} is given in Table~\ref{tab:vector_mesons}. For the mass parameters, the main change concerns the improved functional form of the representation including $\rho$--$\omega$ interference, which leads to an increase in $\mw$ of $0.06\MeV$, while the change in $\mphi$ is much smaller. In both cases, the precision hardly changes when including the BaBar data~\cite{BABAR:2021cde}, ultimately due to the necessity of the energy rescaling~\eqref{rescaling}. In contrast, the uncertainties in the widths decrease appreciably when including Ref.~\cite{BABAR:2021cde}, about a factor $2$ for $\Gw$ and a factor $4$ for $\Gphi$. In the latter case, the determination from $e^+e^-\to 3\pi$ is now competitive with $e^+e^-\to\bar K K$, which dominates the corresponding PDG average. We find agreement with the PDG values in all cases, albeit for $\mw$ only due to the scale factor $S=2.0$ included in the PDG uncertainty, reflecting the conflict between $e^+e^-\to 3\pi$ and $\bar p p\to \omega\pi^0\pi^0$ alluded to above.
For $e^+e^-\to\pi^0\gamma$ and $e^+e^-\to\bar K K$ we observe mostly good agreement as well, except for $\mw$ in the $\pi^0\gamma$ channel, which comes out slightly lower than in $3\pi$, and $\Gphi$ in the $\bar K K$ channel, in which case the $3\pi$ and $\bar K K$ determinations are not compatible within uncertainties.

\begin{table}[t]
\small
	\centering	\renewcommand{\arraystretch}{1.3}
	\begin{tabular}{lccccc}
	\toprule
	& $e^+e^-\to \pi^0\gamma$ & $e^+e^-\to\bar K K$ & \multicolumn{2}{c}{$e^+e^-\to 3\pi$} &\\
	& Ref.~\cite{Hoid:2020xjs} & Ref.~\cite{Stamen:2022uqh} & Ref.~\cite{Hoferichter:2019gzf} & this work & PDG~\cite{ParticleDataGroup:2022pth}\\\midrule
	$\mw \ [\text{MeV}]$ & $782.584(28)$ & -- & $782.631(28)$ & $782.697(32)$ & $782.53(13)$\\
	$\Gw \ [\text{MeV}]$ & $8.65(6)$ & -- & $8.71(6)$ & $8.711(26)$ & $8.74(13)$\\
	$\mphi \ [\text{MeV}]$ & $1019.205(55)$ & $1019.219(4)$ & $1019.196(21)$ & $1019.211(17)$ & $1019.201(16)$\\
	$\Gphi \ [\text{MeV}]$ & $4.07(13)$ & $4.207(8)$ & $4.23(4)$ & $4.270(13)$ & $4.249(13)$\\
	\bottomrule
	\renewcommand{\arraystretch}{1.0}
	\end{tabular}
	\caption{VP-subtracted resonance parameters of $\omega$ and $\phi$ in comparison to our previous determinations from $e^+e^-\to 3\pi$~\cite{Hoferichter:2019gzf}, $e^+e^-\to\pi^0\gamma$~\cite{Hoid:2020xjs}, and $e^+e^-\to\bar K K$~\cite{Stamen:2022uqh}. The last column gives the PDG values~\cite{ParticleDataGroup:2022pth}, with VP removed using the corrections from Ref.~\cite{Holz:2022hwz}.}
	\label{tab:vector_mesons}
\end{table}

\section{Consequences for the anomalous magnetic moment of the muon}
\label{sec:amu}

As key application, we reevaluate the $3\pi$ contribution to HVP, including the separate effects from radiative corrections and $\rho$--$\omega$ mixing. Defining the latter contributions as the leading term in the corresponding IB parameters $e^2$ and $\epsrw$,  we find
\begin{align}
\label{central}
 a_\mu^{3\pi}|_{\leq 1.8\GeV}&=45.91(37)(35)(13)[53]\times 10^{-10},\notag\\
 a_\mu^\text{FSR}[3\pi]&=0.509(4)(6)(6)[9]\times 10^{-10},\notag\\ a_\mu^{\rho\text{--}\omega}[3\pi]&=-2.68(36)(22)(56)[70]\times 10^{-10},
\end{align}
where the errors again refer to statistics, truncation of the conformal polynomial, dependence on $\delta_\eps$, and quadratic sum, respectively.\footnote{We emphasize that the error for FSR does not include an estimate for the subleading, non-IR-enhanced terms. In the $2\pi$ case, such corrections amount to $3\%$~\cite{Moussallam:2013una}, which would translate here to an additional uncertainty of $0.015\times 10^{-10}$.} For the total contribution, both the statistical and systematic errors have decreased by almost a factor $2$ compared to $a_\mu^{3\pi}|_{\leq 1.8\GeV}=46.2(6)(6)\times 10^{-10}$~\cite{Hoferichter:2019gzf}, which traces back to including the BaBar data~\cite{BABAR:2021cde} and to the improved dispersive representation constructed in this paper. Concerning the IB corrections, the FSR piece comports with the naive scaling expectation $a_\mu^\text{FSR}[3\pi]\simeq a_\mu^\text{FSR}[2\pi] a_\mu^{3\pi}/a_\mu^{2\pi}\simeq 0.4\times 10^{-10}$,  while $a_\mu^{\rho\text{--}\omega}[3\pi]$ indeed comes out large and negative, canceling a significant portion of $a_\mu^{\rho\text{--}\omega}[2\pi]=3.68(17)\times 10^{-10}$~\cite{Colangelo:2022prz}, for the reasons anticipated in Sec.~\ref{sec:rhoomega}. The sensitivity to the assumed line shape is clearly reflected by the uncertainties quoted in Eq.~\eqref{central},  as $a_\mu^{\rho\text{--}\omega}[3\pi]$ is the only quantity for which the error derived from the variation in $\delta_\eps$ dominates. Finally, we also provide the decomposition of the total HVP integral onto the Euclidean-time windows from Ref.~\cite{RBC:2018dos}, see Table~\ref{tab:windows}. 

\begin{table}[t]
\small
	\centering	\renewcommand{\arraystretch}{1.3}
	\begin{tabular}{lccc}
	\toprule
	& $10^{10}\times a_\mu^{3\pi}|_{\leq 1.8\GeV}$ & $10^{10}\times a_\mu^\text{FSR}[3\pi]$ & $10^{10}\times a_\mu^{\rho\text{--}\omega}[3\pi]$\\\midrule
short-distance window & $2.51(2)(1)(0)[2]$ & $0.026(0)(0)(0)[0]$ & $-0.13(2)(1)(3)[3]$\\
intermediate window & $18.27(15)(12)(5)[20]$ & $0.199(2)(2)(2)[4]$ & $-1.03(14)(9)(21)[27]$\\
long-distance window & $25.13(20)(22)(8)[31]$ & $0.284(2)(4)(3)[6]$ & $-1.52(20)(12)(33)[40]$\\
total & $45.91(37)(35)(13)[53]$ & $0.509(4)(6)(6)[9]$ & $-2.68(36)(22)(56)[70]$\\\bottomrule
\renewcommand{\arraystretch}{1.0}
	\end{tabular}
	\caption{Decomposition of the total $3\pi$ HVP contribution as well as the FSR and $\rho$--$\omega$ components onto the short-distance, intermediate, and long-distance  windows from Ref.~\cite{RBC:2018dos}.}
	\label{tab:windows}
\end{table}

\section{Conclusions}
\label{sec:conclusions}

In this work, we developed the necessary formalism to describe the leading isospin-breaking effects in $e^+e^-\to 3\pi$, which originate from infrared-enhanced radiative corrections and the interference of $\rho$ and $\omega$ resonances. For the former, we made use of the fact that the dominant effects arise as a remnant of the cancellation of infrared singularities between certain virtual-photon diagrams and bremsstrahlung corrections, leading to a generalization of the standard inclusive FSR correction factor $\eta_{2\pi}(q^2)$ for $e^+e^-\to \pi^+\pi^-$, see Sec.~\ref{sec:EM} for the derivation of the resulting $\eta_{3\pi}(q^2)$. For $\rho$--$\omega$ mixing, we presented an implementation based on a coupled-channel system for $e^+e^-$, $\pi^+\pi^-$, and $3\pi$, preserving analyticity and unitarity properties and predicting the line shape of the $\rho$ in a way consistent with the dispersive representation of the self energies in the multichannel system.  
In particular, our approach allows us to make the connection with the $\rho$--$\omega$ mixing parameter $\epsrw$ in $e^+e^-\to\pi^+\pi^-$ manifest.  

Based on this improved dispersive representation of the $e^+e^-\to 3\pi$ cross section, we performed a phenomenological analysis including the latest data from the BaBar experiment. First, we observed that including $\rho$--$\omega$ mixing in the description markedly improves the fit quality, demonstrating that the effect is visible in the data and can be distinguished from background despite the broad nature of the $\rho$. For the quantitative analysis, the main uncertainty arises from the line shape of the $\rho$--$\omega$ mixing contribution, which, in our framework, can be estimated by investigating the dependence on a small phase of $\epsrw$, as generated by radiative decay channels. We found that $\epsrw$ comes out slightly smaller than expected from $e^+e^-\to 2\pi$, yet in view of the associated uncertainties still indicating a remarkable consistency between the two channels. As a first application, we provided the resonance parameters of $\omega$ and $\phi$ that correspond to the global fit to the $3\pi$ data base, see Eq.~\eqref{resonance_parameters} for the final results. 

The main application concerns the $3\pi$ channel in the hadronic-vacuum-polarization contribution to  the anomalous magnetic moment of the muon. First, with new data from BaBar and our improved dispersive representation, both the statistical and systematic uncertainties reduce by almost a factor $2$, see Eq.~\eqref{central} for the key results. Moreover, we can quantify the contribution of the $3\pi\gamma$ channel, estimated as the combined effect of the dominant infrared-enhanced radiative corrections, as well as the impact of the $\rho$--$\omega$ interference. While the former scales as expected from the total size of the $2\pi$ and $3\pi$ channels, the latter is large and negative, canceling a substantial part of the $\rho$--$\omega$ mixing contribution in the $2\pi$ channel. This cancellation can be understood in terms of narrow-width arguments, see Sec.~\ref{sec:rhoomega}, and likely points to a general interplay between the two channels. Our results corroborate the evaluation of the $3\pi$ channel with reduced uncertainties, and provide crucial input to a phenomenological analysis of isospin-breaking effects in the hadronic-vacuum-polarization contribution to  the anomalous magnetic moment of the muon~\cite{Hoferichter:2022iqe,Hoferichter:2023sli}.

\acknowledgments   
We thank M.~Davier and V.~Druzhinin for helpful communication on Ref.~\cite{BABAR:2021cde}, Dominik Stamen for providing $3\pi$ KT basis functions, and Janak Prabhu for collaboration on the electromagnetic corrections in an early stage of this project.
Financial support by the DFG through the funds provided to the Sino--German Collaborative
Research Center TRR110 ``Symmetries and the Emergence of Structure in QCD''
(DFG Project-ID 196253076 -- TRR 110) and the SNSF (Project No.\ PCEFP2\_181117) is gratefully acknowledged. 
MH thanks the INT at the University of Washington for its hospitality and the DOE for partial support (grant No.\ DE-FG02-00ER41132) during
a visit when part of this work was performed.


\bibliographystyle{apsrev4-1_mod_2}
\bibliography{ref}

\begin{thebibliography}{147}%
\makeatletter
\providecommand \@ifxundefined [1]{%
 \@ifx{#1\undefined}
}%
\providecommand \@ifnum [1]{%
 \ifnum #1\expandafter \@firstoftwo
 \else \expandafter \@secondoftwo
 \fi
}%
\providecommand \@ifx [1]{%
 \ifx #1\expandafter \@firstoftwo
 \else \expandafter \@secondoftwo
 \fi
}%
\providecommand \natexlab [1]{#1}%
\providecommand \enquote  [1]{``#1''}%
\providecommand \bibnamefont  [1]{#1}%
\providecommand \bibfnamefont [1]{#1}%
\providecommand \citenamefont [1]{#1}%
\providecommand \href@noop [0]{\@secondoftwo}%
\providecommand \href [0]{\begingroup \@sanitize@url \@href}%
\providecommand \@href[1]{\@@startlink{#1}\@@href}%
\providecommand \@@href[1]{\endgroup#1\@@endlink}%
\providecommand \@sanitize@url [0]{\catcode `\\12\catcode `\$12\catcode
  `\&12\catcode `\#12\catcode `\^12\catcode `\_12\catcode `\%12\relax}%
\providecommand \@@startlink[1]{}%
\providecommand \@@endlink[0]{}%
\providecommand \url  [0]{\begingroup\@sanitize@url \@url }%
\providecommand \@url [1]{\endgroup\@href {#1}{\urlprefix }}%
\providecommand \urlprefix  [0]{URL }%
\providecommand \Eprint [0]{\href }%
\providecommand \doibase [0]{http://dx.doi.org/}%
\providecommand \selectlanguage [0]{\@gobble}%
\providecommand \bibinfo  [0]{\@secondoftwo}%
\providecommand \bibfield  [0]{\@secondoftwo}%
\providecommand \translation [1]{[#1]}%
\providecommand \BibitemOpen [0]{}%
\providecommand \bibitemStop [0]{}%
\providecommand \bibitemNoStop [0]{.\EOS\space}%
\providecommand \EOS [0]{\spacefactor3000\relax}%
\providecommand \BibitemShut  [1]{\csname bibitem#1\endcsname}%
\let\auto@bib@innerbib\@empty
\bibitem [{\citenamefont {Abi}\ \emph {et~al.}(2021)\citenamefont {Abi} \emph
  {et~al.}}]{Muong-2:2021ojo}%
  \BibitemOpen
  \bibfield  {author} {\bibinfo {author} {\bibfnamefont {B.}~\bibnamefont
  {Abi}}  \emph {et~al.} (\bibinfo {collaboration} {Muon $g-2$}),\ }\href
  {\doibase 10.1103/PhysRevLett.126.141801} {\bibfield  {journal} {\bibinfo
  {journal} {Phys. Rev. Lett.}\ }\textbf {\bibinfo {volume} {126}},\ \bibinfo
  {pages} {141801} (\bibinfo {year} {2021})},\ \Eprint
  {http://arxiv.org/abs/2104.03281} {arXiv:2104.03281 [hep-ex]}\BibitemShut
  {NoStop}%
\bibitem [{\citenamefont {Albahri}\ \emph
  {et~al.}(2021{\natexlab{a}})\citenamefont {Albahri} \emph
  {et~al.}}]{Muong-2:2021ovs}%
  \BibitemOpen
  \bibfield  {author} {\bibinfo {author} {\bibfnamefont {T.}~\bibnamefont
  {Albahri}}  \emph {et~al.} (\bibinfo {collaboration} {Muon $g-2$}),\ }\href
  {\doibase 10.1103/PhysRevA.103.042208} {\bibfield  {journal} {\bibinfo
  {journal} {Phys. Rev. A}\ }\textbf {\bibinfo {volume} {103}},\ \bibinfo
  {pages} {042208} (\bibinfo {year} {2021}{\natexlab{a}})},\ \Eprint
  {http://arxiv.org/abs/2104.03201} {arXiv:2104.03201 [hep-ex]}\BibitemShut
  {NoStop}%
\bibitem [{\citenamefont {Albahri}\ \emph
  {et~al.}(2021{\natexlab{b}})\citenamefont {Albahri} \emph
  {et~al.}}]{Muong-2:2021xzz}%
  \BibitemOpen
  \bibfield  {author} {\bibinfo {author} {\bibfnamefont {T.}~\bibnamefont
  {Albahri}}  \emph {et~al.} (\bibinfo {collaboration} {Muon $g-2$}),\ }\href
  {\doibase 10.1103/PhysRevAccelBeams.24.044002} {\bibfield  {journal}
  {\bibinfo  {journal} {Phys. Rev. Accel. Beams}\ }\textbf {\bibinfo {volume}
  {24}},\ \bibinfo {pages} {044002} (\bibinfo {year} {2021}{\natexlab{b}})},\
  \Eprint {http://arxiv.org/abs/2104.03240} {arXiv:2104.03240
  [physics.acc-ph]}\BibitemShut {NoStop}%
\bibitem [{\citenamefont {Albahri}\ \emph
  {et~al.}(2021{\natexlab{c}})\citenamefont {Albahri} \emph
  {et~al.}}]{Muong-2:2021vma}%
  \BibitemOpen
  \bibfield  {author} {\bibinfo {author} {\bibfnamefont {T.}~\bibnamefont
  {Albahri}}  \emph {et~al.} (\bibinfo {collaboration} {Muon $g-2$}),\ }\href
  {\doibase 10.1103/PhysRevD.103.072002} {\bibfield  {journal} {\bibinfo
  {journal} {Phys. Rev. D}\ }\textbf {\bibinfo {volume} {103}},\ \bibinfo
  {pages} {072002} (\bibinfo {year} {2021}{\natexlab{c}})},\ \Eprint
  {http://arxiv.org/abs/2104.03247} {arXiv:2104.03247 [hep-ex]}\BibitemShut
  {NoStop}%
\bibitem [{\citenamefont {Bennett}\ \emph {et~al.}(2006)\citenamefont {Bennett}
  \emph {et~al.}}]{Muong-2:2006rrc}%
  \BibitemOpen
  \bibfield  {author} {\bibinfo {author} {\bibfnamefont {G.~W.}\ \bibnamefont
  {Bennett}}  \emph {et~al.} (\bibinfo {collaboration} {Muon $g-2$}),\ }\href
  {\doibase 10.1103/PhysRevD.73.072003} {\bibfield  {journal} {\bibinfo
  {journal} {Phys. Rev. D}\ }\textbf {\bibinfo {volume} {73}},\ \bibinfo
  {pages} {072003} (\bibinfo {year} {2006})},\ \Eprint
  {http://arxiv.org/abs/hep-ex/0602035} {arXiv:hep-ex/0602035}\BibitemShut
  {NoStop}%
\bibitem [{\citenamefont {Aoyama}\ \emph {et~al.}(2020)\citenamefont {Aoyama}
  \emph {et~al.}}]{Aoyama:2020ynm}%
  \BibitemOpen
  \bibfield  {author} {\bibinfo {author} {\bibfnamefont {T.}~\bibnamefont
  {Aoyama}}  \emph {et~al.},\ }\href {\doibase 10.1016/j.physrep.2020.07.006}
  {\bibfield  {journal} {\bibinfo  {journal} {Phys. Rept.}\ }\textbf {\bibinfo
  {volume} {887}},\ \bibinfo {pages} {1} (\bibinfo {year} {2020})},\ \Eprint
  {http://arxiv.org/abs/2006.04822} {arXiv:2006.04822 [hep-ph]}\BibitemShut
  {NoStop}%
\bibitem [{\citenamefont {Aoyama}\ \emph {et~al.}(2012)\citenamefont {Aoyama},
  \citenamefont {Hayakawa}, \citenamefont {Kinoshita},\ and\ \citenamefont
  {Nio}}]{Aoyama:2012wk}%
  \BibitemOpen
  \bibfield  {author} {\bibinfo {author} {\bibfnamefont {T.}~\bibnamefont
  {Aoyama}}, \bibinfo {author} {\bibfnamefont {M.}~\bibnamefont {Hayakawa}},
  \bibinfo {author} {\bibfnamefont {T.}~\bibnamefont {Kinoshita}}, and \bibinfo
  {author} {\bibfnamefont {M.}~\bibnamefont {Nio}},\ }\href {\doibase
  10.1103/PhysRevLett.109.111808} {\bibfield  {journal} {\bibinfo  {journal}
  {Phys. Rev. Lett.}\ }\textbf {\bibinfo {volume} {109}},\ \bibinfo {pages}
  {111808} (\bibinfo {year} {2012})},\ \Eprint {http://arxiv.org/abs/1205.5370}
  {arXiv:1205.5370 [hep-ph]}\BibitemShut {NoStop}%
\bibitem [{\citenamefont {Aoyama}\ \emph {et~al.}(2019)\citenamefont {Aoyama},
  \citenamefont {Kinoshita},\ and\ \citenamefont {Nio}}]{Aoyama:2019ryr}%
  \BibitemOpen
  \bibfield  {author} {\bibinfo {author} {\bibfnamefont {T.}~\bibnamefont
  {Aoyama}}, \bibinfo {author} {\bibfnamefont {T.}~\bibnamefont {Kinoshita}},
  and \bibinfo {author} {\bibfnamefont {M.}~\bibnamefont {Nio}},\ }\href
  {\doibase 10.3390/atoms7010028} {\bibfield  {journal} {\bibinfo  {journal}
  {Atoms}\ }\textbf {\bibinfo {volume} {7}},\ \bibinfo {pages} {28} (\bibinfo
  {year} {2019})}\BibitemShut {NoStop}%
\bibitem [{\citenamefont {Czarnecki}\ \emph {et~al.}(2003)\citenamefont
  {Czarnecki}, \citenamefont {Marciano},\ and\ \citenamefont
  {Vainshtein}}]{Czarnecki:2002nt}%
  \BibitemOpen
  \bibfield  {author} {\bibinfo {author} {\bibfnamefont {A.}~\bibnamefont
  {Czarnecki}}, \bibinfo {author} {\bibfnamefont {W.~J.}\ \bibnamefont
  {Marciano}}, and \bibinfo {author} {\bibfnamefont {A.}~\bibnamefont
  {Vainshtein}},\ }\href {\doibase 10.1103/PhysRevD.67.073006} {\bibfield
  {journal} {\bibinfo  {journal} {Phys. Rev. D}\ }\textbf {\bibinfo {volume}
  {67}},\ \bibinfo {pages} {073006} (\bibinfo {year} {2003})},\ \bibinfo {note}
  {[Erratum: Phys. Rev. D {\bf 73}, 119901 (2006)]},\ \Eprint
  {http://arxiv.org/abs/hep-ph/0212229} {arXiv:hep-ph/0212229}\BibitemShut
  {NoStop}%
\bibitem [{\citenamefont {Gnendiger}\ \emph {et~al.}(2013)\citenamefont
  {Gnendiger}, \citenamefont {St\"ockinger},\ and\ \citenamefont
  {St\"ockinger-Kim}}]{Gnendiger:2013pva}%
  \BibitemOpen
  \bibfield  {author} {\bibinfo {author} {\bibfnamefont {C.}~\bibnamefont
  {Gnendiger}}, \bibinfo {author} {\bibfnamefont {D.}~\bibnamefont
  {St\"ockinger}}, and \bibinfo {author} {\bibfnamefont {H.}~\bibnamefont
  {St\"ockinger-Kim}},\ }\href {\doibase 10.1103/PhysRevD.88.053005} {\bibfield
   {journal} {\bibinfo  {journal} {Phys. Rev. D}\ }\textbf {\bibinfo {volume}
  {88}},\ \bibinfo {pages} {053005} (\bibinfo {year} {2013})},\ \Eprint
  {http://arxiv.org/abs/1306.5546} {arXiv:1306.5546 [hep-ph]}\BibitemShut
  {NoStop}%
\bibitem [{\citenamefont {Davier}\ \emph {et~al.}(2017)\citenamefont {Davier},
  \citenamefont {Hoecker}, \citenamefont {Malaescu},\ and\ \citenamefont
  {Zhang}}]{Davier:2017zfy}%
  \BibitemOpen
  \bibfield  {author} {\bibinfo {author} {\bibfnamefont {M.}~\bibnamefont
  {Davier}}, \bibinfo {author} {\bibfnamefont {A.}~\bibnamefont {Hoecker}},
  \bibinfo {author} {\bibfnamefont {B.}~\bibnamefont {Malaescu}}, and \bibinfo
  {author} {\bibfnamefont {Z.}~\bibnamefont {Zhang}},\ }\href {\doibase
  10.1140/epjc/s10052-017-5161-6} {\bibfield  {journal} {\bibinfo  {journal}
  {Eur. Phys. J. C}\ }\textbf {\bibinfo {volume} {77}},\ \bibinfo {pages} {827}
  (\bibinfo {year} {2017})},\ \Eprint {http://arxiv.org/abs/1706.09436}
  {arXiv:1706.09436 [hep-ph]}\BibitemShut {NoStop}%
\bibitem [{\citenamefont {Keshavarzi}\ \emph {et~al.}(2018)\citenamefont
  {Keshavarzi}, \citenamefont {Nomura},\ and\ \citenamefont
  {Teubner}}]{Keshavarzi:2018mgv}%
  \BibitemOpen
  \bibfield  {author} {\bibinfo {author} {\bibfnamefont {A.}~\bibnamefont
  {Keshavarzi}}, \bibinfo {author} {\bibfnamefont {D.}~\bibnamefont {Nomura}},
  and \bibinfo {author} {\bibfnamefont {T.}~\bibnamefont {Teubner}},\ }\href
  {\doibase 10.1103/PhysRevD.97.114025} {\bibfield  {journal} {\bibinfo
  {journal} {Phys. Rev. D}\ }\textbf {\bibinfo {volume} {97}},\ \bibinfo
  {pages} {114025} (\bibinfo {year} {2018})},\ \Eprint
  {http://arxiv.org/abs/1802.02995} {arXiv:1802.02995 [hep-ph]}\BibitemShut
  {NoStop}%
\bibitem [{\citenamefont {Colangelo}\ \emph {et~al.}(2019)\citenamefont
  {Colangelo}, \citenamefont {Hoferichter},\ and\ \citenamefont
  {Stoffer}}]{Colangelo:2018mtw}%
  \BibitemOpen
  \bibfield  {author} {\bibinfo {author} {\bibfnamefont {G.}~\bibnamefont
  {Colangelo}}, \bibinfo {author} {\bibfnamefont {M.}~\bibnamefont
  {Hoferichter}}, and \bibinfo {author} {\bibfnamefont {P.}~\bibnamefont
  {Stoffer}},\ }\href {\doibase 10.1007/JHEP02(2019)006} {\bibfield  {journal}
  {\bibinfo  {journal} {JHEP}\ }\textbf {\bibinfo {volume} {02}},\ \bibinfo
  {pages} {006} (\bibinfo {year} {2019})},\ \Eprint
  {http://arxiv.org/abs/1810.00007} {arXiv:1810.00007 [hep-ph]}\BibitemShut
  {NoStop}%
\bibitem [{\citenamefont {Hoferichter}\ \emph {et~al.}(2019)\citenamefont
  {Hoferichter}, \citenamefont {Hoid},\ and\ \citenamefont
  {Kubis}}]{Hoferichter:2019gzf}%
  \BibitemOpen
  \bibfield  {author} {\bibinfo {author} {\bibfnamefont {M.}~\bibnamefont
  {Hoferichter}}, \bibinfo {author} {\bibfnamefont {B.-L.}\ \bibnamefont
  {Hoid}}, and \bibinfo {author} {\bibfnamefont {B.}~\bibnamefont {Kubis}},\
  }\href {\doibase 10.1007/JHEP08(2019)137} {\bibfield  {journal} {\bibinfo
  {journal} {JHEP}\ }\textbf {\bibinfo {volume} {08}},\ \bibinfo {pages} {137}
  (\bibinfo {year} {2019})},\ \Eprint {http://arxiv.org/abs/1907.01556}
  {arXiv:1907.01556 [hep-ph]}\BibitemShut {NoStop}%
\bibitem [{\citenamefont {Davier}\ \emph {et~al.}(2020)\citenamefont {Davier},
  \citenamefont {Hoecker}, \citenamefont {Malaescu},\ and\ \citenamefont
  {Zhang}}]{Davier:2019can}%
  \BibitemOpen
  \bibfield  {author} {\bibinfo {author} {\bibfnamefont {M.}~\bibnamefont
  {Davier}}, \bibinfo {author} {\bibfnamefont {A.}~\bibnamefont {Hoecker}},
  \bibinfo {author} {\bibfnamefont {B.}~\bibnamefont {Malaescu}}, and \bibinfo
  {author} {\bibfnamefont {Z.}~\bibnamefont {Zhang}},\ }\href {\doibase
  10.1140/epjc/s10052-020-7792-2} {\bibfield  {journal} {\bibinfo  {journal}
  {Eur. Phys. J. C}\ }\textbf {\bibinfo {volume} {80}},\ \bibinfo {pages} {241}
  (\bibinfo {year} {2020})},\ \bibinfo {note} {[Erratum: Eur. Phys. J. C {\bf
  80}, 410 (2020)]},\ \Eprint {http://arxiv.org/abs/1908.00921}
  {arXiv:1908.00921 [hep-ph]}\BibitemShut {NoStop}%
\bibitem [{\citenamefont {Keshavarzi}\ \emph
  {et~al.}(2020{\natexlab{a}})\citenamefont {Keshavarzi}, \citenamefont
  {Nomura},\ and\ \citenamefont {Teubner}}]{Keshavarzi:2019abf}%
  \BibitemOpen
  \bibfield  {author} {\bibinfo {author} {\bibfnamefont {A.}~\bibnamefont
  {Keshavarzi}}, \bibinfo {author} {\bibfnamefont {D.}~\bibnamefont {Nomura}},
  and \bibinfo {author} {\bibfnamefont {T.}~\bibnamefont {Teubner}},\ }\href
  {\doibase 10.1103/PhysRevD.101.014029} {\bibfield  {journal} {\bibinfo
  {journal} {Phys. Rev. D}\ }\textbf {\bibinfo {volume} {101}},\ \bibinfo
  {pages} {014029} (\bibinfo {year} {2020}{\natexlab{a}})},\ \Eprint
  {http://arxiv.org/abs/1911.00367} {arXiv:1911.00367 [hep-ph]}\BibitemShut
  {NoStop}%
\bibitem [{\citenamefont {Hoid}\ \emph {et~al.}(2020)\citenamefont {Hoid},
  \citenamefont {Hoferichter},\ and\ \citenamefont {Kubis}}]{Hoid:2020xjs}%
  \BibitemOpen
  \bibfield  {author} {\bibinfo {author} {\bibfnamefont {B.-L.}\ \bibnamefont
  {Hoid}}, \bibinfo {author} {\bibfnamefont {M.}~\bibnamefont {Hoferichter}},
  and \bibinfo {author} {\bibfnamefont {B.}~\bibnamefont {Kubis}},\ }\href
  {\doibase 10.1140/epjc/s10052-020-08550-2} {\bibfield  {journal} {\bibinfo
  {journal} {Eur. Phys. J. C}\ }\textbf {\bibinfo {volume} {80}},\ \bibinfo
  {pages} {988} (\bibinfo {year} {2020})},\ \Eprint
  {http://arxiv.org/abs/2007.12696} {arXiv:2007.12696 [hep-ph]}\BibitemShut
  {NoStop}%
\bibitem [{\citenamefont {Kurz}\ \emph {et~al.}(2014)\citenamefont {Kurz},
  \citenamefont {Liu}, \citenamefont {Marquard},\ and\ \citenamefont
  {Steinhauser}}]{Kurz:2014wya}%
  \BibitemOpen
  \bibfield  {author} {\bibinfo {author} {\bibfnamefont {A.}~\bibnamefont
  {Kurz}}, \bibinfo {author} {\bibfnamefont {T.}~\bibnamefont {Liu}}, \bibinfo
  {author} {\bibfnamefont {P.}~\bibnamefont {Marquard}}, and \bibinfo {author}
  {\bibfnamefont {M.}~\bibnamefont {Steinhauser}},\ }\href {\doibase
  10.1016/j.physletb.2014.05.043} {\bibfield  {journal} {\bibinfo  {journal}
  {Phys. Lett. B}\ }\textbf {\bibinfo {volume} {734}},\ \bibinfo {pages} {144}
  (\bibinfo {year} {2014})},\ \Eprint {http://arxiv.org/abs/1403.6400}
  {arXiv:1403.6400 [hep-ph]}\BibitemShut {NoStop}%
\bibitem [{\citenamefont {Melnikov}\ and\ \citenamefont
  {Vainshtein}(2004)}]{Melnikov:2003xd}%
  \BibitemOpen
  \bibfield  {author} {\bibinfo {author} {\bibfnamefont {K.}~\bibnamefont
  {Melnikov}} and \bibinfo {author} {\bibfnamefont {A.}~\bibnamefont
  {Vainshtein}},\ }\href {\doibase 10.1103/PhysRevD.70.113006} {\bibfield
  {journal} {\bibinfo  {journal} {Phys. Rev. D}\ }\textbf {\bibinfo {volume}
  {70}},\ \bibinfo {pages} {113006} (\bibinfo {year} {2004})},\ \Eprint
  {http://arxiv.org/abs/hep-ph/0312226} {arXiv:hep-ph/0312226}\BibitemShut
  {NoStop}%
\bibitem [{\citenamefont {Colangelo}\ \emph
  {et~al.}(2014{\natexlab{a}})\citenamefont {Colangelo}, \citenamefont
  {Hoferichter}, \citenamefont {Procura},\ and\ \citenamefont
  {Stoffer}}]{Colangelo:2014dfa}%
  \BibitemOpen
  \bibfield  {author} {\bibinfo {author} {\bibfnamefont {G.}~\bibnamefont
  {Colangelo}}, \bibinfo {author} {\bibfnamefont {M.}~\bibnamefont
  {Hoferichter}}, \bibinfo {author} {\bibfnamefont {M.}~\bibnamefont
  {Procura}}, and \bibinfo {author} {\bibfnamefont {P.}~\bibnamefont
  {Stoffer}},\ }\href {\doibase 10.1007/JHEP09(2014)091} {\bibfield  {journal}
  {\bibinfo  {journal} {JHEP}\ }\textbf {\bibinfo {volume} {09}},\ \bibinfo
  {pages} {091} (\bibinfo {year} {2014}{\natexlab{a}})},\ \Eprint
  {http://arxiv.org/abs/1402.7081} {arXiv:1402.7081 [hep-ph]}\BibitemShut
  {NoStop}%
\bibitem [{\citenamefont {Colangelo}\ \emph
  {et~al.}(2014{\natexlab{b}})\citenamefont {Colangelo}, \citenamefont
  {Hoferichter}, \citenamefont {Kubis}, \citenamefont {Procura},\ and\
  \citenamefont {Stoffer}}]{Colangelo:2014pva}%
  \BibitemOpen
  \bibfield  {author} {\bibinfo {author} {\bibfnamefont {G.}~\bibnamefont
  {Colangelo}}, \bibinfo {author} {\bibfnamefont {M.}~\bibnamefont
  {Hoferichter}}, \bibinfo {author} {\bibfnamefont {B.}~\bibnamefont {Kubis}},
  \bibinfo {author} {\bibfnamefont {M.}~\bibnamefont {Procura}}, and \bibinfo
  {author} {\bibfnamefont {P.}~\bibnamefont {Stoffer}},\ }\href {\doibase
  10.1016/j.physletb.2014.09.021} {\bibfield  {journal} {\bibinfo  {journal}
  {Phys. Lett. B}\ }\textbf {\bibinfo {volume} {738}},\ \bibinfo {pages} {6}
  (\bibinfo {year} {2014}{\natexlab{b}})},\ \Eprint
  {http://arxiv.org/abs/1408.2517} {arXiv:1408.2517 [hep-ph]}\BibitemShut
  {NoStop}%
\bibitem [{\citenamefont {Colangelo}\ \emph {et~al.}(2015)\citenamefont
  {Colangelo}, \citenamefont {Hoferichter}, \citenamefont {Procura},\ and\
  \citenamefont {Stoffer}}]{Colangelo:2015ama}%
  \BibitemOpen
  \bibfield  {author} {\bibinfo {author} {\bibfnamefont {G.}~\bibnamefont
  {Colangelo}}, \bibinfo {author} {\bibfnamefont {M.}~\bibnamefont
  {Hoferichter}}, \bibinfo {author} {\bibfnamefont {M.}~\bibnamefont
  {Procura}}, and \bibinfo {author} {\bibfnamefont {P.}~\bibnamefont
  {Stoffer}},\ }\href {\doibase 10.1007/JHEP09(2015)074} {\bibfield  {journal}
  {\bibinfo  {journal} {JHEP}\ }\textbf {\bibinfo {volume} {09}},\ \bibinfo
  {pages} {074} (\bibinfo {year} {2015})},\ \Eprint
  {http://arxiv.org/abs/1506.01386} {arXiv:1506.01386 [hep-ph]}\BibitemShut
  {NoStop}%
\bibitem [{\citenamefont {Masjuan}\ and\ \citenamefont
  {S{\'a}nchez-Puertas}(2017)}]{Masjuan:2017tvw}%
  \BibitemOpen
  \bibfield  {author} {\bibinfo {author} {\bibfnamefont {P.}~\bibnamefont
  {Masjuan}} and \bibinfo {author} {\bibfnamefont {P.}~\bibnamefont
  {S{\'a}nchez-Puertas}},\ }\href {\doibase 10.1103/PhysRevD.95.054026}
  {\bibfield  {journal} {\bibinfo  {journal} {Phys. Rev. D}\ }\textbf {\bibinfo
  {volume} {95}},\ \bibinfo {pages} {054026} (\bibinfo {year} {2017})},\
  \Eprint {http://arxiv.org/abs/1701.05829} {arXiv:1701.05829
  [hep-ph]}\BibitemShut {NoStop}%
\bibitem [{\citenamefont {Colangelo}\ \emph
  {et~al.}(2017{\natexlab{a}})\citenamefont {Colangelo}, \citenamefont
  {Hoferichter}, \citenamefont {Procura},\ and\ \citenamefont
  {Stoffer}}]{Colangelo:2017qdm}%
  \BibitemOpen
  \bibfield  {author} {\bibinfo {author} {\bibfnamefont {G.}~\bibnamefont
  {Colangelo}}, \bibinfo {author} {\bibfnamefont {M.}~\bibnamefont
  {Hoferichter}}, \bibinfo {author} {\bibfnamefont {M.}~\bibnamefont
  {Procura}}, and \bibinfo {author} {\bibfnamefont {P.}~\bibnamefont
  {Stoffer}},\ }\href {\doibase 10.1103/PhysRevLett.118.232001} {\bibfield
  {journal} {\bibinfo  {journal} {Phys. Rev. Lett.}\ }\textbf {\bibinfo
  {volume} {118}},\ \bibinfo {pages} {232001} (\bibinfo {year}
  {2017}{\natexlab{a}})},\ \Eprint {http://arxiv.org/abs/1701.06554}
  {arXiv:1701.06554 [hep-ph]}\BibitemShut {NoStop}%
\bibitem [{\citenamefont {Colangelo}\ \emph
  {et~al.}(2017{\natexlab{b}})\citenamefont {Colangelo}, \citenamefont
  {Hoferichter}, \citenamefont {Procura},\ and\ \citenamefont
  {Stoffer}}]{Colangelo:2017fiz}%
  \BibitemOpen
  \bibfield  {author} {\bibinfo {author} {\bibfnamefont {G.}~\bibnamefont
  {Colangelo}}, \bibinfo {author} {\bibfnamefont {M.}~\bibnamefont
  {Hoferichter}}, \bibinfo {author} {\bibfnamefont {M.}~\bibnamefont
  {Procura}}, and \bibinfo {author} {\bibfnamefont {P.}~\bibnamefont
  {Stoffer}},\ }\href {\doibase 10.1007/JHEP04(2017)161} {\bibfield  {journal}
  {\bibinfo  {journal} {JHEP}\ }\textbf {\bibinfo {volume} {04}},\ \bibinfo
  {pages} {161} (\bibinfo {year} {2017}{\natexlab{b}})},\ \Eprint
  {http://arxiv.org/abs/1702.07347} {arXiv:1702.07347 [hep-ph]}\BibitemShut
  {NoStop}%
\bibitem [{\citenamefont {Hoferichter}\ \emph
  {et~al.}(2018{\natexlab{a}})\citenamefont {Hoferichter}, \citenamefont
  {Hoid}, \citenamefont {Kubis}, \citenamefont {Leupold},\ and\ \citenamefont
  {Schneider}}]{Hoferichter:2018dmo}%
  \BibitemOpen
  \bibfield  {author} {\bibinfo {author} {\bibfnamefont {M.}~\bibnamefont
  {Hoferichter}}, \bibinfo {author} {\bibfnamefont {B.-L.}\ \bibnamefont
  {Hoid}}, \bibinfo {author} {\bibfnamefont {B.}~\bibnamefont {Kubis}},
  \bibinfo {author} {\bibfnamefont {S.}~\bibnamefont {Leupold}}, and \bibinfo
  {author} {\bibfnamefont {S.~P.}\ \bibnamefont {Schneider}},\ }\href {\doibase
  10.1103/PhysRevLett.121.112002} {\bibfield  {journal} {\bibinfo  {journal}
  {Phys. Rev. Lett.}\ }\textbf {\bibinfo {volume} {121}},\ \bibinfo {pages}
  {112002} (\bibinfo {year} {2018}{\natexlab{a}})},\ \Eprint
  {http://arxiv.org/abs/1805.01471} {arXiv:1805.01471 [hep-ph]}\BibitemShut
  {NoStop}%
\bibitem [{\citenamefont {Hoferichter}\ \emph
  {et~al.}(2018{\natexlab{b}})\citenamefont {Hoferichter}, \citenamefont
  {Hoid}, \citenamefont {Kubis}, \citenamefont {Leupold},\ and\ \citenamefont
  {Schneider}}]{Hoferichter:2018kwz}%
  \BibitemOpen
  \bibfield  {author} {\bibinfo {author} {\bibfnamefont {M.}~\bibnamefont
  {Hoferichter}}, \bibinfo {author} {\bibfnamefont {B.-L.}\ \bibnamefont
  {Hoid}}, \bibinfo {author} {\bibfnamefont {B.}~\bibnamefont {Kubis}},
  \bibinfo {author} {\bibfnamefont {S.}~\bibnamefont {Leupold}}, and \bibinfo
  {author} {\bibfnamefont {S.~P.}\ \bibnamefont {Schneider}},\ }\href {\doibase
  10.1007/JHEP10(2018)141} {\bibfield  {journal} {\bibinfo  {journal} {JHEP}\
  }\textbf {\bibinfo {volume} {10}},\ \bibinfo {pages} {141} (\bibinfo {year}
  {2018}{\natexlab{b}})},\ \Eprint {http://arxiv.org/abs/1808.04823}
  {arXiv:1808.04823 [hep-ph]}\BibitemShut {NoStop}%
\bibitem [{\citenamefont {G\'erardin}\ \emph {et~al.}(2019)\citenamefont
  {G\'erardin}, \citenamefont {Meyer},\ and\ \citenamefont
  {Nyffeler}}]{Gerardin:2019vio}%
  \BibitemOpen
  \bibfield  {author} {\bibinfo {author} {\bibfnamefont {A.}~\bibnamefont
  {G\'erardin}}, \bibinfo {author} {\bibfnamefont {H.~B.}\ \bibnamefont
  {Meyer}}, and \bibinfo {author} {\bibfnamefont {A.}~\bibnamefont
  {Nyffeler}},\ }\href {\doibase 10.1103/PhysRevD.100.034520} {\bibfield
  {journal} {\bibinfo  {journal} {Phys. Rev. D}\ }\textbf {\bibinfo {volume}
  {100}},\ \bibinfo {pages} {034520} (\bibinfo {year} {2019})},\ \Eprint
  {http://arxiv.org/abs/1903.09471} {arXiv:1903.09471 [hep-lat]}\BibitemShut
  {NoStop}%
\bibitem [{\citenamefont {Bijnens}\ \emph {et~al.}(2019)\citenamefont
  {Bijnens}, \citenamefont {Hermansson-Truedsson},\ and\ \citenamefont
  {Rodr\'\i{}guez-S\'anchez}}]{Bijnens:2019ghy}%
  \BibitemOpen
  \bibfield  {author} {\bibinfo {author} {\bibfnamefont {J.}~\bibnamefont
  {Bijnens}}, \bibinfo {author} {\bibfnamefont {N.}~\bibnamefont
  {Hermansson-Truedsson}}, and \bibinfo {author} {\bibfnamefont
  {A.}~\bibnamefont {Rodr\'\i{}guez-S\'anchez}},\ }\href {\doibase
  10.1016/j.physletb.2019.134994} {\bibfield  {journal} {\bibinfo  {journal}
  {Phys. Lett. B}\ }\textbf {\bibinfo {volume} {798}},\ \bibinfo {pages}
  {134994} (\bibinfo {year} {2019})},\ \Eprint
  {http://arxiv.org/abs/1908.03331} {arXiv:1908.03331 [hep-ph]}\BibitemShut
  {NoStop}%
\bibitem [{\citenamefont {Colangelo}\ \emph
  {et~al.}(2020{\natexlab{a}})\citenamefont {Colangelo}, \citenamefont
  {Hagelstein}, \citenamefont {Hoferichter}, \citenamefont {Laub},\ and\
  \citenamefont {Stoffer}}]{Colangelo:2019lpu}%
  \BibitemOpen
  \bibfield  {author} {\bibinfo {author} {\bibfnamefont {G.}~\bibnamefont
  {Colangelo}}, \bibinfo {author} {\bibfnamefont {F.}~\bibnamefont
  {Hagelstein}}, \bibinfo {author} {\bibfnamefont {M.}~\bibnamefont
  {Hoferichter}}, \bibinfo {author} {\bibfnamefont {L.}~\bibnamefont {Laub}},
  and \bibinfo {author} {\bibfnamefont {P.}~\bibnamefont {Stoffer}},\ }\href
  {\doibase 10.1103/PhysRevD.101.051501} {\bibfield  {journal} {\bibinfo
  {journal} {Phys. Rev. D}\ }\textbf {\bibinfo {volume} {101}},\ \bibinfo
  {pages} {051501} (\bibinfo {year} {2020}{\natexlab{a}})},\ \Eprint
  {http://arxiv.org/abs/1910.11881} {arXiv:1910.11881 [hep-ph]}\BibitemShut
  {NoStop}%
\bibitem [{\citenamefont {Colangelo}\ \emph
  {et~al.}(2020{\natexlab{b}})\citenamefont {Colangelo}, \citenamefont
  {Hagelstein}, \citenamefont {Hoferichter}, \citenamefont {Laub},\ and\
  \citenamefont {Stoffer}}]{Colangelo:2019uex}%
  \BibitemOpen
  \bibfield  {author} {\bibinfo {author} {\bibfnamefont {G.}~\bibnamefont
  {Colangelo}}, \bibinfo {author} {\bibfnamefont {F.}~\bibnamefont
  {Hagelstein}}, \bibinfo {author} {\bibfnamefont {M.}~\bibnamefont
  {Hoferichter}}, \bibinfo {author} {\bibfnamefont {L.}~\bibnamefont {Laub}},
  and \bibinfo {author} {\bibfnamefont {P.}~\bibnamefont {Stoffer}},\ }\href
  {\doibase 10.1007/JHEP03(2020)101} {\bibfield  {journal} {\bibinfo  {journal}
  {JHEP}\ }\textbf {\bibinfo {volume} {03}},\ \bibinfo {pages} {101} (\bibinfo
  {year} {2020}{\natexlab{b}})},\ \Eprint {http://arxiv.org/abs/1910.13432}
  {arXiv:1910.13432 [hep-ph]}\BibitemShut {NoStop}%
\bibitem [{\citenamefont {Blum}\ \emph {et~al.}(2020)\citenamefont {Blum},
  \citenamefont {Christ}, \citenamefont {Hayakawa}, \citenamefont {Izubuchi},
  \citenamefont {Jin}, \citenamefont {Jung},\ and\ \citenamefont
  {Lehner}}]{Blum:2019ugy}%
  \BibitemOpen
  \bibfield  {author} {\bibinfo {author} {\bibfnamefont {T.}~\bibnamefont
  {Blum}}, \bibinfo {author} {\bibfnamefont {N.}~\bibnamefont {Christ}},
  \bibinfo {author} {\bibfnamefont {M.}~\bibnamefont {Hayakawa}}, \bibinfo
  {author} {\bibfnamefont {T.}~\bibnamefont {Izubuchi}}, \bibinfo {author}
  {\bibfnamefont {L.}~\bibnamefont {Jin}}, \bibinfo {author} {\bibfnamefont
  {C.}~\bibnamefont {Jung}}, and \bibinfo {author} {\bibfnamefont
  {C.}~\bibnamefont {Lehner}} (\bibinfo {collaboration} {RBC, UKQCD}),\ }\href
  {\doibase 10.1103/PhysRevLett.124.132002} {\bibfield  {journal} {\bibinfo
  {journal} {Phys. Rev. Lett.}\ }\textbf {\bibinfo {volume} {124}},\ \bibinfo
  {pages} {132002} (\bibinfo {year} {2020})},\ \Eprint
  {http://arxiv.org/abs/1911.08123} {arXiv:1911.08123 [hep-lat]}\BibitemShut
  {NoStop}%
\bibitem [{\citenamefont {Colangelo}\ \emph
  {et~al.}(2014{\natexlab{c}})\citenamefont {Colangelo}, \citenamefont
  {Hoferichter}, \citenamefont {Nyffeler}, \citenamefont {Passera},\ and\
  \citenamefont {Stoffer}}]{Colangelo:2014qya}%
  \BibitemOpen
  \bibfield  {author} {\bibinfo {author} {\bibfnamefont {G.}~\bibnamefont
  {Colangelo}}, \bibinfo {author} {\bibfnamefont {M.}~\bibnamefont
  {Hoferichter}}, \bibinfo {author} {\bibfnamefont {A.}~\bibnamefont
  {Nyffeler}}, \bibinfo {author} {\bibfnamefont {M.}~\bibnamefont {Passera}},
  and \bibinfo {author} {\bibfnamefont {P.}~\bibnamefont {Stoffer}},\ }\href
  {\doibase 10.1016/j.physletb.2014.06.012} {\bibfield  {journal} {\bibinfo
  {journal} {Phys. Lett. B}\ }\textbf {\bibinfo {volume} {735}},\ \bibinfo
  {pages} {90} (\bibinfo {year} {2014}{\natexlab{c}})},\ \Eprint
  {http://arxiv.org/abs/1403.7512} {arXiv:1403.7512 [hep-ph]}\BibitemShut
  {NoStop}%
\bibitem [{\citenamefont {Chao}\ \emph {et~al.}(2021)\citenamefont {Chao},
  \citenamefont {Hudspith}, \citenamefont {G\'erardin}, \citenamefont {Green},
  \citenamefont {Meyer},\ and\ \citenamefont {Ottnad}}]{Chao:2021tvp}%
  \BibitemOpen
  \bibfield  {author} {\bibinfo {author} {\bibfnamefont {E.-H.}\ \bibnamefont
  {Chao}}, \bibinfo {author} {\bibfnamefont {R.~J.}\ \bibnamefont {Hudspith}},
  \bibinfo {author} {\bibfnamefont {A.}~\bibnamefont {G\'erardin}}, \bibinfo
  {author} {\bibfnamefont {J.~R.}\ \bibnamefont {Green}}, \bibinfo {author}
  {\bibfnamefont {H.~B.}\ \bibnamefont {Meyer}}, and \bibinfo {author}
  {\bibfnamefont {K.}~\bibnamefont {Ottnad}},\ }\href {\doibase
  10.1140/epjc/s10052-021-09455-4} {\bibfield  {journal} {\bibinfo  {journal}
  {Eur. Phys. J. C}\ }\textbf {\bibinfo {volume} {81}},\ \bibinfo {pages} {651}
  (\bibinfo {year} {2021})},\ \Eprint {http://arxiv.org/abs/2104.02632}
  {arXiv:2104.02632 [hep-lat]}\BibitemShut {NoStop}%
\bibitem [{\citenamefont {Chao}\ \emph {et~al.}(2022)\citenamefont {Chao},
  \citenamefont {Hudspith}, \citenamefont {G\'erardin}, \citenamefont {Green},\
  and\ \citenamefont {Meyer}}]{Chao:2022xzg}%
  \BibitemOpen
  \bibfield  {author} {\bibinfo {author} {\bibfnamefont {E.-H.}\ \bibnamefont
  {Chao}}, \bibinfo {author} {\bibfnamefont {R.~J.}\ \bibnamefont {Hudspith}},
  \bibinfo {author} {\bibfnamefont {A.}~\bibnamefont {G\'erardin}}, \bibinfo
  {author} {\bibfnamefont {J.~R.}\ \bibnamefont {Green}}, and \bibinfo {author}
  {\bibfnamefont {H.~B.}\ \bibnamefont {Meyer}},\ }\href {\doibase
  10.1140/epjc/s10052-022-10589-2} {\bibfield  {journal} {\bibinfo  {journal}
  {Eur. Phys. J. C}\ }\textbf {\bibinfo {volume} {82}},\ \bibinfo {pages} {664}
  (\bibinfo {year} {2022})},\ \Eprint {http://arxiv.org/abs/2204.08844}
  {arXiv:2204.08844 [hep-lat]}\BibitemShut {NoStop}%
\bibitem [{\citenamefont {Blum}\ \emph
  {et~al.}(2023{\natexlab{a}})\citenamefont {Blum}, \citenamefont {Christ},
  \citenamefont {Hayakawa}, \citenamefont {Izubuchi}, \citenamefont {Jin},
  \citenamefont {Jung}, \citenamefont {Lehner},\ and\ \citenamefont
  {Tu}}]{Blum:2023vlm}%
  \BibitemOpen
  \bibfield  {author} {\bibinfo {author} {\bibfnamefont {T.}~\bibnamefont
  {Blum}}, \bibinfo {author} {\bibfnamefont {N.}~\bibnamefont {Christ}},
  \bibinfo {author} {\bibfnamefont {M.}~\bibnamefont {Hayakawa}}, \bibinfo
  {author} {\bibfnamefont {T.}~\bibnamefont {Izubuchi}}, \bibinfo {author}
  {\bibfnamefont {L.}~\bibnamefont {Jin}}, \bibinfo {author} {\bibfnamefont
  {C.}~\bibnamefont {Jung}}, \bibinfo {author} {\bibfnamefont {C.}~\bibnamefont
  {Lehner}}, and \bibinfo {author} {\bibfnamefont {C.}~\bibnamefont {Tu}}
  (\bibinfo {collaboration} {RBC, UKQCD}),\ }\href@noop {} {\  (\bibinfo {year}
  {2023}{\natexlab{a}})},\ \Eprint {http://arxiv.org/abs/2304.04423}
  {arXiv:2304.04423 [hep-lat]}\BibitemShut {NoStop}%
\bibitem [{\citenamefont {Alexandrou}\ \emph {et~al.}(2022)\citenamefont
  {Alexandrou} \emph {et~al.}}]{Alexandrou:2022qyf}%
  \BibitemOpen
  \bibfield  {author} {\bibinfo {author} {\bibfnamefont {C.}~\bibnamefont
  {Alexandrou}}  \emph {et~al.} (\bibinfo {collaboration} {ETM}),\ }\href@noop
  {} {\  (\bibinfo {year} {2022})},\ \Eprint {http://arxiv.org/abs/2212.06704}
  {arXiv:2212.06704 [hep-lat]}\BibitemShut {NoStop}%
\bibitem [{\citenamefont {G\'erardin}\ \emph {et~al.}(2023)\citenamefont
  {G\'erardin} \emph {et~al.}}]{Gerardin:2023naa}%
  \BibitemOpen
  \bibfield  {author} {\bibinfo {author} {\bibfnamefont {A.}~\bibnamefont
  {G\'erardin}}  \emph {et~al.} (\bibinfo {collaboration} {BMWc}),\ }\href@noop
  {} {\  (\bibinfo {year} {2023})},\ \Eprint {http://arxiv.org/abs/2305.04570}
  {arXiv:2305.04570 [hep-lat]}\BibitemShut {NoStop}%
\bibitem [{\citenamefont {Hoferichter}\ and\ \citenamefont
  {Stoffer}(2020)}]{Hoferichter:2020lap}%
  \BibitemOpen
  \bibfield  {author} {\bibinfo {author} {\bibfnamefont {M.}~\bibnamefont
  {Hoferichter}} and \bibinfo {author} {\bibfnamefont {P.}~\bibnamefont
  {Stoffer}},\ }\href {\doibase 10.1007/JHEP05(2020)159} {\bibfield  {journal}
  {\bibinfo  {journal} {JHEP}\ }\textbf {\bibinfo {volume} {05}},\ \bibinfo
  {pages} {159} (\bibinfo {year} {2020})},\ \Eprint
  {http://arxiv.org/abs/2004.06127} {arXiv:2004.06127 [hep-ph]}\BibitemShut
  {NoStop}%
\bibitem [{\citenamefont {L\"udtke}\ and\ \citenamefont
  {Procura}(2020)}]{Ludtke:2020moa}%
  \BibitemOpen
  \bibfield  {author} {\bibinfo {author} {\bibfnamefont {J.}~\bibnamefont
  {L\"udtke}} and \bibinfo {author} {\bibfnamefont {M.}~\bibnamefont
  {Procura}},\ }\href {\doibase 10.1140/epjc/s10052-020-08611-6} {\bibfield
  {journal} {\bibinfo  {journal} {Eur. Phys. J. C}\ }\textbf {\bibinfo {volume}
  {80}},\ \bibinfo {pages} {1108} (\bibinfo {year} {2020})},\ \Eprint
  {http://arxiv.org/abs/2006.00007} {arXiv:2006.00007 [hep-ph]}\BibitemShut
  {NoStop}%
\bibitem [{\citenamefont {Bijnens}\ \emph {et~al.}(2020)\citenamefont
  {Bijnens}, \citenamefont {Hermansson-Truedsson}, \citenamefont {Laub},\ and\
  \citenamefont {Rodr\'iguez-S\'anchez}}]{Bijnens:2020xnl}%
  \BibitemOpen
  \bibfield  {author} {\bibinfo {author} {\bibfnamefont {J.}~\bibnamefont
  {Bijnens}}, \bibinfo {author} {\bibfnamefont {N.}~\bibnamefont
  {Hermansson-Truedsson}}, \bibinfo {author} {\bibfnamefont {L.}~\bibnamefont
  {Laub}}, and \bibinfo {author} {\bibfnamefont {A.}~\bibnamefont
  {Rodr\'iguez-S\'anchez}},\ }\href {\doibase 10.1007/JHEP10(2020)203}
  {\bibfield  {journal} {\bibinfo  {journal} {JHEP}\ }\textbf {\bibinfo
  {volume} {10}},\ \bibinfo {pages} {203} (\bibinfo {year} {2020})},\ \Eprint
  {http://arxiv.org/abs/2008.13487} {arXiv:2008.13487 [hep-ph]}\BibitemShut
  {NoStop}%
\bibitem [{\citenamefont {Bijnens}\ \emph {et~al.}(2021)\citenamefont
  {Bijnens}, \citenamefont {Hermansson-Truedsson}, \citenamefont {Laub},\ and\
  \citenamefont {Rodr\'iguez-S\'anchez}}]{Bijnens:2021jqo}%
  \BibitemOpen
  \bibfield  {author} {\bibinfo {author} {\bibfnamefont {J.}~\bibnamefont
  {Bijnens}}, \bibinfo {author} {\bibfnamefont {N.}~\bibnamefont
  {Hermansson-Truedsson}}, \bibinfo {author} {\bibfnamefont {L.}~\bibnamefont
  {Laub}}, and \bibinfo {author} {\bibfnamefont {A.}~\bibnamefont
  {Rodr\'iguez-S\'anchez}},\ }\href {\doibase 10.1007/JHEP04(2021)240}
  {\bibfield  {journal} {\bibinfo  {journal} {JHEP}\ }\textbf {\bibinfo
  {volume} {04}},\ \bibinfo {pages} {240} (\bibinfo {year} {2021})},\ \Eprint
  {http://arxiv.org/abs/2101.09169} {arXiv:2101.09169 [hep-ph]}\BibitemShut
  {NoStop}%
\bibitem [{\citenamefont {Zanke}\ \emph {et~al.}(2021)\citenamefont {Zanke},
  \citenamefont {Hoferichter},\ and\ \citenamefont {Kubis}}]{Zanke:2021wiq}%
  \BibitemOpen
  \bibfield  {author} {\bibinfo {author} {\bibfnamefont {M.}~\bibnamefont
  {Zanke}}, \bibinfo {author} {\bibfnamefont {M.}~\bibnamefont {Hoferichter}},
  and \bibinfo {author} {\bibfnamefont {B.}~\bibnamefont {Kubis}},\ }\href
  {\doibase 10.1007/JHEP07(2021)106} {\bibfield  {journal} {\bibinfo  {journal}
  {JHEP}\ }\textbf {\bibinfo {volume} {07}},\ \bibinfo {pages} {106} (\bibinfo
  {year} {2021})},\ \Eprint {http://arxiv.org/abs/2103.09829} {arXiv:2103.09829
  [hep-ph]}\BibitemShut {NoStop}%
\bibitem [{\citenamefont {Danilkin}\ \emph {et~al.}(2021)\citenamefont
  {Danilkin}, \citenamefont {Hoferichter},\ and\ \citenamefont
  {Stoffer}}]{Danilkin:2021icn}%
  \BibitemOpen
  \bibfield  {author} {\bibinfo {author} {\bibfnamefont {I.}~\bibnamefont
  {Danilkin}}, \bibinfo {author} {\bibfnamefont {M.}~\bibnamefont
  {Hoferichter}}, and \bibinfo {author} {\bibfnamefont {P.}~\bibnamefont
  {Stoffer}},\ }\href {\doibase 10.1016/j.physletb.2021.136502} {\bibfield
  {journal} {\bibinfo  {journal} {Phys. Lett. B}\ }\textbf {\bibinfo {volume}
  {820}},\ \bibinfo {pages} {136502} (\bibinfo {year} {2021})},\ \Eprint
  {http://arxiv.org/abs/2105.01666} {arXiv:2105.01666 [hep-ph]}\BibitemShut
  {NoStop}%
\bibitem [{\citenamefont {Colangelo}\ \emph
  {et~al.}(2021{\natexlab{a}})\citenamefont {Colangelo}, \citenamefont
  {Hagelstein}, \citenamefont {Hoferichter}, \citenamefont {Laub},\ and\
  \citenamefont {Stoffer}}]{Colangelo:2021nkr}%
  \BibitemOpen
  \bibfield  {author} {\bibinfo {author} {\bibfnamefont {G.}~\bibnamefont
  {Colangelo}}, \bibinfo {author} {\bibfnamefont {F.}~\bibnamefont
  {Hagelstein}}, \bibinfo {author} {\bibfnamefont {M.}~\bibnamefont
  {Hoferichter}}, \bibinfo {author} {\bibfnamefont {L.}~\bibnamefont {Laub}},
  and \bibinfo {author} {\bibfnamefont {P.}~\bibnamefont {Stoffer}},\ }\href
  {\doibase 10.1140/epjc/s10052-021-09513-x} {\bibfield  {journal} {\bibinfo
  {journal} {Eur. Phys. J. C}\ }\textbf {\bibinfo {volume} {81}},\ \bibinfo
  {pages} {702} (\bibinfo {year} {2021}{\natexlab{a}})},\ \Eprint
  {http://arxiv.org/abs/2106.13222} {arXiv:2106.13222 [hep-ph]}\BibitemShut
  {NoStop}%
\bibitem [{\citenamefont {Holz}\ \emph {et~al.}(2022)\citenamefont {Holz},
  \citenamefont {Hanhart}, \citenamefont {Hoferichter},\ and\ \citenamefont
  {Kubis}}]{Holz:2022hwz}%
  \BibitemOpen
  \bibfield  {author} {\bibinfo {author} {\bibfnamefont {S.}~\bibnamefont
  {Holz}}, \bibinfo {author} {\bibfnamefont {C.}~\bibnamefont {Hanhart}},
  \bibinfo {author} {\bibfnamefont {M.}~\bibnamefont {Hoferichter}}, and
  \bibinfo {author} {\bibfnamefont {B.}~\bibnamefont {Kubis}},\ }\href
  {\doibase 10.1140/epjc/s10052-022-10247-7} {\bibfield  {journal} {\bibinfo
  {journal} {Eur. Phys. J. C}\ }\textbf {\bibinfo {volume} {82}},\ \bibinfo
  {pages} {434} (\bibinfo {year} {2022})},\ \bibinfo {note} {[Addendum: Eur.
  Phys. J. C {\bf 82}, 1159 (2022)]},\ \Eprint
  {http://arxiv.org/abs/2202.05846} {arXiv:2202.05846 [hep-ph]}\BibitemShut
  {NoStop}%
\bibitem [{\citenamefont {Leutgeb}\ \emph {et~al.}(2023)\citenamefont
  {Leutgeb}, \citenamefont {Mager},\ and\ \citenamefont
  {Rebhan}}]{Leutgeb:2022lqw}%
  \BibitemOpen
  \bibfield  {author} {\bibinfo {author} {\bibfnamefont {J.}~\bibnamefont
  {Leutgeb}}, \bibinfo {author} {\bibfnamefont {J.}~\bibnamefont {Mager}}, and
  \bibinfo {author} {\bibfnamefont {A.}~\bibnamefont {Rebhan}},\ }\href
  {\doibase 10.1103/PhysRevD.107.054021} {\bibfield  {journal} {\bibinfo
  {journal} {Phys. Rev. D}\ }\textbf {\bibinfo {volume} {107}},\ \bibinfo
  {pages} {054021} (\bibinfo {year} {2023})},\ \Eprint
  {http://arxiv.org/abs/2211.16562} {arXiv:2211.16562 [hep-ph]}\BibitemShut
  {NoStop}%
\bibitem [{\citenamefont {Bijnens}\ \emph {et~al.}(2023)\citenamefont
  {Bijnens}, \citenamefont {Hermansson-Truedsson},\ and\ \citenamefont
  {Rodr\'\i{}guez-S\'anchez}}]{Bijnens:2022itw}%
  \BibitemOpen
  \bibfield  {author} {\bibinfo {author} {\bibfnamefont {J.}~\bibnamefont
  {Bijnens}}, \bibinfo {author} {\bibfnamefont {N.}~\bibnamefont
  {Hermansson-Truedsson}}, and \bibinfo {author} {\bibfnamefont
  {A.}~\bibnamefont {Rodr\'\i{}guez-S\'anchez}},\ }\href {\doibase
  10.1007/JHEP02(2023)167} {\bibfield  {journal} {\bibinfo  {journal} {JHEP}\
  }\textbf {\bibinfo {volume} {02}},\ \bibinfo {pages} {167} (\bibinfo {year}
  {2023})},\ \Eprint {http://arxiv.org/abs/2211.17183} {arXiv:2211.17183
  [hep-ph]}\BibitemShut {NoStop}%
\bibitem [{\citenamefont {L\"udtke}\ \emph {et~al.}(2023)\citenamefont
  {L\"udtke}, \citenamefont {Procura},\ and\ \citenamefont
  {Stoffer}}]{Ludtke:2023hvz}%
  \BibitemOpen
  \bibfield  {author} {\bibinfo {author} {\bibfnamefont {J.}~\bibnamefont
  {L\"udtke}}, \bibinfo {author} {\bibfnamefont {M.}~\bibnamefont {Procura}},
  and \bibinfo {author} {\bibfnamefont {P.}~\bibnamefont {Stoffer}},\ }\href
  {\doibase 10.1007/JHEP04(2023)125} {\bibfield  {journal} {\bibinfo  {journal}
  {JHEP}\ }\textbf {\bibinfo {volume} {04}},\ \bibinfo {pages} {125} (\bibinfo
  {year} {2023})},\ \Eprint {http://arxiv.org/abs/2302.12264} {arXiv:2302.12264
  [hep-ph]}\BibitemShut {NoStop}%
\bibitem [{\citenamefont {Hoferichter}\ \emph
  {et~al.}(2023{\natexlab{a}})\citenamefont {Hoferichter}, \citenamefont
  {Kubis},\ and\ \citenamefont {Zanke}}]{Hoferichter:2023tgp}%
  \BibitemOpen
  \bibfield  {author} {\bibinfo {author} {\bibfnamefont {M.}~\bibnamefont
  {Hoferichter}}, \bibinfo {author} {\bibfnamefont {B.}~\bibnamefont {Kubis}},
  and \bibinfo {author} {\bibfnamefont {M.}~\bibnamefont {Zanke}},\ }\href@noop
  {} {\  (\bibinfo {year} {2023}{\natexlab{a}})},\ \Eprint
  {http://arxiv.org/abs/2307.14413} {arXiv:2307.14413 [hep-ph]}\BibitemShut
  {NoStop}%
\bibitem [{\citenamefont {Grange}\ \emph {et~al.}(2015)\citenamefont {Grange}
  \emph {et~al.}}]{Muong-2:2015xgu}%
  \BibitemOpen
  \bibfield  {author} {\bibinfo {author} {\bibfnamefont {J.}~\bibnamefont
  {Grange}}  \emph {et~al.} (\bibinfo {collaboration} {Muon $g-2$}),\
  }\href@noop {} {\  (\bibinfo {year} {2015})},\ \Eprint
  {http://arxiv.org/abs/1501.06858} {arXiv:1501.06858
  [physics.ins-det]}\BibitemShut {NoStop}%
\bibitem [{\citenamefont {Colangelo}\ \emph
  {et~al.}(2022{\natexlab{a}})\citenamefont {Colangelo} \emph
  {et~al.}}]{Colangelo:2022jxc}%
  \BibitemOpen
  \bibfield  {author} {\bibinfo {author} {\bibfnamefont {G.}~\bibnamefont
  {Colangelo}}  \emph {et~al.},\ }\href@noop {} {\  (\bibinfo {year}
  {2022}{\natexlab{a}})},\ \Eprint {http://arxiv.org/abs/2203.15810}
  {arXiv:2203.15810 [hep-ph]}\BibitemShut {NoStop}%
\bibitem [{\citenamefont {Calmet}\ \emph {et~al.}(1976)\citenamefont {Calmet},
  \citenamefont {Narison}, \citenamefont {Perrottet},\ and\ \citenamefont
  {de~Rafael}}]{Calmet:1976kd}%
  \BibitemOpen
  \bibfield  {author} {\bibinfo {author} {\bibfnamefont {J.}~\bibnamefont
  {Calmet}}, \bibinfo {author} {\bibfnamefont {S.}~\bibnamefont {Narison}},
  \bibinfo {author} {\bibfnamefont {M.}~\bibnamefont {Perrottet}}, and \bibinfo
  {author} {\bibfnamefont {E.}~\bibnamefont {de~Rafael}},\ }\href {\doibase
  10.1016/0370-2693(76)90150-7} {\bibfield  {journal} {\bibinfo  {journal}
  {Phys. Lett. B}\ }\textbf {\bibinfo {volume} {61}},\ \bibinfo {pages} {283}
  (\bibinfo {year} {1976})}\BibitemShut {NoStop}%
\bibitem [{\citenamefont {Hoferichter}\ and\ \citenamefont
  {Teubner}(2022)}]{Hoferichter:2021wyj}%
  \BibitemOpen
  \bibfield  {author} {\bibinfo {author} {\bibfnamefont {M.}~\bibnamefont
  {Hoferichter}} and \bibinfo {author} {\bibfnamefont {T.}~\bibnamefont
  {Teubner}},\ }\href {\doibase 10.1103/PhysRevLett.128.112002} {\bibfield
  {journal} {\bibinfo  {journal} {Phys. Rev. Lett.}\ }\textbf {\bibinfo
  {volume} {128}},\ \bibinfo {pages} {112002} (\bibinfo {year} {2022})},\
  \Eprint {http://arxiv.org/abs/2112.06929} {arXiv:2112.06929
  [hep-ph]}\BibitemShut {NoStop}%
\bibitem [{\citenamefont {Borsanyi}\ \emph {et~al.}(2021)\citenamefont
  {Borsanyi} \emph {et~al.}}]{Borsanyi:2020mff}%
  \BibitemOpen
  \bibfield  {author} {\bibinfo {author} {\bibfnamefont {S.}~\bibnamefont
  {Borsanyi}}  \emph {et~al.} (\bibinfo {collaboration} {BMWc}),\ }\href
  {\doibase 10.1038/s41586-021-03418-1} {\bibfield  {journal} {\bibinfo
  {journal} {Nature}\ }\textbf {\bibinfo {volume} {593}},\ \bibinfo {pages}
  {51} (\bibinfo {year} {2021})},\ \Eprint {http://arxiv.org/abs/2002.12347}
  {arXiv:2002.12347 [hep-lat]}\BibitemShut {NoStop}%
\bibitem [{\citenamefont {Blum}\ \emph {et~al.}(2018)\citenamefont {Blum} \emph
  {et~al.}}]{RBC:2018dos}%
  \BibitemOpen
  \bibfield  {author} {\bibinfo {author} {\bibfnamefont {T.}~\bibnamefont
  {Blum}}  \emph {et~al.} (\bibinfo {collaboration} {RBC, UKQCD}),\ }\href
  {\doibase 10.1103/PhysRevLett.121.022003} {\bibfield  {journal} {\bibinfo
  {journal} {Phys. Rev. Lett.}\ }\textbf {\bibinfo {volume} {121}},\ \bibinfo
  {pages} {022003} (\bibinfo {year} {2018})},\ \Eprint
  {http://arxiv.org/abs/1801.07224} {arXiv:1801.07224 [hep-lat]}\BibitemShut
  {NoStop}%
\bibitem [{\citenamefont {C\`e}\ \emph
  {et~al.}(2022{\natexlab{a}})\citenamefont {C\`e} \emph
  {et~al.}}]{Ce:2022kxy}%
  \BibitemOpen
  \bibfield  {author} {\bibinfo {author} {\bibfnamefont {M.}~\bibnamefont
  {C\`e}}  \emph {et~al.},\ }\href {\doibase 10.1103/PhysRevD.106.114502}
  {\bibfield  {journal} {\bibinfo  {journal} {Phys. Rev. D}\ }\textbf {\bibinfo
  {volume} {106}},\ \bibinfo {pages} {114502} (\bibinfo {year}
  {2022}{\natexlab{a}})},\ \Eprint {http://arxiv.org/abs/2206.06582}
  {arXiv:2206.06582 [hep-lat]}\BibitemShut {NoStop}%
\bibitem [{\citenamefont {Alexandrou}\ \emph
  {et~al.}(2023{\natexlab{a}})\citenamefont {Alexandrou} \emph
  {et~al.}}]{ExtendedTwistedMass:2022jpw}%
  \BibitemOpen
  \bibfield  {author} {\bibinfo {author} {\bibfnamefont {C.}~\bibnamefont
  {Alexandrou}}  \emph {et~al.} (\bibinfo {collaboration} {ETM}),\ }\href
  {\doibase 10.1103/PhysRevD.107.074506} {\bibfield  {journal} {\bibinfo
  {journal} {Phys. Rev. D}\ }\textbf {\bibinfo {volume} {107}},\ \bibinfo
  {pages} {074506} (\bibinfo {year} {2023}{\natexlab{a}})},\ \Eprint
  {http://arxiv.org/abs/2206.15084} {arXiv:2206.15084 [hep-lat]}\BibitemShut
  {NoStop}%
\bibitem [{\citenamefont {Bazavov}\ \emph {et~al.}(2023)\citenamefont {Bazavov}
  \emph {et~al.}}]{FermilabLatticeHPQCD:2023jof}%
  \BibitemOpen
  \bibfield  {author} {\bibinfo {author} {\bibfnamefont {A.}~\bibnamefont
  {Bazavov}}  \emph {et~al.} (\bibinfo {collaboration} {Fermilab Lattice,
  HPQCD, MILC}),\ }\href {\doibase 10.1103/PhysRevD.107.114514} {\bibfield
  {journal} {\bibinfo  {journal} {Phys. Rev. D}\ }\textbf {\bibinfo {volume}
  {107}},\ \bibinfo {pages} {114514} (\bibinfo {year} {2023})},\ \Eprint
  {http://arxiv.org/abs/2301.08274} {arXiv:2301.08274 [hep-lat]}\BibitemShut
  {NoStop}%
\bibitem [{\citenamefont {Blum}\ \emph
  {et~al.}(2023{\natexlab{b}})\citenamefont {Blum} \emph
  {et~al.}}]{Blum:2023qou}%
  \BibitemOpen
  \bibfield  {author} {\bibinfo {author} {\bibfnamefont {T.}~\bibnamefont
  {Blum}}  \emph {et~al.} (\bibinfo {collaboration} {RBC, UKQCD}),\ }\href@noop
  {} {\  (\bibinfo {year} {2023}{\natexlab{b}})},\ \Eprint
  {http://arxiv.org/abs/2301.08696} {arXiv:2301.08696 [hep-lat]}\BibitemShut
  {NoStop}%
\bibitem [{\citenamefont {Colangelo}\ \emph
  {et~al.}(2022{\natexlab{b}})\citenamefont {Colangelo}, \citenamefont
  {El-Khadra}, \citenamefont {Hoferichter}, \citenamefont {Keshavarzi},
  \citenamefont {Lehner}, \citenamefont {Stoffer},\ and\ \citenamefont
  {Teubner}}]{Colangelo:2022vok}%
  \BibitemOpen
  \bibfield  {author} {\bibinfo {author} {\bibfnamefont {G.}~\bibnamefont
  {Colangelo}}, \bibinfo {author} {\bibfnamefont {A.~X.}\ \bibnamefont
  {El-Khadra}}, \bibinfo {author} {\bibfnamefont {M.}~\bibnamefont
  {Hoferichter}}, \bibinfo {author} {\bibfnamefont {A.}~\bibnamefont
  {Keshavarzi}}, \bibinfo {author} {\bibfnamefont {C.}~\bibnamefont {Lehner}},
  \bibinfo {author} {\bibfnamefont {P.}~\bibnamefont {Stoffer}}, and \bibinfo
  {author} {\bibfnamefont {T.}~\bibnamefont {Teubner}},\ }\href {\doibase
  10.1016/j.physletb.2022.137313} {\bibfield  {journal} {\bibinfo  {journal}
  {Phys. Lett. B}\ }\textbf {\bibinfo {volume} {833}},\ \bibinfo {pages}
  {137313} (\bibinfo {year} {2022}{\natexlab{b}})},\ \Eprint
  {http://arxiv.org/abs/2205.12963} {arXiv:2205.12963 [hep-ph]}\BibitemShut
  {NoStop}%
\bibitem [{\citenamefont {Achasov}\ \emph {et~al.}(2021)\citenamefont {Achasov}
  \emph {et~al.}}]{Achasov:2020iys}%
  \BibitemOpen
  \bibfield  {author} {\bibinfo {author} {\bibfnamefont {M.~N.}\ \bibnamefont
  {Achasov}}  \emph {et~al.} (\bibinfo {collaboration} {SND}),\ }\href
  {\doibase 10.1007/JHEP01(2021)113} {\bibfield  {journal} {\bibinfo  {journal}
  {JHEP}\ }\textbf {\bibinfo {volume} {01}},\ \bibinfo {pages} {113} (\bibinfo
  {year} {2021})},\ \Eprint {http://arxiv.org/abs/2004.00263} {arXiv:2004.00263
  [hep-ex]}\BibitemShut {NoStop}%
\bibitem [{\citenamefont {Ignatov}\ \emph {et~al.}(2023)\citenamefont {Ignatov}
  \emph {et~al.}}]{CMD-3:2023alj}%
  \BibitemOpen
  \bibfield  {author} {\bibinfo {author} {\bibfnamefont {F.~V.}\ \bibnamefont
  {Ignatov}}  \emph {et~al.} (\bibinfo {collaboration} {CMD-3}),\ }\href@noop
  {} {\  (\bibinfo {year} {2023})},\ \Eprint {http://arxiv.org/abs/2302.08834}
  {arXiv:2302.08834 [hep-ex]}\BibitemShut {NoStop}%
\bibitem [{\citenamefont {Akhmetshin}\ \emph {et~al.}(2007)\citenamefont
  {Akhmetshin} \emph {et~al.}}]{Akhmetshin:2006bx}%
  \BibitemOpen
  \bibfield  {author} {\bibinfo {author} {\bibfnamefont {R.~R.}\ \bibnamefont
  {Akhmetshin}}  \emph {et~al.} (\bibinfo {collaboration} {CMD-2}),\ }\href
  {\doibase 10.1016/j.physletb.2007.01.073} {\bibfield  {journal} {\bibinfo
  {journal} {Phys. Lett. B}\ }\textbf {\bibinfo {volume} {648}},\ \bibinfo
  {pages} {28} (\bibinfo {year} {2007})},\ \Eprint
  {http://arxiv.org/abs/hep-ex/0610021} {arXiv:hep-ex/0610021}\BibitemShut
  {NoStop}%
\bibitem [{\citenamefont {Achasov}\ \emph {et~al.}(2006)\citenamefont {Achasov}
  \emph {et~al.}}]{Achasov:2006vp}%
  \BibitemOpen
  \bibfield  {author} {\bibinfo {author} {\bibfnamefont {M.~N.}\ \bibnamefont
  {Achasov}}  \emph {et~al.} (\bibinfo {collaboration} {SND}),\ }\href
  {\doibase 10.1134/S106377610609007X} {\bibfield  {journal} {\bibinfo
  {journal} {J. Exp. Theor. Phys.}\ }\textbf {\bibinfo {volume} {103}},\
  \bibinfo {pages} {380} (\bibinfo {year} {2006})},\ \Eprint
  {http://arxiv.org/abs/hep-ex/0605013} {arXiv:hep-ex/0605013}\BibitemShut
  {NoStop}%
\bibitem [{\citenamefont {Lees}\ \emph {et~al.}(2012)\citenamefont {Lees} \emph
  {et~al.}}]{Lees:2012cj}%
  \BibitemOpen
  \bibfield  {author} {\bibinfo {author} {\bibfnamefont {J.~P.}\ \bibnamefont
  {Lees}}  \emph {et~al.} (\bibinfo {collaboration} {BaBar}),\ }\href {\doibase
  10.1103/PhysRevD.86.032013} {\bibfield  {journal} {\bibinfo  {journal} {Phys.
  Rev. D}\ }\textbf {\bibinfo {volume} {86}},\ \bibinfo {pages} {032013}
  (\bibinfo {year} {2012})},\ \Eprint {http://arxiv.org/abs/1205.2228}
  {arXiv:1205.2228 [hep-ex]}\BibitemShut {NoStop}%
\bibitem [{\citenamefont {Anastasi}\ \emph {et~al.}(2018)\citenamefont
  {Anastasi} \emph {et~al.}}]{Anastasi:2017eio}%
  \BibitemOpen
  \bibfield  {author} {\bibinfo {author} {\bibfnamefont {A.}~\bibnamefont
  {Anastasi}}  \emph {et~al.} (\bibinfo {collaboration} {KLOE-2}),\ }\href
  {\doibase 10.1007/JHEP03(2018)173} {\bibfield  {journal} {\bibinfo  {journal}
  {JHEP}\ }\textbf {\bibinfo {volume} {03}},\ \bibinfo {pages} {173} (\bibinfo
  {year} {2018})},\ \Eprint {http://arxiv.org/abs/1711.03085} {arXiv:1711.03085
  [hep-ex]}\BibitemShut {NoStop}%
\bibitem [{\citenamefont {Ablikim}\ \emph {et~al.}(2016)\citenamefont {Ablikim}
  \emph {et~al.}}]{Ablikim:2015orh}%
  \BibitemOpen
  \bibfield  {author} {\bibinfo {author} {\bibfnamefont {M.}~\bibnamefont
  {Ablikim}}  \emph {et~al.} (\bibinfo {collaboration} {BESIII}),\ }\href
  {\doibase 10.1016/j.physletb.2015.11.043} {\bibfield  {journal} {\bibinfo
  {journal} {Phys. Lett. B}\ }\textbf {\bibinfo {volume} {753}},\ \bibinfo
  {pages} {629} (\bibinfo {year} {2016})},\ \bibinfo {note} {[Erratum: Phys.
  Lett. B {\bf 812}, 135982 (2021)]},\ \Eprint
  {http://arxiv.org/abs/1507.08188} {arXiv:1507.08188 [hep-ex]}\BibitemShut
  {NoStop}%
\bibitem [{\citenamefont {Di~Luzio}\ \emph {et~al.}(2022)\citenamefont
  {Di~Luzio}, \citenamefont {Masiero}, \citenamefont {Paradisi},\ and\
  \citenamefont {Passera}}]{DiLuzio:2021uty}%
  \BibitemOpen
  \bibfield  {author} {\bibinfo {author} {\bibfnamefont {L.}~\bibnamefont
  {Di~Luzio}}, \bibinfo {author} {\bibfnamefont {A.}~\bibnamefont {Masiero}},
  \bibinfo {author} {\bibfnamefont {P.}~\bibnamefont {Paradisi}}, and \bibinfo
  {author} {\bibfnamefont {M.}~\bibnamefont {Passera}},\ }\href {\doibase
  10.1016/j.physletb.2022.137037} {\bibfield  {journal} {\bibinfo  {journal}
  {Phys. Lett. B}\ }\textbf {\bibinfo {volume} {829}},\ \bibinfo {pages}
  {137037} (\bibinfo {year} {2022})},\ \Eprint
  {http://arxiv.org/abs/2112.08312} {arXiv:2112.08312 [hep-ph]}\BibitemShut
  {NoStop}%
\bibitem [{\citenamefont {Darm\'e}\ \emph {et~al.}(2022)\citenamefont
  {Darm\'e}, \citenamefont {Grilli~di Cortona},\ and\ \citenamefont
  {Nardi}}]{Darme:2021huc}%
  \BibitemOpen
  \bibfield  {author} {\bibinfo {author} {\bibfnamefont {L.}~\bibnamefont
  {Darm\'e}}, \bibinfo {author} {\bibfnamefont {G.}~\bibnamefont {Grilli~di
  Cortona}}, and \bibinfo {author} {\bibfnamefont {E.}~\bibnamefont {Nardi}},\
  }\href {\doibase 10.1007/JHEP06(2022)122} {\bibfield  {journal} {\bibinfo
  {journal} {JHEP}\ }\textbf {\bibinfo {volume} {06}},\ \bibinfo {pages} {122}
  (\bibinfo {year} {2022})},\ \Eprint {http://arxiv.org/abs/2112.09139}
  {arXiv:2112.09139 [hep-ph]}\BibitemShut {NoStop}%
\bibitem [{\citenamefont {Crivellin}\ and\ \citenamefont
  {Hoferichter}(2023)}]{Crivellin:2022gfu}%
  \BibitemOpen
  \bibfield  {author} {\bibinfo {author} {\bibfnamefont {A.}~\bibnamefont
  {Crivellin}} and \bibinfo {author} {\bibfnamefont {M.}~\bibnamefont
  {Hoferichter}},\ }\href {\doibase 10.1103/PhysRevD.108.013005} {\bibfield
  {journal} {\bibinfo  {journal} {Phys. Rev. D}\ }\textbf {\bibinfo {volume}
  {108}},\ \bibinfo {pages} {013005} (\bibinfo {year} {2023})},\ \Eprint
  {http://arxiv.org/abs/2211.12516} {arXiv:2211.12516 [hep-ph]}\BibitemShut
  {NoStop}%
\bibitem [{\citenamefont {Coyle}\ and\ \citenamefont
  {Wagner}(2023)}]{Coyle:2023nmi}%
  \BibitemOpen
  \bibfield  {author} {\bibinfo {author} {\bibfnamefont {N.~M.}\ \bibnamefont
  {Coyle}} and \bibinfo {author} {\bibfnamefont {C.~E.~M.}\ \bibnamefont
  {Wagner}},\ }\href@noop {} {\  (\bibinfo {year} {2023})},\ \Eprint
  {http://arxiv.org/abs/2305.02354} {arXiv:2305.02354 [hep-ph]}\BibitemShut
  {NoStop}%
\bibitem [{\citenamefont {Passera}\ \emph {et~al.}(2008)\citenamefont
  {Passera}, \citenamefont {Marciano},\ and\ \citenamefont
  {Sirlin}}]{Passera:2008jk}%
  \BibitemOpen
  \bibfield  {author} {\bibinfo {author} {\bibfnamefont {M.}~\bibnamefont
  {Passera}}, \bibinfo {author} {\bibfnamefont {W.~J.}\ \bibnamefont
  {Marciano}}, and \bibinfo {author} {\bibfnamefont {A.}~\bibnamefont
  {Sirlin}},\ }\href {\doibase 10.1103/PhysRevD.78.013009} {\bibfield
  {journal} {\bibinfo  {journal} {Phys. Rev. D}\ }\textbf {\bibinfo {volume}
  {78}},\ \bibinfo {pages} {013009} (\bibinfo {year} {2008})},\ \Eprint
  {http://arxiv.org/abs/0804.1142} {arXiv:0804.1142 [hep-ph]}\BibitemShut
  {NoStop}%
\bibitem [{\citenamefont {Crivellin}\ \emph {et~al.}(2020)\citenamefont
  {Crivellin}, \citenamefont {Hoferichter}, \citenamefont {Manzari},\ and\
  \citenamefont {Montull}}]{Crivellin:2020zul}%
  \BibitemOpen
  \bibfield  {author} {\bibinfo {author} {\bibfnamefont {A.}~\bibnamefont
  {Crivellin}}, \bibinfo {author} {\bibfnamefont {M.}~\bibnamefont
  {Hoferichter}}, \bibinfo {author} {\bibfnamefont {C.~A.}\ \bibnamefont
  {Manzari}}, and \bibinfo {author} {\bibfnamefont {M.}~\bibnamefont
  {Montull}},\ }\href {\doibase 10.1103/PhysRevLett.125.091801} {\bibfield
  {journal} {\bibinfo  {journal} {Phys. Rev. Lett.}\ }\textbf {\bibinfo
  {volume} {125}},\ \bibinfo {pages} {091801} (\bibinfo {year} {2020})},\
  \Eprint {http://arxiv.org/abs/2003.04886} {arXiv:2003.04886
  [hep-ph]}\BibitemShut {NoStop}%
\bibitem [{\citenamefont {Keshavarzi}\ \emph
  {et~al.}(2020{\natexlab{b}})\citenamefont {Keshavarzi}, \citenamefont
  {Marciano}, \citenamefont {Passera},\ and\ \citenamefont
  {Sirlin}}]{Keshavarzi:2020bfy}%
  \BibitemOpen
  \bibfield  {author} {\bibinfo {author} {\bibfnamefont {A.}~\bibnamefont
  {Keshavarzi}}, \bibinfo {author} {\bibfnamefont {W.~J.}\ \bibnamefont
  {Marciano}}, \bibinfo {author} {\bibfnamefont {M.}~\bibnamefont {Passera}},
  and \bibinfo {author} {\bibfnamefont {A.}~\bibnamefont {Sirlin}},\ }\href
  {\doibase 10.1103/PhysRevD.102.033002} {\bibfield  {journal} {\bibinfo
  {journal} {Phys. Rev. D}\ }\textbf {\bibinfo {volume} {102}},\ \bibinfo
  {pages} {033002} (\bibinfo {year} {2020}{\natexlab{b}})},\ \Eprint
  {http://arxiv.org/abs/2006.12666} {arXiv:2006.12666 [hep-ph]}\BibitemShut
  {NoStop}%
\bibitem [{\citenamefont {Malaescu}\ and\ \citenamefont
  {Schott}(2021)}]{Malaescu:2020zuc}%
  \BibitemOpen
  \bibfield  {author} {\bibinfo {author} {\bibfnamefont {B.}~\bibnamefont
  {Malaescu}} and \bibinfo {author} {\bibfnamefont {M.}~\bibnamefont
  {Schott}},\ }\href {\doibase 10.1140/epjc/s10052-021-08848-9} {\bibfield
  {journal} {\bibinfo  {journal} {Eur. Phys. J. C}\ }\textbf {\bibinfo {volume}
  {81}},\ \bibinfo {pages} {46} (\bibinfo {year} {2021})},\ \Eprint
  {http://arxiv.org/abs/2008.08107} {arXiv:2008.08107 [hep-ph]}\BibitemShut
  {NoStop}%
\bibitem [{\citenamefont {Colangelo}\ \emph
  {et~al.}(2021{\natexlab{b}})\citenamefont {Colangelo}, \citenamefont
  {Hoferichter},\ and\ \citenamefont {Stoffer}}]{Colangelo:2020lcg}%
  \BibitemOpen
  \bibfield  {author} {\bibinfo {author} {\bibfnamefont {G.}~\bibnamefont
  {Colangelo}}, \bibinfo {author} {\bibfnamefont {M.}~\bibnamefont
  {Hoferichter}}, and \bibinfo {author} {\bibfnamefont {P.}~\bibnamefont
  {Stoffer}},\ }\href {\doibase 10.1016/j.physletb.2021.136073} {\bibfield
  {journal} {\bibinfo  {journal} {Phys. Lett. B}\ }\textbf {\bibinfo {volume}
  {814}},\ \bibinfo {pages} {136073} (\bibinfo {year} {2021}{\natexlab{b}})},\
  \Eprint {http://arxiv.org/abs/2010.07943} {arXiv:2010.07943
  [hep-ph]}\BibitemShut {NoStop}%
\bibitem [{\citenamefont {C\`e}\ \emph
  {et~al.}(2022{\natexlab{b}})\citenamefont {C\`e} \emph
  {et~al.}}]{Ce:2022eix}%
  \BibitemOpen
  \bibfield  {author} {\bibinfo {author} {\bibfnamefont {M.}~\bibnamefont
  {C\`e}}  \emph {et~al.},\ }\href {\doibase 10.1007/JHEP08(2022)220}
  {\bibfield  {journal} {\bibinfo  {journal} {JHEP}\ }\textbf {\bibinfo
  {volume} {08}},\ \bibinfo {pages} {220} (\bibinfo {year}
  {2022}{\natexlab{b}})},\ \Eprint {http://arxiv.org/abs/2203.08676}
  {arXiv:2203.08676 [hep-lat]}\BibitemShut {NoStop}%
\bibitem [{\citenamefont {Colangelo}\ \emph
  {et~al.}(2022{\natexlab{c}})\citenamefont {Colangelo}, \citenamefont
  {Hoferichter}, \citenamefont {Monnard},\ and\ \citenamefont {Ruiz~de
  Elvira}}]{Colangelo:2022lzg}%
  \BibitemOpen
  \bibfield  {author} {\bibinfo {author} {\bibfnamefont {G.}~\bibnamefont
  {Colangelo}}, \bibinfo {author} {\bibfnamefont {M.}~\bibnamefont
  {Hoferichter}}, \bibinfo {author} {\bibfnamefont {J.}~\bibnamefont
  {Monnard}}, and \bibinfo {author} {\bibfnamefont {J.}~\bibnamefont {Ruiz~de
  Elvira}},\ }\href {\doibase 10.1007/JHEP08(2022)295} {\bibfield  {journal}
  {\bibinfo  {journal} {JHEP}\ }\textbf {\bibinfo {volume} {08}},\ \bibinfo
  {pages} {295} (\bibinfo {year} {2022}{\natexlab{c}})},\ \Eprint
  {http://arxiv.org/abs/2207.03495} {arXiv:2207.03495 [hep-ph]}\BibitemShut
  {NoStop}%
\bibitem [{\citenamefont {Chanturia}(2022)}]{Chanturia:2022rcz}%
  \BibitemOpen
  \bibfield  {author} {\bibinfo {author} {\bibfnamefont {G.}~\bibnamefont
  {Chanturia}},\ }\href {\doibase 10.22323/1.412.0030} {\bibfield  {journal}
  {\bibinfo  {journal} {PoS}\ }\textbf {\bibinfo {volume} {Regio2021}},\
  \bibinfo {pages} {030} (\bibinfo {year} {2022})}\BibitemShut {NoStop}%
\bibitem [{\citenamefont {Colangelo}\ \emph
  {et~al.}(2022{\natexlab{d}})\citenamefont {Colangelo}, \citenamefont
  {Hoferichter}, \citenamefont {Kubis},\ and\ \citenamefont
  {Stoffer}}]{Colangelo:2022prz}%
  \BibitemOpen
  \bibfield  {author} {\bibinfo {author} {\bibfnamefont {G.}~\bibnamefont
  {Colangelo}}, \bibinfo {author} {\bibfnamefont {M.}~\bibnamefont
  {Hoferichter}}, \bibinfo {author} {\bibfnamefont {B.}~\bibnamefont {Kubis}},
  and \bibinfo {author} {\bibfnamefont {P.}~\bibnamefont {Stoffer}},\ }\href
  {\doibase 10.1007/JHEP10(2022)032} {\bibfield  {journal} {\bibinfo  {journal}
  {JHEP}\ }\textbf {\bibinfo {volume} {10}},\ \bibinfo {pages} {032} (\bibinfo
  {year} {2022}{\natexlab{d}})},\ \Eprint {http://arxiv.org/abs/2208.08993}
  {arXiv:2208.08993 [hep-ph]}\BibitemShut {NoStop}%
\bibitem [{\citenamefont {Stamen}\ \emph {et~al.}(2022)\citenamefont {Stamen},
  \citenamefont {Hariharan}, \citenamefont {Hoferichter}, \citenamefont
  {Kubis},\ and\ \citenamefont {Stoffer}}]{Stamen:2022uqh}%
  \BibitemOpen
  \bibfield  {author} {\bibinfo {author} {\bibfnamefont {D.}~\bibnamefont
  {Stamen}}, \bibinfo {author} {\bibfnamefont {D.}~\bibnamefont {Hariharan}},
  \bibinfo {author} {\bibfnamefont {M.}~\bibnamefont {Hoferichter}}, \bibinfo
  {author} {\bibfnamefont {B.}~\bibnamefont {Kubis}}, and \bibinfo {author}
  {\bibfnamefont {P.}~\bibnamefont {Stoffer}},\ }\href {\doibase
  10.1140/epjc/s10052-022-10348-3} {\bibfield  {journal} {\bibinfo  {journal}
  {Eur. Phys. J. C}\ }\textbf {\bibinfo {volume} {82}},\ \bibinfo {pages} {432}
  (\bibinfo {year} {2022})},\ \Eprint {http://arxiv.org/abs/2202.11106}
  {arXiv:2202.11106 [hep-ph]}\BibitemShut {NoStop}%
\bibitem [{\citenamefont {Alexandrou}\ \emph
  {et~al.}(2023{\natexlab{b}})\citenamefont {Alexandrou} \emph
  {et~al.}}]{ExtendedTwistedMassCollaborationETMC:2022sta}%
  \BibitemOpen
  \bibfield  {author} {\bibinfo {author} {\bibfnamefont {C.}~\bibnamefont
  {Alexandrou}}  \emph {et~al.} (\bibinfo {collaboration} {ETM}),\ }\href
  {\doibase 10.1103/PhysRevLett.130.241901} {\bibfield  {journal} {\bibinfo
  {journal} {Phys. Rev. Lett.}\ }\textbf {\bibinfo {volume} {130}},\ \bibinfo
  {pages} {241901} (\bibinfo {year} {2023}{\natexlab{b}})},\ \Eprint
  {http://arxiv.org/abs/2212.08467} {arXiv:2212.08467 [hep-lat]}\BibitemShut
  {NoStop}%
\bibitem [{\citenamefont {Hoferichter}\ \emph {et~al.}(2022)\citenamefont
  {Hoferichter}, \citenamefont {Colangelo}, \citenamefont {Hoid}, \citenamefont
  {Kubis}, \citenamefont {Ruiz~de Elvira}, \citenamefont {Stamen},\ and\
  \citenamefont {Stoffer}}]{Hoferichter:2022iqe}%
  \BibitemOpen
  \bibfield  {author} {\bibinfo {author} {\bibfnamefont {M.}~\bibnamefont
  {Hoferichter}}, \bibinfo {author} {\bibfnamefont {G.}~\bibnamefont
  {Colangelo}}, \bibinfo {author} {\bibfnamefont {B.-L.}\ \bibnamefont {Hoid}},
  \bibinfo {author} {\bibfnamefont {B.}~\bibnamefont {Kubis}}, \bibinfo
  {author} {\bibfnamefont {J.}~\bibnamefont {Ruiz~de Elvira}}, \bibinfo
  {author} {\bibfnamefont {D.}~\bibnamefont {Stamen}}, and \bibinfo {author}
  {\bibfnamefont {P.}~\bibnamefont {Stoffer}},\ }\href {\doibase
  10.22323/1.430.0316} {\bibfield  {journal} {\bibinfo  {journal} {PoS}\
  }\textbf {\bibinfo {volume} {LATTICE2022}},\ \bibinfo {pages} {316} (\bibinfo
  {year} {2022})},\ \Eprint {http://arxiv.org/abs/2210.11904} {arXiv:2210.11904
  [hep-ph]}\BibitemShut {NoStop}%
\bibitem [{\citenamefont {James}\ \emph {et~al.}(2022)\citenamefont {James},
  \citenamefont {Lewis},\ and\ \citenamefont {Maltman}}]{James:2021sor}%
  \BibitemOpen
  \bibfield  {author} {\bibinfo {author} {\bibfnamefont {C.~L.}\ \bibnamefont
  {James}}, \bibinfo {author} {\bibfnamefont {R.}~\bibnamefont {Lewis}}, and
  \bibinfo {author} {\bibfnamefont {K.}~\bibnamefont {Maltman}},\ }\href
  {\doibase 10.1103/PhysRevD.105.053010} {\bibfield  {journal} {\bibinfo
  {journal} {Phys. Rev. D}\ }\textbf {\bibinfo {volume} {105}},\ \bibinfo
  {pages} {053010} (\bibinfo {year} {2022})},\ \Eprint
  {http://arxiv.org/abs/2109.13729} {arXiv:2109.13729 [hep-ph]}\BibitemShut
  {NoStop}%
\bibitem [{\citenamefont {Lees}\ \emph {et~al.}(2021)\citenamefont {Lees} \emph
  {et~al.}}]{BABAR:2021cde}%
  \BibitemOpen
  \bibfield  {author} {\bibinfo {author} {\bibfnamefont {J.~P.}\ \bibnamefont
  {Lees}}  \emph {et~al.} (\bibinfo {collaboration} {BaBar}),\ }\href {\doibase
  10.1103/PhysRevD.104.112003} {\bibfield  {journal} {\bibinfo  {journal}
  {Phys. Rev. D}\ }\textbf {\bibinfo {volume} {104}},\ \bibinfo {pages}
  {112003} (\bibinfo {year} {2021})},\ \Eprint
  {http://arxiv.org/abs/2110.00520} {arXiv:2110.00520 [hep-ex]}\BibitemShut
  {NoStop}%
\bibitem [{\citenamefont {Boito}\ \emph {et~al.}(2022)\citenamefont {Boito},
  \citenamefont {Golterman}, \citenamefont {Maltman},\ and\ \citenamefont
  {Peris}}]{Boito:2022rkw}%
  \BibitemOpen
  \bibfield  {author} {\bibinfo {author} {\bibfnamefont {D.}~\bibnamefont
  {Boito}}, \bibinfo {author} {\bibfnamefont {M.}~\bibnamefont {Golterman}},
  \bibinfo {author} {\bibfnamefont {K.}~\bibnamefont {Maltman}}, and \bibinfo
  {author} {\bibfnamefont {S.}~\bibnamefont {Peris}},\ }\href {\doibase
  10.1103/PhysRevD.105.093003} {\bibfield  {journal} {\bibinfo  {journal}
  {Phys. Rev. D}\ }\textbf {\bibinfo {volume} {105}},\ \bibinfo {pages}
  {093003} (\bibinfo {year} {2022})},\ \Eprint
  {http://arxiv.org/abs/2203.05070} {arXiv:2203.05070 [hep-ph]}\BibitemShut
  {NoStop}%
\bibitem [{\citenamefont {Boito}\ \emph {et~al.}(2023)\citenamefont {Boito},
  \citenamefont {Golterman}, \citenamefont {Maltman},\ and\ \citenamefont
  {Peris}}]{Boito:2022dry}%
  \BibitemOpen
  \bibfield  {author} {\bibinfo {author} {\bibfnamefont {D.}~\bibnamefont
  {Boito}}, \bibinfo {author} {\bibfnamefont {M.}~\bibnamefont {Golterman}},
  \bibinfo {author} {\bibfnamefont {K.}~\bibnamefont {Maltman}}, and \bibinfo
  {author} {\bibfnamefont {S.}~\bibnamefont {Peris}},\ }\href {\doibase
  10.1103/PhysRevD.107.074001} {\bibfield  {journal} {\bibinfo  {journal}
  {Phys. Rev. D}\ }\textbf {\bibinfo {volume} {107}},\ \bibinfo {pages}
  {074001} (\bibinfo {year} {2023})},\ \Eprint
  {http://arxiv.org/abs/2211.11055} {arXiv:2211.11055 [hep-ph]}\BibitemShut
  {NoStop}%
\bibitem [{\citenamefont {Benton}\ \emph {et~al.}(2023)\citenamefont {Benton},
  \citenamefont {Boito}, \citenamefont {Golterman}, \citenamefont {Keshavarzi},
  \citenamefont {Maltman},\ and\ \citenamefont {Peris}}]{Benton:2023dci}%
  \BibitemOpen
  \bibfield  {author} {\bibinfo {author} {\bibfnamefont {G.}~\bibnamefont
  {Benton}}, \bibinfo {author} {\bibfnamefont {D.}~\bibnamefont {Boito}},
  \bibinfo {author} {\bibfnamefont {M.}~\bibnamefont {Golterman}}, \bibinfo
  {author} {\bibfnamefont {A.}~\bibnamefont {Keshavarzi}}, \bibinfo {author}
  {\bibfnamefont {K.}~\bibnamefont {Maltman}}, and \bibinfo {author}
  {\bibfnamefont {S.}~\bibnamefont {Peris}},\ }\href@noop {} {\  (\bibinfo
  {year} {2023})},\ \Eprint {http://arxiv.org/abs/2306.16808} {arXiv:2306.16808
  [hep-ph]}\BibitemShut {NoStop}%
\bibitem [{\citenamefont {Hoefer}\ \emph {et~al.}(2002)\citenamefont {Hoefer},
  \citenamefont {Gluza},\ and\ \citenamefont {Jegerlehner}}]{Hoefer:2001mx}%
  \BibitemOpen
  \bibfield  {author} {\bibinfo {author} {\bibfnamefont {A.}~\bibnamefont
  {Hoefer}}, \bibinfo {author} {\bibfnamefont {J.}~\bibnamefont {Gluza}}, and
  \bibinfo {author} {\bibfnamefont {F.}~\bibnamefont {Jegerlehner}},\ }\href
  {\doibase 10.1007/s100520200916} {\bibfield  {journal} {\bibinfo  {journal}
  {Eur. Phys. J. C}\ }\textbf {\bibinfo {volume} {24}},\ \bibinfo {pages} {51}
  (\bibinfo {year} {2002})},\ \Eprint {http://arxiv.org/abs/hep-ph/0107154}
  {arXiv:hep-ph/0107154}\BibitemShut {NoStop}%
\bibitem [{\citenamefont {Czy{\.z}}\ \emph {et~al.}(2005)\citenamefont
  {Czy{\.z}}, \citenamefont {Grzeli{\'n}ska}, \citenamefont {K{\"u}hn},\ and\
  \citenamefont {Rodrigo}}]{Czyz:2004rj}%
  \BibitemOpen
  \bibfield  {author} {\bibinfo {author} {\bibfnamefont {H.}~\bibnamefont
  {Czy{\.z}}}, \bibinfo {author} {\bibfnamefont {A.}~\bibnamefont
  {Grzeli{\'n}ska}}, \bibinfo {author} {\bibfnamefont {J.~H.}\ \bibnamefont
  {K{\"u}hn}}, and \bibinfo {author} {\bibfnamefont {G.}~\bibnamefont
  {Rodrigo}},\ }\href {\doibase 10.1140/epjc/s2004-02103-1} {\bibfield
  {journal} {\bibinfo  {journal} {Eur. Phys. J. C}\ }\textbf {\bibinfo {volume}
  {39}},\ \bibinfo {pages} {411} (\bibinfo {year} {2005})},\ \Eprint
  {http://arxiv.org/abs/hep-ph/0404078} {arXiv:hep-ph/0404078}\BibitemShut
  {NoStop}%
\bibitem [{\citenamefont {Gluza}\ \emph {et~al.}(2003)\citenamefont {Gluza},
  \citenamefont {Hoefer}, \citenamefont {Jadach},\ and\ \citenamefont
  {Jegerlehner}}]{Gluza:2002ui}%
  \BibitemOpen
  \bibfield  {author} {\bibinfo {author} {\bibfnamefont {J.}~\bibnamefont
  {Gluza}}, \bibinfo {author} {\bibfnamefont {A.}~\bibnamefont {Hoefer}},
  \bibinfo {author} {\bibfnamefont {S.}~\bibnamefont {Jadach}}, and \bibinfo
  {author} {\bibfnamefont {F.}~\bibnamefont {Jegerlehner}},\ }\href {\doibase
  10.1140/epjc/s2003-01146-0} {\bibfield  {journal} {\bibinfo  {journal} {Eur.
  Phys. J. C}\ }\textbf {\bibinfo {volume} {28}},\ \bibinfo {pages} {261}
  (\bibinfo {year} {2003})},\ \Eprint {http://arxiv.org/abs/hep-ph/0212386}
  {arXiv:hep-ph/0212386}\BibitemShut {NoStop}%
\bibitem [{\citenamefont {Bystritskiy}\ \emph {et~al.}(2005)\citenamefont
  {Bystritskiy}, \citenamefont {Kuraev}, \citenamefont {Fedotovich},\ and\
  \citenamefont {Ignatov}}]{Bystritskiy:2005ib}%
  \BibitemOpen
  \bibfield  {author} {\bibinfo {author} {\bibfnamefont {Y.~M.}\ \bibnamefont
  {Bystritskiy}}, \bibinfo {author} {\bibfnamefont {E.~A.}\ \bibnamefont
  {Kuraev}}, \bibinfo {author} {\bibfnamefont {G.~V.}\ \bibnamefont
  {Fedotovich}}, and \bibinfo {author} {\bibfnamefont {F.~V.}\ \bibnamefont
  {Ignatov}},\ }\href {\doibase 10.1103/PhysRevD.72.114019} {\bibfield
  {journal} {\bibinfo  {journal} {Phys. Rev. D}\ }\textbf {\bibinfo {volume}
  {72}},\ \bibinfo {pages} {114019} (\bibinfo {year} {2005})},\ \Eprint
  {http://arxiv.org/abs/hep-ph/0505236} {arXiv:hep-ph/0505236}\BibitemShut
  {NoStop}%
\bibitem [{\citenamefont {Wess}\ and\ \citenamefont
  {Zumino}(1971)}]{Wess:1971yu}%
  \BibitemOpen
  \bibfield  {author} {\bibinfo {author} {\bibfnamefont {J.}~\bibnamefont
  {Wess}} and \bibinfo {author} {\bibfnamefont {B.}~\bibnamefont {Zumino}},\
  }\href {\doibase 10.1016/0370-2693(71)90582-X} {\bibfield  {journal}
  {\bibinfo  {journal} {Phys. Lett. B}\ }\textbf {\bibinfo {volume} {37}},\
  \bibinfo {pages} {95} (\bibinfo {year} {1971})}\BibitemShut {NoStop}%
\bibitem [{\citenamefont {Witten}(1983)}]{Witten:1983tw}%
  \BibitemOpen
  \bibfield  {author} {\bibinfo {author} {\bibfnamefont {E.}~\bibnamefont
  {Witten}},\ }\href {\doibase 10.1016/0550-3213(83)90063-9} {\bibfield
  {journal} {\bibinfo  {journal} {Nucl. Phys. B}\ }\textbf {\bibinfo {volume}
  {223}},\ \bibinfo {pages} {422} (\bibinfo {year} {1983})}\BibitemShut
  {NoStop}%
\bibitem [{\citenamefont {Ametller}\ \emph {et~al.}(2001)\citenamefont
  {Ametller}, \citenamefont {Knecht},\ and\ \citenamefont
  {Talavera}}]{Ametller:2001yk}%
  \BibitemOpen
  \bibfield  {author} {\bibinfo {author} {\bibfnamefont {L.}~\bibnamefont
  {Ametller}}, \bibinfo {author} {\bibfnamefont {M.}~\bibnamefont {Knecht}},
  and \bibinfo {author} {\bibfnamefont {P.}~\bibnamefont {Talavera}},\ }\href
  {\doibase 10.1103/PhysRevD.64.094009} {\bibfield  {journal} {\bibinfo
  {journal} {Phys. Rev. D}\ }\textbf {\bibinfo {volume} {64}},\ \bibinfo
  {pages} {094009} (\bibinfo {year} {2001})},\ \Eprint
  {http://arxiv.org/abs/hep-ph/0107127} {arXiv:hep-ph/0107127}\BibitemShut
  {NoStop}%
\bibitem [{\citenamefont {Ahmedov}\ \emph {et~al.}(2002)\citenamefont
  {Ahmedov}, \citenamefont {Fedotovich}, \citenamefont {Kuraev},\ and\
  \citenamefont {Silagadze}}]{Ahmedov:2002tg}%
  \BibitemOpen
  \bibfield  {author} {\bibinfo {author} {\bibfnamefont {A.~I.}\ \bibnamefont
  {Ahmedov}}, \bibinfo {author} {\bibfnamefont {G.~V.}\ \bibnamefont
  {Fedotovich}}, \bibinfo {author} {\bibfnamefont {E.~A.}\ \bibnamefont
  {Kuraev}}, and \bibinfo {author} {\bibfnamefont {Z.~K.}\ \bibnamefont
  {Silagadze}},\ }\href {\doibase 10.1134/1.1755389} {\bibfield  {journal}
  {\bibinfo  {journal} {JHEP}\ }\textbf {\bibinfo {volume} {09}},\ \bibinfo
  {pages} {008} (\bibinfo {year} {2002})},\ \Eprint
  {http://arxiv.org/abs/hep-ph/0201157} {arXiv:hep-ph/0201157}\BibitemShut
  {NoStop}%
\bibitem [{\citenamefont {Bakmaev}\ \emph {et~al.}(2006)\citenamefont
  {Bakmaev}, \citenamefont {Bystritskiy},\ and\ \citenamefont
  {Kuraev}}]{Bakmaev:2005sg}%
  \BibitemOpen
  \bibfield  {author} {\bibinfo {author} {\bibfnamefont {S.}~\bibnamefont
  {Bakmaev}}, \bibinfo {author} {\bibfnamefont {Y.~M.}\ \bibnamefont
  {Bystritskiy}}, and \bibinfo {author} {\bibfnamefont {E.~A.}\ \bibnamefont
  {Kuraev}},\ }\href {\doibase 10.1103/PhysRevD.73.034010} {\bibfield
  {journal} {\bibinfo  {journal} {Phys. Rev. D}\ }\textbf {\bibinfo {volume}
  {73}},\ \bibinfo {pages} {034010} (\bibinfo {year} {2006})},\ \Eprint
  {http://arxiv.org/abs/hep-ph/0507219} {arXiv:hep-ph/0507219}\BibitemShut
  {NoStop}%
\bibitem [{\citenamefont {Moussallam}(2013)}]{Moussallam:2013una}%
  \BibitemOpen
  \bibfield  {author} {\bibinfo {author} {\bibfnamefont {B.}~\bibnamefont
  {Moussallam}},\ }\href {\doibase 10.1140/epjc/s10052-013-2539-y} {\bibfield
  {journal} {\bibinfo  {journal} {Eur. Phys. J. C}\ }\textbf {\bibinfo {volume}
  {73}},\ \bibinfo {pages} {2539} (\bibinfo {year} {2013})},\ \Eprint
  {http://arxiv.org/abs/1305.3143} {arXiv:1305.3143 [hep-ph]}\BibitemShut
  {NoStop}%
\bibitem [{\citenamefont {Khuri}\ and\ \citenamefont
  {Treiman}(1960)}]{Khuri:1960zz}%
  \BibitemOpen
  \bibfield  {author} {\bibinfo {author} {\bibfnamefont {N.~N.}\ \bibnamefont
  {Khuri}} and \bibinfo {author} {\bibfnamefont {S.~B.}\ \bibnamefont
  {Treiman}},\ }\href {\doibase 10.1103/PhysRev.119.1115} {\bibfield  {journal}
  {\bibinfo  {journal} {Phys. Rev.}\ }\textbf {\bibinfo {volume} {119}},\
  \bibinfo {pages} {1115} (\bibinfo {year} {1960})}\BibitemShut {NoStop}%
\bibitem [{\citenamefont {Hoferichter}\ \emph {et~al.}(2014)\citenamefont
  {Hoferichter}, \citenamefont {Kubis}, \citenamefont {Leupold}, \citenamefont
  {Niecknig},\ and\ \citenamefont {Schneider}}]{Hoferichter:2014vra}%
  \BibitemOpen
  \bibfield  {author} {\bibinfo {author} {\bibfnamefont {M.}~\bibnamefont
  {Hoferichter}}, \bibinfo {author} {\bibfnamefont {B.}~\bibnamefont {Kubis}},
  \bibinfo {author} {\bibfnamefont {S.}~\bibnamefont {Leupold}}, \bibinfo
  {author} {\bibfnamefont {F.}~\bibnamefont {Niecknig}}, and \bibinfo {author}
  {\bibfnamefont {S.~P.}\ \bibnamefont {Schneider}},\ }\href {\doibase
  10.1140/epjc/s10052-014-3180-0} {\bibfield  {journal} {\bibinfo  {journal}
  {Eur. Phys. J. C}\ }\textbf {\bibinfo {volume} {74}},\ \bibinfo {pages}
  {3180} (\bibinfo {year} {2014})},\ \Eprint {http://arxiv.org/abs/1410.4691}
  {arXiv:1410.4691 [hep-ph]}\BibitemShut {NoStop}%
\bibitem [{\citenamefont {Adler}\ \emph {et~al.}(1971)\citenamefont {Adler},
  \citenamefont {Lee}, \citenamefont {Treiman},\ and\ \citenamefont
  {Zee}}]{Adler:1971nq}%
  \BibitemOpen
  \bibfield  {author} {\bibinfo {author} {\bibfnamefont {S.~L.}\ \bibnamefont
  {Adler}}, \bibinfo {author} {\bibfnamefont {B.~W.}\ \bibnamefont {Lee}},
  \bibinfo {author} {\bibfnamefont {S.~B.}\ \bibnamefont {Treiman}}, and
  \bibinfo {author} {\bibfnamefont {A.}~\bibnamefont {Zee}},\ }\href {\doibase
  10.1103/PhysRevD.4.3497} {\bibfield  {journal} {\bibinfo  {journal} {Phys.
  Rev. D}\ }\textbf {\bibinfo {volume} {4}},\ \bibinfo {pages} {3497} (\bibinfo
  {year} {1971})}\BibitemShut {NoStop}%
\bibitem [{\citenamefont {Terent'ev}(1972)}]{Terentev:1971cso}%
  \BibitemOpen
  \bibfield  {author} {\bibinfo {author} {\bibfnamefont {M.~V.}\ \bibnamefont
  {Terent'ev}},\ }\href {\doibase 10.1016/0370-2693(72)90171-2} {\bibfield
  {journal} {\bibinfo  {journal} {Phys. Lett. B}\ }\textbf {\bibinfo {volume}
  {38}},\ \bibinfo {pages} {419} (\bibinfo {year} {1972})}\BibitemShut
  {NoStop}%
\bibitem [{\citenamefont {Aviv}\ and\ \citenamefont {Zee}(1972)}]{Aviv:1971hq}%
  \BibitemOpen
  \bibfield  {author} {\bibinfo {author} {\bibfnamefont {R.}~\bibnamefont
  {Aviv}} and \bibinfo {author} {\bibfnamefont {A.}~\bibnamefont {Zee}},\
  }\href {\doibase 10.1103/PhysRevD.5.2372} {\bibfield  {journal} {\bibinfo
  {journal} {Phys. Rev. D}\ }\textbf {\bibinfo {volume} {5}},\ \bibinfo {pages}
  {2372} (\bibinfo {year} {1972})}\BibitemShut {NoStop}%
\bibitem [{\citenamefont {Aitchison}\ and\ \citenamefont
  {Golding}(1978)}]{Aitchison:1977ej}%
  \BibitemOpen
  \bibfield  {author} {\bibinfo {author} {\bibfnamefont {I.~J.~R.}\
  \bibnamefont {Aitchison}} and \bibinfo {author} {\bibfnamefont {R.~J.~A.}\
  \bibnamefont {Golding}},\ }\href {\doibase 10.1088/0305-4616/4/1/007}
  {\bibfield  {journal} {\bibinfo  {journal} {J. Phys. G}\ }\textbf {\bibinfo
  {volume} {4}},\ \bibinfo {pages} {43} (\bibinfo {year} {1978})}\BibitemShut
  {NoStop}%
\bibitem [{\citenamefont {Niecknig}\ \emph {et~al.}(2012)\citenamefont
  {Niecknig}, \citenamefont {Kubis},\ and\ \citenamefont
  {Schneider}}]{Niecknig:2012sj}%
  \BibitemOpen
  \bibfield  {author} {\bibinfo {author} {\bibfnamefont {F.}~\bibnamefont
  {Niecknig}}, \bibinfo {author} {\bibfnamefont {B.}~\bibnamefont {Kubis}}, and
  \bibinfo {author} {\bibfnamefont {S.~P.}\ \bibnamefont {Schneider}},\ }\href
  {\doibase 10.1140/epjc/s10052-012-2014-1} {\bibfield  {journal} {\bibinfo
  {journal} {Eur. Phys. J. C}\ }\textbf {\bibinfo {volume} {72}},\ \bibinfo
  {pages} {2014} (\bibinfo {year} {2012})},\ \Eprint
  {http://arxiv.org/abs/1203.2501} {arXiv:1203.2501 [hep-ph]}\BibitemShut
  {NoStop}%
\bibitem [{\citenamefont {Schneider}\ \emph {et~al.}(2012)\citenamefont
  {Schneider}, \citenamefont {Kubis},\ and\ \citenamefont
  {Niecknig}}]{Schneider:2012ez}%
  \BibitemOpen
  \bibfield  {author} {\bibinfo {author} {\bibfnamefont {S.~P.}\ \bibnamefont
  {Schneider}}, \bibinfo {author} {\bibfnamefont {B.}~\bibnamefont {Kubis}},
  and \bibinfo {author} {\bibfnamefont {F.}~\bibnamefont {Niecknig}},\ }\href
  {\doibase 10.1103/PhysRevD.86.054013} {\bibfield  {journal} {\bibinfo
  {journal} {Phys. Rev. D}\ }\textbf {\bibinfo {volume} {86}},\ \bibinfo
  {pages} {054013} (\bibinfo {year} {2012})},\ \Eprint
  {http://arxiv.org/abs/1206.3098} {arXiv:1206.3098 [hep-ph]}\BibitemShut
  {NoStop}%
\bibitem [{\citenamefont {Hoferichter}\ \emph {et~al.}(2012)\citenamefont
  {Hoferichter}, \citenamefont {Kubis},\ and\ \citenamefont
  {Sakkas}}]{Hoferichter:2012pm}%
  \BibitemOpen
  \bibfield  {author} {\bibinfo {author} {\bibfnamefont {M.}~\bibnamefont
  {Hoferichter}}, \bibinfo {author} {\bibfnamefont {B.}~\bibnamefont {Kubis}},
  and \bibinfo {author} {\bibfnamefont {D.}~\bibnamefont {Sakkas}},\ }\href
  {\doibase 10.1103/PhysRevD.86.116009} {\bibfield  {journal} {\bibinfo
  {journal} {Phys. Rev. D}\ }\textbf {\bibinfo {volume} {86}},\ \bibinfo
  {pages} {116009} (\bibinfo {year} {2012})},\ \Eprint
  {http://arxiv.org/abs/1210.6793} {arXiv:1210.6793 [hep-ph]}\BibitemShut
  {NoStop}%
\bibitem [{\citenamefont {Danilkin}\ \emph {et~al.}(2015)\citenamefont
  {Danilkin} \emph {et~al.}}]{Danilkin:2014cra}%
  \BibitemOpen
  \bibfield  {author} {\bibinfo {author} {\bibfnamefont {I.~V.}\ \bibnamefont
  {Danilkin}}  \emph {et~al.},\ }\href {\doibase 10.1103/PhysRevD.91.094029}
  {\bibfield  {journal} {\bibinfo  {journal} {Phys. Rev. D}\ }\textbf {\bibinfo
  {volume} {91}},\ \bibinfo {pages} {094029} (\bibinfo {year} {2015})},\
  \Eprint {http://arxiv.org/abs/1409.7708} {arXiv:1409.7708
  [hep-ph]}\BibitemShut {NoStop}%
\bibitem [{\citenamefont {Dax}\ \emph {et~al.}(2018)\citenamefont {Dax},
  \citenamefont {Isken},\ and\ \citenamefont {Kubis}}]{Dax:2018rvs}%
  \BibitemOpen
  \bibfield  {author} {\bibinfo {author} {\bibfnamefont {M.}~\bibnamefont
  {Dax}}, \bibinfo {author} {\bibfnamefont {T.}~\bibnamefont {Isken}}, and
  \bibinfo {author} {\bibfnamefont {B.}~\bibnamefont {Kubis}},\ }\href
  {\doibase 10.1140/epjc/s10052-018-6346-3} {\bibfield  {journal} {\bibinfo
  {journal} {Eur. Phys. J. C}\ }\textbf {\bibinfo {volume} {78}},\ \bibinfo
  {pages} {859} (\bibinfo {year} {2018})},\ \Eprint
  {http://arxiv.org/abs/1808.08957} {arXiv:1808.08957 [hep-ph]}\BibitemShut
  {NoStop}%
\bibitem [{\citenamefont {Jacob}\ and\ \citenamefont
  {Wick}(1959)}]{Jacob:1959at}%
  \BibitemOpen
  \bibfield  {author} {\bibinfo {author} {\bibfnamefont {M.}~\bibnamefont
  {Jacob}} and \bibinfo {author} {\bibfnamefont {G.~C.}\ \bibnamefont {Wick}},\
  }\href {\doibase 10.1016/0003-4916(59)90051-X} {\bibfield  {journal}
  {\bibinfo  {journal} {Annals Phys.}\ }\textbf {\bibinfo {volume} {7}},\
  \bibinfo {pages} {404} (\bibinfo {year} {1959})},\ \bibinfo {note} {[Annals
  Phys. {\bf 281}, 774 (2000)]}\BibitemShut {NoStop}%
\bibitem [{\citenamefont {Hoferichter}\ \emph {et~al.}(2017)\citenamefont
  {Hoferichter}, \citenamefont {Kubis},\ and\ \citenamefont
  {Zanke}}]{Hoferichter:2017ftn}%
  \BibitemOpen
  \bibfield  {author} {\bibinfo {author} {\bibfnamefont {M.}~\bibnamefont
  {Hoferichter}}, \bibinfo {author} {\bibfnamefont {B.}~\bibnamefont {Kubis}},
  and \bibinfo {author} {\bibfnamefont {M.}~\bibnamefont {Zanke}},\ }\href
  {\doibase 10.1103/PhysRevD.96.114016} {\bibfield  {journal} {\bibinfo
  {journal} {Phys. Rev. D}\ }\textbf {\bibinfo {volume} {96}},\ \bibinfo
  {pages} {114016} (\bibinfo {year} {2017})},\ \Eprint
  {http://arxiv.org/abs/1710.00824} {arXiv:1710.00824 [hep-ph]}\BibitemShut
  {NoStop}%
\bibitem [{\citenamefont {Bijnens}\ \emph {et~al.}(1990)\citenamefont
  {Bijnens}, \citenamefont {Bramon},\ and\ \citenamefont
  {Cornet}}]{Bijnens:1989ff}%
  \BibitemOpen
  \bibfield  {author} {\bibinfo {author} {\bibfnamefont {J.}~\bibnamefont
  {Bijnens}}, \bibinfo {author} {\bibfnamefont {A.}~\bibnamefont {Bramon}}, and
  \bibinfo {author} {\bibfnamefont {F.}~\bibnamefont {Cornet}},\ }\href
  {\doibase 10.1016/0370-2693(90)91212-T} {\bibfield  {journal} {\bibinfo
  {journal} {Phys. Lett. B}\ }\textbf {\bibinfo {volume} {237}},\ \bibinfo
  {pages} {488} (\bibinfo {year} {1990})}\BibitemShut {NoStop}%
\bibitem [{\citenamefont {Brice\~no}\ \emph {et~al.}(2016)\citenamefont
  {Brice\~no}, \citenamefont {Dudek}, \citenamefont {Edwards}, \citenamefont
  {Shultz}, \citenamefont {Thomas},\ and\ \citenamefont
  {Wilson}}]{Briceno:2016kkp}%
  \BibitemOpen
  \bibfield  {author} {\bibinfo {author} {\bibfnamefont {R.~A.}\ \bibnamefont
  {Brice\~no}}, \bibinfo {author} {\bibfnamefont {J.~J.}\ \bibnamefont
  {Dudek}}, \bibinfo {author} {\bibfnamefont {R.~G.}\ \bibnamefont {Edwards}},
  \bibinfo {author} {\bibfnamefont {C.~J.}\ \bibnamefont {Shultz}}, \bibinfo
  {author} {\bibfnamefont {C.~E.}\ \bibnamefont {Thomas}}, and \bibinfo
  {author} {\bibfnamefont {D.~J.}\ \bibnamefont {Wilson}} (\bibinfo
  {collaboration} {HadSpec}),\ }\href {\doibase 10.1103/PhysRevD.93.114508}
  {\bibfield  {journal} {\bibinfo  {journal} {Phys. Rev. D}\ }\textbf {\bibinfo
  {volume} {93}},\ \bibinfo {pages} {114508} (\bibinfo {year} {2016})},\
  \bibinfo {note} {[Erratum: Phys. Rev. D {\bf 105}, 079902 (2022)]},\ \Eprint
  {http://arxiv.org/abs/1604.03530} {arXiv:1604.03530 [hep-ph]}\BibitemShut
  {NoStop}%
\bibitem [{\citenamefont {Alexandrou}\ \emph {et~al.}(2018)\citenamefont
  {Alexandrou} \emph {et~al.}}]{Alexandrou:2018jbt}%
  \BibitemOpen
  \bibfield  {author} {\bibinfo {author} {\bibfnamefont {C.}~\bibnamefont
  {Alexandrou}}  \emph {et~al.},\ }\href {\doibase 10.1103/PhysRevD.98.074502}
  {\bibfield  {journal} {\bibinfo  {journal} {Phys. Rev. D}\ }\textbf {\bibinfo
  {volume} {98}},\ \bibinfo {pages} {074502} (\bibinfo {year} {2018})},\
  \bibinfo {note} {[Erratum: Phys. Rev. D {\bf 105}, 019902 (2022)]},\ \Eprint
  {http://arxiv.org/abs/1807.08357} {arXiv:1807.08357 [hep-lat]}\BibitemShut
  {NoStop}%
\bibitem [{\citenamefont {Niehus}\ \emph {et~al.}(2021)\citenamefont {Niehus},
  \citenamefont {Hoferichter},\ and\ \citenamefont {Kubis}}]{Niehus:2021iin}%
  \BibitemOpen
  \bibfield  {author} {\bibinfo {author} {\bibfnamefont {M.}~\bibnamefont
  {Niehus}}, \bibinfo {author} {\bibfnamefont {M.}~\bibnamefont {Hoferichter}},
  and \bibinfo {author} {\bibfnamefont {B.}~\bibnamefont {Kubis}},\ }\href
  {\doibase 10.1007/JHEP12(2021)038} {\bibfield  {journal} {\bibinfo  {journal}
  {JHEP}\ }\textbf {\bibinfo {volume} {12}},\ \bibinfo {pages} {038} (\bibinfo
  {year} {2021})},\ \Eprint {http://arxiv.org/abs/2110.11372} {arXiv:2110.11372
  [hep-ph]}\BibitemShut {NoStop}%
\bibitem [{\citenamefont {Workman}\ \emph {et~al.}(2022)\citenamefont {Workman}
  \emph {et~al.}}]{ParticleDataGroup:2022pth}%
  \BibitemOpen
  \bibfield  {author} {\bibinfo {author} {\bibfnamefont {R.~L.}\ \bibnamefont
  {Workman}}  \emph {et~al.} (\bibinfo {collaboration} {Particle Data Group}),\
  }\href {\doibase 10.1093/ptep/ptac097} {\bibfield  {journal} {\bibinfo
  {journal} {PTEP}\ }\textbf {\bibinfo {volume} {2022}},\ \bibinfo {pages}
  {083C01} (\bibinfo {year} {2022})}\BibitemShut {NoStop}%
\bibitem [{\citenamefont {Campanario}\ \emph {et~al.}(2019)\citenamefont
  {Campanario}, \citenamefont {Czy\.z}, \citenamefont {Gluza}, \citenamefont
  {Jeli\'nski}, \citenamefont {Rodrigo}, \citenamefont {Tracz},\ and\
  \citenamefont {Zhuridov}}]{Campanario:2019mjh}%
  \BibitemOpen
  \bibfield  {author} {\bibinfo {author} {\bibfnamefont {F.}~\bibnamefont
  {Campanario}}, \bibinfo {author} {\bibfnamefont {H.}~\bibnamefont {Czy\.z}},
  \bibinfo {author} {\bibfnamefont {J.}~\bibnamefont {Gluza}}, \bibinfo
  {author} {\bibfnamefont {T.}~\bibnamefont {Jeli\'nski}}, \bibinfo {author}
  {\bibfnamefont {G.}~\bibnamefont {Rodrigo}}, \bibinfo {author} {\bibfnamefont
  {S.}~\bibnamefont {Tracz}}, and \bibinfo {author} {\bibfnamefont
  {D.}~\bibnamefont {Zhuridov}},\ }\href {\doibase 10.1103/PhysRevD.100.076004}
  {\bibfield  {journal} {\bibinfo  {journal} {Phys. Rev. D}\ }\textbf {\bibinfo
  {volume} {100}},\ \bibinfo {pages} {076004} (\bibinfo {year} {2019})},\
  \Eprint {http://arxiv.org/abs/1903.10197} {arXiv:1903.10197
  [hep-ph]}\BibitemShut {NoStop}%
\bibitem [{\citenamefont {Ignatov}\ and\ \citenamefont
  {Lee}(2022)}]{Ignatov:2022iou}%
  \BibitemOpen
  \bibfield  {author} {\bibinfo {author} {\bibfnamefont {F.}~\bibnamefont
  {Ignatov}} and \bibinfo {author} {\bibfnamefont {R.~N.}\ \bibnamefont
  {Lee}},\ }\href {\doibase 10.1016/j.physletb.2022.137283} {\bibfield
  {journal} {\bibinfo  {journal} {Phys. Lett. B}\ }\textbf {\bibinfo {volume}
  {833}},\ \bibinfo {pages} {137283} (\bibinfo {year} {2022})},\ \Eprint
  {http://arxiv.org/abs/2204.12235} {arXiv:2204.12235 [hep-ph]}\BibitemShut
  {NoStop}%
\bibitem [{\citenamefont {Monnard}(2020)}]{JMPhDThesis}%
  \BibitemOpen
  \bibfield  {author} {\bibinfo {author} {\bibfnamefont {J.}~\bibnamefont
  {Monnard}},\ }\href@noop {} {Ph.D. thesis},\ \bibinfo  {school} {Bern
  University} (\bibinfo {year} {2020}),\ \bibinfo {note}
  {\url{https://boristheses.unibe.ch/2825/}}\BibitemShut {NoStop}%
\bibitem [{\citenamefont {Abbiendi}\ \emph {et~al.}(2022)\citenamefont
  {Abbiendi} \emph {et~al.}}]{Abbiendi:2022liz}%
  \BibitemOpen
  \bibfield  {author} {\bibinfo {author} {\bibfnamefont {G.}~\bibnamefont
  {Abbiendi}}  \emph {et~al.},\ }\href@noop {} {\  (\bibinfo {year} {2022})},\
  \Eprint {http://arxiv.org/abs/2201.12102} {arXiv:2201.12102
  [hep-ph]}\BibitemShut {NoStop}%
\bibitem [{\citenamefont {Stamen}\ \emph {et~al.}(2023)\citenamefont {Stamen},
  \citenamefont {Isken}, \citenamefont {Kubis}, \citenamefont {Mikhasenko},\
  and\ \citenamefont {Niehus}}]{Stamen:2022eda}%
  \BibitemOpen
  \bibfield  {author} {\bibinfo {author} {\bibfnamefont {D.}~\bibnamefont
  {Stamen}}, \bibinfo {author} {\bibfnamefont {T.}~\bibnamefont {Isken}},
  \bibinfo {author} {\bibfnamefont {B.}~\bibnamefont {Kubis}}, \bibinfo
  {author} {\bibfnamefont {M.}~\bibnamefont {Mikhasenko}}, and \bibinfo
  {author} {\bibfnamefont {M.}~\bibnamefont {Niehus}},\ }\href {\doibase
  10.1140/epjc/s10052-023-11665-x} {\bibfield  {journal} {\bibinfo  {journal}
  {Eur. Phys. J. C}\ }\textbf {\bibinfo {volume} {83}},\ \bibinfo {pages} {510}
  (\bibinfo {year} {2023})},\ \bibinfo {note} {[Erratum: Eur. Phys. J. C {\bf
  83}, 586 (2023)]},\ \Eprint {http://arxiv.org/abs/2212.11767}
  {arXiv:2212.11767 [hep-ph]}\BibitemShut {NoStop}%
\bibitem [{\citenamefont {Hanhart}(2012)}]{Hanhart:2012wi}%
  \BibitemOpen
  \bibfield  {author} {\bibinfo {author} {\bibfnamefont {C.}~\bibnamefont
  {Hanhart}},\ }\href {\doibase 10.1016/j.physletb.2012.07.038} {\bibfield
  {journal} {\bibinfo  {journal} {Phys. Lett. B}\ }\textbf {\bibinfo {volume}
  {715}},\ \bibinfo {pages} {170} (\bibinfo {year} {2012})},\ \Eprint
  {http://arxiv.org/abs/1203.6839} {arXiv:1203.6839 [hep-ph]}\BibitemShut
  {NoStop}%
\bibitem [{\citenamefont {Ropertz}\ \emph {et~al.}(2018)\citenamefont
  {Ropertz}, \citenamefont {Hanhart},\ and\ \citenamefont
  {Kubis}}]{Ropertz:2018stk}%
  \BibitemOpen
  \bibfield  {author} {\bibinfo {author} {\bibfnamefont {S.}~\bibnamefont
  {Ropertz}}, \bibinfo {author} {\bibfnamefont {C.}~\bibnamefont {Hanhart}},
  and \bibinfo {author} {\bibfnamefont {B.}~\bibnamefont {Kubis}},\ }\href
  {\doibase 10.1140/epjc/s10052-018-6416-6} {\bibfield  {journal} {\bibinfo
  {journal} {Eur. Phys. J. C}\ }\textbf {\bibinfo {volume} {78}},\ \bibinfo
  {pages} {1000} (\bibinfo {year} {2018})},\ \Eprint
  {http://arxiv.org/abs/1809.06867} {arXiv:1809.06867 [hep-ph]}\BibitemShut
  {NoStop}%
\bibitem [{\citenamefont {von Detten}\ \emph {et~al.}(2021)\citenamefont {von
  Detten}, \citenamefont {No\"el}, \citenamefont {Hanhart}, \citenamefont
  {Hoferichter},\ and\ \citenamefont {Kubis}}]{VonDetten:2021rax}%
  \BibitemOpen
  \bibfield  {author} {\bibinfo {author} {\bibfnamefont {L.}~\bibnamefont {von
  Detten}}, \bibinfo {author} {\bibfnamefont {F.}~\bibnamefont {No\"el}},
  \bibinfo {author} {\bibfnamefont {C.}~\bibnamefont {Hanhart}}, \bibinfo
  {author} {\bibfnamefont {M.}~\bibnamefont {Hoferichter}}, and \bibinfo
  {author} {\bibfnamefont {B.}~\bibnamefont {Kubis}},\ }\href {\doibase
  10.1140/epjc/s10052-021-09169-7} {\bibfield  {journal} {\bibinfo  {journal}
  {Eur. Phys. J. C}\ }\textbf {\bibinfo {volume} {81}},\ \bibinfo {pages} {420}
  (\bibinfo {year} {2021})},\ \Eprint {http://arxiv.org/abs/2103.01966}
  {arXiv:2103.01966 [hep-ph]}\BibitemShut {NoStop}%
\bibitem [{\citenamefont {Omn{\`e}s}(1958)}]{Omnes:1958hv}%
  \BibitemOpen
  \bibfield  {author} {\bibinfo {author} {\bibfnamefont {R.}~\bibnamefont
  {Omn{\`e}s}},\ }\href {\doibase 10.1007/BF02747746} {\bibfield  {journal}
  {\bibinfo  {journal} {Nuovo Cim.}\ }\textbf {\bibinfo {volume} {8}},\
  \bibinfo {pages} {316} (\bibinfo {year} {1958})}\BibitemShut {NoStop}%
\bibitem [{\citenamefont {Sakurai}(1969)}]{Sakurai:1969}%
  \BibitemOpen
  \bibfield  {author} {\bibinfo {author} {\bibfnamefont {J.~J.}\ \bibnamefont
  {Sakurai}},\ }\href@noop {} {\emph {\bibinfo {title} {{Currents and
  Mesons}}}}\ (\bibinfo  {publisher} {University of Chicago Press},\ \bibinfo
  {address} {Chicago},\ \bibinfo {year} {1969})\BibitemShut {NoStop}%
\bibitem [{\citenamefont {Klingl}\ \emph {et~al.}(1996)\citenamefont {Klingl},
  \citenamefont {Kaiser},\ and\ \citenamefont {Weise}}]{Klingl:1996by}%
  \BibitemOpen
  \bibfield  {author} {\bibinfo {author} {\bibfnamefont {F.}~\bibnamefont
  {Klingl}}, \bibinfo {author} {\bibfnamefont {N.}~\bibnamefont {Kaiser}}, and
  \bibinfo {author} {\bibfnamefont {W.}~\bibnamefont {Weise}},\ }\href
  {\doibase 10.1007/s002180050167} {\bibfield  {journal} {\bibinfo  {journal}
  {Z. Phys. A}\ }\textbf {\bibinfo {volume} {356}},\ \bibinfo {pages} {193}
  (\bibinfo {year} {1996})},\ \Eprint {http://arxiv.org/abs/hep-ph/9607431}
  {arXiv:hep-ph/9607431}\BibitemShut {NoStop}%
\bibitem [{\citenamefont {Bouchiat}\ and\ \citenamefont
  {Michel}(1961)}]{Bouchiat:1961lbg}%
  \BibitemOpen
  \bibfield  {author} {\bibinfo {author} {\bibfnamefont {C.}~\bibnamefont
  {Bouchiat}} and \bibinfo {author} {\bibfnamefont {L.}~\bibnamefont
  {Michel}},\ }\href {\doibase 10.1051/jphysrad:01961002202012101} {\bibfield
  {journal} {\bibinfo  {journal} {J. Phys. Radium}\ }\textbf {\bibinfo {volume}
  {22}},\ \bibinfo {pages} {121} (\bibinfo {year} {1961})}\BibitemShut
  {NoStop}%
\bibitem [{\citenamefont {Brodsky}\ and\ \citenamefont
  {de~Rafael}(1968)}]{Brodsky:1967sr}%
  \BibitemOpen
  \bibfield  {author} {\bibinfo {author} {\bibfnamefont {S.~J.}\ \bibnamefont
  {Brodsky}} and \bibinfo {author} {\bibfnamefont {E.}~\bibnamefont
  {de~Rafael}},\ }\href {\doibase 10.1103/PhysRev.168.1620} {\bibfield
  {journal} {\bibinfo  {journal} {Phys. Rev.}\ }\textbf {\bibinfo {volume}
  {168}},\ \bibinfo {pages} {1620} (\bibinfo {year} {1968})}\BibitemShut
  {NoStop}%
\bibitem [{\citenamefont {Aul'chenko}\ \emph {et~al.}(2015)\citenamefont
  {Aul'chenko} \emph {et~al.}}]{Aulchenko:2015mwt}%
  \BibitemOpen
  \bibfield  {author} {\bibinfo {author} {\bibfnamefont {V.~M.}\ \bibnamefont
  {Aul'chenko}}  \emph {et~al.} (\bibinfo {collaboration} {SND}),\ }\href
  {\doibase 10.1134/S1063776115060023} {\bibfield  {journal} {\bibinfo
  {journal} {J. Exp. Theor. Phys.}\ }\textbf {\bibinfo {volume} {121}},\
  \bibinfo {pages} {27} (\bibinfo {year} {2015})},\ \bibinfo {note} {[Zh. Eksp.
  Teor. Fiz. {\bf 148}, 34 (2015)]}\BibitemShut {NoStop}%
\bibitem [{\citenamefont {Achasov}\ \emph {et~al.}(2020)\citenamefont {Achasov}
  \emph {et~al.}}]{SND:2020ajg}%
  \BibitemOpen
  \bibfield  {author} {\bibinfo {author} {\bibfnamefont {M.~N.}\ \bibnamefont
  {Achasov}}  \emph {et~al.} (\bibinfo {collaboration} {SND}),\ }\href
  {\doibase 10.1140/epjc/s10052-020-08524-4} {\bibfield  {journal} {\bibinfo
  {journal} {Eur. Phys. J. C}\ }\textbf {\bibinfo {volume} {80}},\ \bibinfo
  {pages} {993} (\bibinfo {year} {2020})},\ \Eprint
  {http://arxiv.org/abs/2007.14595} {arXiv:2007.14595 [hep-ex]}\BibitemShut
  {NoStop}%
\bibitem [{\citenamefont {Aubert}\ \emph {et~al.}(2004)\citenamefont {Aubert}
  \emph {et~al.}}]{BaBar:2004ytv}%
  \BibitemOpen
  \bibfield  {author} {\bibinfo {author} {\bibfnamefont {B.}~\bibnamefont
  {Aubert}}  \emph {et~al.} (\bibinfo {collaboration} {BaBar}),\ }\href
  {\doibase 10.1103/PhysRevD.70.072004} {\bibfield  {journal} {\bibinfo
  {journal} {Phys. Rev. D}\ }\textbf {\bibinfo {volume} {70}},\ \bibinfo
  {pages} {072004} (\bibinfo {year} {2004})},\ \Eprint
  {http://arxiv.org/abs/hep-ex/0408078} {arXiv:hep-ex/0408078}\BibitemShut
  {NoStop}%
\bibitem [{\citenamefont {Achasov}\ \emph {et~al.}(2001)\citenamefont {Achasov}
  \emph {et~al.}}]{Achasov:2000am}%
  \BibitemOpen
  \bibfield  {author} {\bibinfo {author} {\bibfnamefont {M.~N.}\ \bibnamefont
  {Achasov}}  \emph {et~al.} (\bibinfo {collaboration} {SND}),\ }\href
  {\doibase 10.1103/PhysRevD.63.072002} {\bibfield  {journal} {\bibinfo
  {journal} {Phys. Rev. D}\ }\textbf {\bibinfo {volume} {63}},\ \bibinfo
  {pages} {072002} (\bibinfo {year} {2001})},\ \Eprint
  {http://arxiv.org/abs/hep-ex/0009036} {arXiv:hep-ex/0009036}\BibitemShut
  {NoStop}%
\bibitem [{\citenamefont {Achasov}\ \emph {et~al.}(2002)\citenamefont {Achasov}
  \emph {et~al.}}]{Achasov:2002ud}%
  \BibitemOpen
  \bibfield  {author} {\bibinfo {author} {\bibfnamefont {M.~N.}\ \bibnamefont
  {Achasov}}  \emph {et~al.} (\bibinfo {collaboration} {SND}),\ }\href
  {\doibase 10.1103/PhysRevD.66.032001} {\bibfield  {journal} {\bibinfo
  {journal} {Phys. Rev. D}\ }\textbf {\bibinfo {volume} {66}},\ \bibinfo
  {pages} {032001} (\bibinfo {year} {2002})},\ \Eprint
  {http://arxiv.org/abs/hep-ex/0201040} {arXiv:hep-ex/0201040}\BibitemShut
  {NoStop}%
\bibitem [{\citenamefont {Achasov}\ \emph {et~al.}(2003)\citenamefont {Achasov}
  \emph {et~al.}}]{Achasov:2003ir}%
  \BibitemOpen
  \bibfield  {author} {\bibinfo {author} {\bibfnamefont {M.~N.}\ \bibnamefont
  {Achasov}}  \emph {et~al.} (\bibinfo {collaboration} {SND}),\ }\href
  {\doibase 10.1103/PhysRevD.68.052006} {\bibfield  {journal} {\bibinfo
  {journal} {Phys. Rev. D}\ }\textbf {\bibinfo {volume} {68}},\ \bibinfo
  {pages} {052006} (\bibinfo {year} {2003})},\ \Eprint
  {http://arxiv.org/abs/hep-ex/0305049} {arXiv:hep-ex/0305049}\BibitemShut
  {NoStop}%
\bibitem [{\citenamefont {Akhmetshin}\ \emph {et~al.}(1995)\citenamefont
  {Akhmetshin} \emph {et~al.}}]{Akhmetshin:1995vz}%
  \BibitemOpen
  \bibfield  {author} {\bibinfo {author} {\bibfnamefont {R.~R.}\ \bibnamefont
  {Akhmetshin}}  \emph {et~al.} (\bibinfo {collaboration} {CMD-2}),\ }\href
  {\doibase 10.1016/0370-2693(95)01394-6} {\bibfield  {journal} {\bibinfo
  {journal} {Phys. Lett. B}\ }\textbf {\bibinfo {volume} {364}},\ \bibinfo
  {pages} {199} (\bibinfo {year} {1995})}\BibitemShut {NoStop}%
\bibitem [{\citenamefont {Akhmetshin}\ \emph {et~al.}(1998)\citenamefont
  {Akhmetshin} \emph {et~al.}}]{Akhmetshin:1998se}%
  \BibitemOpen
  \bibfield  {author} {\bibinfo {author} {\bibfnamefont {R.~R.}\ \bibnamefont
  {Akhmetshin}}  \emph {et~al.} (\bibinfo {collaboration} {CMD-2}),\ }\href
  {\doibase 10.1016/S0370-2693(98)00826-0} {\bibfield  {journal} {\bibinfo
  {journal} {Phys. Lett. B}\ }\textbf {\bibinfo {volume} {434}},\ \bibinfo
  {pages} {426} (\bibinfo {year} {1998})}\BibitemShut {NoStop}%
\bibitem [{\citenamefont {Akhmetshin}\ \emph {et~al.}(2004)\citenamefont
  {Akhmetshin} \emph {et~al.}}]{Akhmetshin:2003zn}%
  \BibitemOpen
  \bibfield  {author} {\bibinfo {author} {\bibfnamefont {R.~R.}\ \bibnamefont
  {Akhmetshin}}  \emph {et~al.} (\bibinfo {collaboration} {CMD-2}),\ }\href
  {\doibase 10.1016/j.physletb.2003.10.108} {\bibfield  {journal} {\bibinfo
  {journal} {Phys. Lett. B}\ }\textbf {\bibinfo {volume} {578}},\ \bibinfo
  {pages} {285} (\bibinfo {year} {2004})},\ \Eprint
  {http://arxiv.org/abs/hep-ex/0308008} {arXiv:hep-ex/0308008}\BibitemShut
  {NoStop}%
\bibitem [{\citenamefont {Akhmetshin}\ \emph {et~al.}(2006)\citenamefont
  {Akhmetshin} \emph {et~al.}}]{Akhmetshin:2006sc}%
  \BibitemOpen
  \bibfield  {author} {\bibinfo {author} {\bibfnamefont {R.~R.}\ \bibnamefont
  {Akhmetshin}}  \emph {et~al.} (\bibinfo {collaboration} {CMD-2}),\ }\href
  {\doibase 10.1016/j.physletb.2006.09.041} {\bibfield  {journal} {\bibinfo
  {journal} {Phys. Lett. B}\ }\textbf {\bibinfo {volume} {642}},\ \bibinfo
  {pages} {203} (\bibinfo {year} {2006})}\BibitemShut {NoStop}%
\bibitem [{\citenamefont {Cordier}\ \emph {et~al.}(1980)\citenamefont {Cordier}
  \emph {et~al.}}]{Cordier:1979qg}%
  \BibitemOpen
  \bibfield  {author} {\bibinfo {author} {\bibfnamefont {A.}~\bibnamefont
  {Cordier}}  \emph {et~al.} (\bibinfo {collaboration} {DM1}),\ }\href
  {\doibase 10.1016/0550-3213(80)90157-1} {\bibfield  {journal} {\bibinfo
  {journal} {Nucl. Phys. B}\ }\textbf {\bibinfo {volume} {172}},\ \bibinfo
  {pages} {13} (\bibinfo {year} {1980})}\BibitemShut {NoStop}%
\bibitem [{\citenamefont {Antonelli}\ \emph {et~al.}(1992)\citenamefont
  {Antonelli} \emph {et~al.}}]{DM2:1992zkc}%
  \BibitemOpen
  \bibfield  {author} {\bibinfo {author} {\bibfnamefont {A.}~\bibnamefont
  {Antonelli}}  \emph {et~al.} (\bibinfo {collaboration} {DM2}),\ }\href
  {\doibase 10.1007/BF01589702} {\bibfield  {journal} {\bibinfo  {journal} {Z.
  Phys. C}\ }\textbf {\bibinfo {volume} {56}},\ \bibinfo {pages} {15} (\bibinfo
  {year} {1992})}\BibitemShut {NoStop}%
\bibitem [{\citenamefont {Dolinsky}\ \emph {et~al.}(1991)\citenamefont
  {Dolinsky} \emph {et~al.}}]{Dolinsky:1991vq}%
  \BibitemOpen
  \bibfield  {author} {\bibinfo {author} {\bibfnamefont {S.~I.}\ \bibnamefont
  {Dolinsky}}  \emph {et~al.},\ }\href {\doibase 10.1016/0370-1573(91)90127-8}
  {\bibfield  {journal} {\bibinfo  {journal} {Phys. Rept.}\ }\textbf {\bibinfo
  {volume} {202}},\ \bibinfo {pages} {99} (\bibinfo {year} {1991})}\BibitemShut
  {NoStop}%
\bibitem [{\citenamefont {D'Agostini}(1994)}]{DAgostini:1993arp}%
  \BibitemOpen
  \bibfield  {author} {\bibinfo {author} {\bibfnamefont {G.}~\bibnamefont
  {D'Agostini}},\ }\href {\doibase 10.1016/0168-9002(94)90719-6} {\bibfield
  {journal} {\bibinfo  {journal} {Nucl. Instrum. Meth. A}\ }\textbf {\bibinfo
  {volume} {346}},\ \bibinfo {pages} {306} (\bibinfo {year}
  {1994})}\BibitemShut {NoStop}%
\bibitem [{\citenamefont {Ball}\ \emph {et~al.}(2010)\citenamefont {Ball},
  \citenamefont {Del~Debbio}, \citenamefont {Forte}, \citenamefont {Guffanti},
  \citenamefont {Latorre}, \citenamefont {Rojo},\ and\ \citenamefont
  {Ubiali}}]{Ball:2009qv}%
  \BibitemOpen
  \bibfield  {author} {\bibinfo {author} {\bibfnamefont {R.~D.}\ \bibnamefont
  {Ball}}, \bibinfo {author} {\bibfnamefont {L.}~\bibnamefont {Del~Debbio}},
  \bibinfo {author} {\bibfnamefont {S.}~\bibnamefont {Forte}}, \bibinfo
  {author} {\bibfnamefont {A.}~\bibnamefont {Guffanti}}, \bibinfo {author}
  {\bibfnamefont {J.~I.}\ \bibnamefont {Latorre}}, \bibinfo {author}
  {\bibfnamefont {J.}~\bibnamefont {Rojo}}, and \bibinfo {author}
  {\bibfnamefont {M.}~\bibnamefont {Ubiali}} (\bibinfo {collaboration}
  {NNPDF}),\ }\href {\doibase 10.1007/JHEP05(2010)075} {\bibfield  {journal}
  {\bibinfo  {journal} {JHEP}\ }\textbf {\bibinfo {volume} {05}},\ \bibinfo
  {pages} {075} (\bibinfo {year} {2010})},\ \Eprint
  {http://arxiv.org/abs/0912.2276} {arXiv:0912.2276 [hep-ph]}\BibitemShut
  {NoStop}%
\bibitem [{\citenamefont {Amsler}\ \emph {et~al.}(1993)\citenamefont {Amsler}
  \emph {et~al.}}]{CrystalBarrel:1993gtk}%
  \BibitemOpen
  \bibfield  {author} {\bibinfo {author} {\bibfnamefont {C.}~\bibnamefont
  {Amsler}}  \emph {et~al.} (\bibinfo {collaboration} {Crystal Barrel}),\
  }\href {\doibase 10.1016/0370-2693(93)90583-4} {\bibfield  {journal}
  {\bibinfo  {journal} {Phys. Lett. B}\ }\textbf {\bibinfo {volume} {311}},\
  \bibinfo {pages} {362} (\bibinfo {year} {1993})}\BibitemShut {NoStop}%
\bibitem [{\citenamefont {Hoferichter}\ \emph
  {et~al.}(2023{\natexlab{b}})\citenamefont {Hoferichter}, \citenamefont
  {Colangelo}, \citenamefont {Hoid}, \citenamefont {Kubis}, \citenamefont
  {Ruiz~de Elvira}, \citenamefont {Schuh}, \citenamefont {Stamen},\ and\
  \citenamefont {Stoffer}}]{Hoferichter:2023sli}%
  \BibitemOpen
  \bibfield  {author} {\bibinfo {author} {\bibfnamefont {M.}~\bibnamefont
  {Hoferichter}}, \bibinfo {author} {\bibfnamefont {G.}~\bibnamefont
  {Colangelo}}, \bibinfo {author} {\bibfnamefont {B.-L.}\ \bibnamefont {Hoid}},
  \bibinfo {author} {\bibfnamefont {B.}~\bibnamefont {Kubis}}, \bibinfo
  {author} {\bibfnamefont {J.}~\bibnamefont {Ruiz~de Elvira}}, \bibinfo
  {author} {\bibfnamefont {D.}~\bibnamefont {Schuh}}, \bibinfo {author}
  {\bibfnamefont {D.}~\bibnamefont {Stamen}}, and \bibinfo {author}
  {\bibfnamefont {P.}~\bibnamefont {Stoffer}},\ }\href@noop {} {\  (\bibinfo
  {year} {2023}{\natexlab{b}})},\ \Eprint {http://arxiv.org/abs/2307.02532}
  {arXiv:2307.02532 [hep-ph]}\BibitemShut {NoStop}%
\end{thebibliography}%
	
\end{document}